\documentclass[a4paper,10pt]{article}

\usepackage{amsfonts,amssymb}
\usepackage{amsmath}
\makeatletter\@addtoreset{equation}{section}\makeatother

\newtheorem{theorem}{Theorem}[section]
\newtheorem{corollary}[theorem]{Corollary}
\newtheorem{lemma}[theorem]{Lemma}

\newtheorem{proposition}[theorem]{Proposition}

\newtheorem{definition}[theorem]{Definition}
\newtheorem{remark}[theorem]{Remark}
\numberwithin{equation}{section}

\title{Phase Transitions and Quantum  Stabilization \\ in  Quantum Anharmonic Crystals\\}
\author{ALINA KARGOL\\[.6cm]
Instytut Matematyki, Uniwersytet Marii Curie-Sk{\l}odowskiej\\[.3cm]
20-031 Lublin, Poland\\[.3cm]
akargol@golem.umcs.lublin.pl \\[.8cm]
YURI KONDRATIEV\\[.6cm]
Fakult\"at f\"ur Mathematik, Universit\"at Bielefeld\\[.3cm]
D-33615 Bielefeld, Germany\\[.3cm]
kondrat@math.uni-bielefeld.de\\[.8cm]
YURI KOZITSKY\\[.6cm]
Instytut Matematyki, Uniwersytet Marii Curie-Sk{\l}odowskiej\\[.3cm]
20-031 Lublin, Poland\\[.3cm]
jkozi@golem.umcs.lublin.pl}

\begin{document}

\maketitle

\begin{abstract}
A unified theory of phase transitions and quantum effects in quantum
anharmonic crystals is presented. In its framework, the relationship
between these two phenomena is analyzed. The theory is based on the
representation of the model Gibbs states in terms of path measures
(Euclidean Gibbs measures). It covers the case of crystals without
translation invariance, as well as the case of asymmetric anharmonic
potentials. The results obtained are compared with those known in
the literature.
\end{abstract}

\tableofcontents

\section{Introduction and Setup}

\label{1s}

In recent years, there appeared a number of publications
 describing infuence of quantum effects on phase transitions in quantum
anharmonic crystals, where the results were obtained by means of
path integrals, see
\cite{[AKK2],[AKKR],[AKKRPRL],[AKKRNN],[KK],Koz,[Koz4],[Minlos],[RZ],[VZ]}. Their common point is a statement that the phase transition (understood in one or another way) is suppressed if the model parameters obey a sufficient condition (more or less explicitely formulated).
The existence of phase transitions in quantum crystals of certain types
was proven earlier, see \cite{[BaK],[BaK0],[DLP],[Kondr],[Pastur]}, also mostly  by means of path integral methods.
At the same time, by now only two works, \cite{[RevMF]} and \cite{[KoT]}, have apperared  where both these phenomena are studied in one and the same context.
In the latter paper,
a more complete and extended version of the
theory of interacting systems of quantum anharmonic oscillators
based on path integral methods has been elaborated, see also \cite{[AKPR],[AKPR1],[AKPR2]} for more recent development,  and \cite{[KoT1]} where the results of \cite{[KoT]} were announced.
 The aim of the present
article is to refine and extend the previous results and to develop a
unified and more or less complete theory of phase transitions and quantum effects in quantum
anharmonic crystals, also in the light of the results of
\cite{[KoT1],[KoT]}. Note that, in particular, with the help of these results we prove here phase transitions
in quantum crystals with asymmetric anharmonic potentials\footnote{This result was announced in \cite{[KaK]}.}, what could hardly be done by other  methods.

The quantum crystal  studied in this article is a system of
interacting quantum anharmonic oscillators indexed by the elements
of a crystal lattice $\mathbb{L}$, which for simplicity we assume to
be a $d$-dimensional simple cubic lattice $\mathbb{Z}^d$. The
quantum anharmonic oscillator is a mathematical model of a quantum
particle moving in a potential field with possibly multiple minima,
which has a sufficient growth at infinity and hence localizes the
particle. Most of the models  of interacting quantum oscillators are
related with solids such as ionic crystals containing localized
light particles oscillating in the field created by heavy ionic
complexes, or quantum crystals consisting entirely of such
particles. For instance, a potential field with multiple minima is
seen by a helium atom located at the center of the crystal cell in
bcc helium, see page 11 in \cite{[Koeler]}. The same situation
exists in other quantum crystals, ${\rm He}$, ${\rm H}_2$ and to
some extent ${\rm Ne}$. An example of the ionic crystal with
localized quantum particles moving in a double-well potential field
is a ${\rm KDP}$-type ferroelectric with hydrogen bounds, in which
such particles are protons or deuterons performing one-dimensional
oscillations along the bounds, see
\cite{[Blinc],[S],[Tokunaga],[Vaks]}. It is believed that in such
substances phase transitions are
triggered by the  ordering of protons.
 Another relevant physical object of this
kind is a system of apex oxygen ions in YBaCuO-type high-temperature
superconductors, see \cite{[Frick],[KMueller],[StasT],[StasT1]}.
Quantum anharmonic oscillators
 are also used in models
describing interaction of vibrating quantum particles with a
radiation (photon) field, see \cite{[Hainzl],[HHS],[Osada]}, or
strong electron-electron correlations caused by the interaction of
electrons with vibrating ions, see \cite{[Freericks],[FreerL]},
responsible for such phenomena as superconductivity, charge density
waves etc. Finally, we mention systems of light atoms, like ${\rm
Li}$, doped into ionic crystals, like ${\rm KCl}$. The quantum
particles in this system are not necessarily regularly distributed.
For more information on this subject, we refer to the survey
\cite{[Horner]}.

To be more concrete we assume that our model describes an ionic
crystal and thus adopt the ferroelectric terminology. In the
corresponding physical substances, the quantum particles carry
electric charge; hence, the displacement of the particle from its
equilibrium point produces dipole moment. Therefore, the main
contribution into the two-particle interaction is proportional to
the product of the displacements of particles and is of long range.
According to these arguments our model  is described by the
following formal Hamiltonian
\begin{equation} \label{U1}
H = - \frac{1}{2} \sum_{\ell,\ell'} J_{\ell\ell'} \cdot(q_{\ell} ,
q_{\ell'}) + \sum_{\ell} H_{\ell}.
\end{equation}
Here the sums run through the lattice  $\mathbb{L}=\mathbb{Z}^d$,
$d\in \mathbb{N}$, the displacement, $q_\ell$, of the oscillator
attached to a given $\ell\in \mathbb{L}$ is a $\nu$-dimensional
vector. In general, we do not assume that the interaction intensities $J_{\ell\ell'}$ have finite range. By $(\cdot, \cdot)$ and $|\cdot|$ we denote the scalar
product and norm in $\mathbb{R}^\nu$, $\mathbb{R}^d$. The one-site
Hamiltonian
\begin{equation} \label{U2} H_\ell =
H_{\ell}^{\rm har} +  V_\ell (q_\ell ) \ \stackrel{\rm def}{=}\
\frac{1}{2m} |p_\ell|^2 + \frac{a}{2} |q_\ell|^2 + V_\ell (q_\ell ),
\quad a>0,
\end{equation}
 describes an isolated quantum anharmonic oscillator. Its part $H^{\rm har}_\ell$ corresponds
to a $\nu$-dimensional harmonic oscillator of rigidity $a$. The mass
parameter $m$ includes Planck's constant, that is,
\begin{equation} \label{In} m = m_{\rm ph}/\hbar^2,
\end{equation}
where $m_{\rm ph}$ is the physical mass of the particle. Therefore,
the commutation relation for the components of the momentum  and displacement
takes the form
\begin{equation} \label{cr}
p_\ell^{(j)} q_{\ell'}^{(j')} - q_{\ell'}^{(j')} p_\ell^{(j)} = -
\imath \delta_{\ell \ell'} \delta_{jj'}, \quad j, j' = 1, \dots ,
\nu.
\end{equation}
 For a
detailed discussion on how to derive a model like (\ref{U1}),
(\ref{U2}) from physical models of concrete substances, we refer the
reader to the survey \cite{[S]}.

The theory of phase transitions is one of the most important and
spectacular parts of  equilibrium statistical mechanics. For
classical lattice models, a complete description of the equilibrium
thermodynamic properties is given by constructing their Gibbs states
as probability measures on appropriate configuration spaces.
Usually, it is made in the Dobrushin-Lanford-Ruelle (DLR) approach
which is now well-elaborated, see Georgii's monograph  \cite{[Ge]}
and the references therein. In general, the quantum case does not
permit such a universal description. For some systems with bounded
one-site Hamiltonians, e.g., quantum spin models, the Gibbs states
are defined as positive normalized functionals on algebras of
quasi-local observables obeyng the condition of equilibrium between
the dynamic and thermodynamic behavior of the model (KMS condition),
see \cite{[BrR]}. However, this algebraic way cannot be applied to
the model (\ref{U1}), (\ref{U2}) since the construction of its
dynamics in  the whole crystal $\mathbb{L}$ is  beyond the technical
possibilities available  by this time. In 1975, an approach
employing path integral methods to describe thermodynamic properties
of models like (\ref{U1}), (\ref{U2}) has been initiated in
\cite{[AH-K]}. Its main idea was to pass from real to imaginary
values of time, similarly as it was done in Euclidean quantum field
theory, see \cite{[GJ],[Sim1]}, and thereby to describe the dynamics
of the model in terms of stochastic processes. Afterwards, this
approach, also called Euclidean, has been developed in a number of
works. Its latest and most general version is presented in
\cite{[KoT1],[KoT]}, where the reader can also find an extensive
bibliography on this subject. The methods developed in these works
will be extensively used in the present study.

Phase transitions are very important  phenomena in the substances
modeled by the Hamiltonians (\ref{U1}), (\ref{U2}). According to their commonly
adopted  physical interpretation, at low
temperatures  the oscillations
of the particles become strongly correlated  that produces
macroscopic ordering. The mathematical theory of phase transitions in
models like (\ref{U1}), (\ref{U2}) is based on quantum versions of
the method of infrared estimates developed in \cite{[FSS]}. The
first publication where the infrared estimates were applied to
quantum spin models seems to be \cite{[DLS]}. After
certain modifications this method, combined with path integral techniques,
was applied in \cite{[BaK],[BaK0],[DLP],[Kondr],[Pastur]} to particular versions
of our model. The
main characteristic feature of these versions was a symmetry, broken
by the phase transition.

In classical systems, ordering  is achieved in competition with
thermal fluctuations only.
 However, in quantum systems  quantum
 effects play a significant disordering role, especially at low
temperatures. This role was first discussed in \cite{[9]}. Later on
a number of publications dedicated to the study of quantum effects
in such systems  had appeared, see e.g., \cite{[Minlos],[VZ]} and
the references therein. For better understanding, illuminating
exactly solvable models of systems of interacting quantum anharmonic
oscillators were introduced and studied, see
\cite{[Plakida1],[STZ],[VZ1],[VZ2]}. In these works, the quantity
$m^{-1} = \hbar^2 / m_{\rm ph}$ was used as a parameter describing
the rate of quantum effects. Such effects became strong in the small
mass limit, which was in agreement with the experimental data, e.g.,
on the isotopic effect in the ferroelectrics with hydrogen bounds,
see \cite{[Blinc],[Vaks]}, see also \cite{[KMueller]} for the data
on the isotopic effect in the YBaCuO-type high-temperature
superconductors. However, in those works no other
quantum effects, e.g., those connected with special properties of the anharmonic potentials, were discussed.
At the same time,
experimental data, see e.g., the table on page 11 in the monograph
\cite{[Blinc]} or the article \cite{[12]}, show that high
hydrostatic pressure applied to KDP-type ferroelectrics prevents
them from ordering. It is believed
that the pressure shortens the hydrogen bounds and thereby changes the anharmonic potential. This
makes the
tunneling motion of the quantum particles more intensive, which is equivalent to diminishing the particle mass. In
\cite{[AKKR],[AKKRPRL],[AKKRNN]}, a  theory of such quantum effects
in the model (\ref{U1}), (\ref{U2}), which explains both mentioned
mechanisms, was buit up. Its main conclusion is that the quantum
dynamical properties, which depend on the mass $m$, the interaction
intensities $J_{\ell\ell'}$, and the anharmonic potentials $V_\ell$,
can be such that the model is stable with respect to phase
transitions at all temperatures.

As was mentioned above, the aim of this article is to present a
 unified description of phase transitions and quantum
stabilization in the model (\ref{U1}), (\ref{U2}), mostly by means of methods developed in \cite{[KoT1],[KoT]}.
We also give here complete proofs of a number of statements announced in
our previous publications. The article is organized as follows. In
Section \ref{2s}, we briefly describe those elements of the
theory developed in \cite{[KoT1],[KoT]} which we then apply in the
subsequent sections. In Section \ref{3s}, we present the theory of
phase transitions in the model (\ref{U1}), (\ref{U2}). We begin by
introducing three definitions of a phase transition in this model
and study the relationships between them. Then we develop a version
of the method of infrared estimates adapted to our model, which is
more transpatrent and appropriate than the one employed in
\cite{[RevMF]}. Afterwards, we obtain a sufficient conditions for the
phase transitions to occur in a number of versions of the model
(\ref{U1}), (\ref{U2}).
This includes also the case of asymmetric anharmonic potentials
$V_\ell$ which was never studied before.
At the end of the section we make some
comments on the results obtained and compare them with similar
results known in the literature. Section \ref{4s} is dedicated to the study of quantum
stabilization, which we understand as the suppression of phase transitions by quantum  effects. Here we discuss the problem of stability of quantum
crystals and the ways of its description. In particular, we
introduce a parameter (quantum rigidity), responsible for the
stability and prove a number of statements about its properties.
Then we show that under the stability condition which we introduce
here the correlations decay `in a proper way', that means the
absence of phase transitions. The relationship between the quantum
stabilization and phase transitions are also analyzed.
In the simplest case, where the model is translation invariant, scalar ($\nu =1$), and with the interaction of nearest neighbor type, this relation looks as follows. The key parameter is $ 8 d m J \vartheta_*^2$, where $d$ is the lattice dimension, $J>0$ is the interaction intensity, and $\vartheta_*>0$ is determined by the anharmonic potential $V$ (the steeper is $V$ the smaller is $\vartheta_*$). Then the quantum stabilization condition (respectively, the phase transition condition) is $ 8 d m J \vartheta_*^2 <1$, see (\ref{De20}), (respectively,
$ 8 d m J \vartheta_*^2 > \phi(d)$, see (\ref{rp52}) and (\ref{DeE})). Here $\phi$ is a function, such that $\phi(d) >1$ and $\phi(d)\rightarrow 1$ as $d \rightarrow + \infty$.
We conclude
the section  by commenting the results obtained therein.

\section{Euclidean Gibbs States}

\label{2s}

The main element of the Euclidean approach is
the description of the equilibrium thermodynamic properties of the
model (\ref{U1}), (\ref{U2}) by means of Euclidean Gibbs states,
which are probability measures on certain configuration spaces.
In this section, we briefly describe the main elements of this approach which are then used in the subsequent parts of the article. For more details, we refer to \cite{[KoT]}.

\subsection{Local Gibbs states}
\label{2.1ss}

Let us begin by specifying the properties of the model described by
the Hamiltonian (\ref{U1}). The general assumptions regarding the
interaction intensities $J_{\ell \ell'}$ are
\begin{equation} \label{a1}
J_{\ell\ell'}= J_{\ell'\ell} \geq 0, \quad J_{\ell\ell}=0, \quad
\hat{J}_0 \ \stackrel{\rm def}{=} \ \sup_{\ell} \sum_{\ell'} J_{\ell
\ell'} < \infty.
\end{equation}
In view of the first of these properties the model is {\it
ferroelectric}. Regarding the anharmonic potentials we assume that
each $V_\ell:\mathbb{R}^\nu \rightarrow \mathbb{R}$ is a continuous
function, which obeys
\begin{equation}
\label{a2} A_V |x|^{2r} + B_V \leq V_\ell (x) \leq V(x),
\end{equation}
with a continuous function $V$ and constants  $r>1$,
$A_V>0$, $B_V\in \mathbb{R}$. In certain cases, we shall include
 an external field term in the form
\begin{equation} \label{4}
V_\ell (x) = V_\ell^{0} (x) - (h , x), \ \  \quad h \in
\mathbb{R}^\nu,
\end{equation}
where $V_\ell^{0}$ is an appropriate function.
\begin{definition} \label{1df}
The model is translation invariant if  $V_\ell = V$ for all $\ell$,
and the interaction intensities $J_{\ell \ell'}$ are invariant under
the translations of $\mathbb{L}$. The model is rotation invariant if
for every orthogonal transformation $U\in O(\nu)$ and every $\ell$,
$V_\ell (U x) = V_\ell(x)$.  The interaction has finite range if
there exists $R>0$ such that $J_{\ell\ell'} = 0$ whenever
$|\ell-\ell'|>R$.
\end{definition}
\noindent If $V_\ell \equiv 0$ for all  $\ell$, one gets a  quantum
harmonic crystal. It is stable if $\hat{J}_0 < a$, see Remark
\ref{apprm} below.

By $\Lambda$ we denote subsets of the lattice $\mathbb{L}$; we write
$\Lambda \Subset \mathbb{L}$ if $\Lambda$ is non-void and finite.
For such $\Lambda$, by $|\Lambda|$ we denote its cardinality. A sequence of subsets $\Lambda \Subset \mathbb{L}$ is called {\it cofinal} if it is ordered by inclusion and exhausts the lattice $\mathbb{L}$.
If we
say that something holds for all $\ell$, we mean it holds for all
$\ell \in \mathbb{L}$; sums like $\sum_{\ell}$ mean $\sum_{\ell \in
\mathbb{L}}$. We also use the notations $\mathbb{R}^+ = [0,
+\infty)$ and $\mathbb{N}_0 = \mathbb{N}\cup \{0\}$, $\mathbb{N}$
being the set of positive integers.

Given $\Lambda \Subset \mathbb{L}$, the local Hamiltonian of the
model is
\begin{equation} \label{a3}
H_\Lambda = - \frac{1}{2} \sum_{\ell ,\ell'\in \Lambda} J_{\ell
\ell'} \cdot (q_\ell , q_{\ell'}) + \sum_{\ell \in \Lambda}H_\ell,
\end{equation}
which by the assumptions made above is a self-adjoint and lower bounded
operator in the physical Hilbert space $L^2 (\mathbb{R}^{\nu
|\Lambda|})$. For every $\beta= 1/ k_{\rm B} T$, $T$ being absolute
temperature, the local Gibbs state in $\Lambda\Subset \mathbb{L}$ is
\begin{equation}
\label{a4} \varrho_\Lambda (A) = {\rm trace}[A\exp(- \beta
H_\Lambda)]/ Z_\Lambda, \quad A\in \mathfrak{C}_\Lambda,
\end{equation}
 where
\begin{equation}
\label{a5} Z_\Lambda = {\rm trace}[\exp(- \beta H_\Lambda)]< \infty
\end{equation}
is the partition function, and $\mathfrak{C}_\Lambda$ is the algebra
of all bounded linear operators on $L^2 (\mathbb{R}^{\nu
|\Lambda|})$. Note that adjective {\it local} will always stand for
a property related with a certain $\Lambda \Subset \mathbb{L}$,
whereas {\it global} will characterize the whole infinite system.

The dynamics of the subsystem located in $\Lambda$ is described by
the time automorphisms
\begin{equation}
\label{a6} \mathfrak{C}_\Lambda \ni A \mapsto \mathfrak{a}_t^\Lambda
(A) = \exp(\imath t H_\Lambda) A \exp(-\imath t H_\Lambda),
\end{equation}
where $t\in \mathbb{R}$ is time. Given $n \in \mathbb{N}$ and $A_1 ,
\dots , A_n\in \mathfrak{C}_\Lambda$, the corresponding {\it Green
function} is
\begin{equation}
\label{a7} G^\Lambda_{A_1 , \dots A_n} (t_1 , \dots , t_n) =
\varrho_\Lambda \left[ \mathfrak{a}^\Lambda_{t_1} (A_1) \cdots
\mathfrak{a}^\Lambda_{t_n}(A_n) \right],
\end{equation}
which is a complex valued function on  $\mathbb{R}^n$. Each such a
function can be looked upon, see \cite{[AH-K],[RevMF]}, as the
restriction of a function $G^\Lambda_{A_1 , \dots A_n}$ analytic in
the domain
\begin{equation}
\label{a8} \mathcal{D}^n_\beta = \{ (z_1 , \dots , z_n)\in
\mathbb{C}^n \ | \ 0 < \Im (z_1) < \cdots < \Im(z_n ) < \beta\},
\end{equation}
and continuous on its closure.  The corresponding statement is known
as {\it the multiple-time analyticity theorem}, see
\cite{[AH-K],[RevMF]}, as well as  \cite{[KL]} for a more general
consideration. For every $n \in \mathbb{N}$, the subset
\begin{equation} \label{a88}
\{ (z_1 , \dots , z_n)\in \mathcal{D}^n_\beta \ | \ \Re(z_1) =
\cdots = \Re(z_n) = 0\}
\end{equation}
is an inner uniqueness set for functions analytic in
$\mathcal{D}^n_\beta$, see pages 101 and 352 in \cite{[Shabat]}. This means that two such functions which
coincide on this set should coincide everywhere on
${\mathcal{D}}^n_\beta$.

For a bounded continuous function
$F:\mathbb{R}^{\nu|\Lambda|}\rightarrow \mathbb{C}$, the
corresponding multiplication operator $F\in \mathfrak{C}_\Lambda$
acts as follows
\[
(F\psi )(x) = F(x) \psi (x), \qquad \psi \in L^2
(\mathbb{R}^{\nu|\Lambda|}).
\]
Let $\mathfrak{F}_\Lambda \subset \mathfrak{C}_\Lambda$ be the set
of all such operators. One can prove (the density theorem, see
\cite{[Koz5],[Koz6]}) that the linear span of the products
\[
\mathfrak{a}^\Lambda_{t_1} (F_1) \cdots \mathfrak{a}^\Lambda_{t_n}
(F_n),
\]
with all possible choices of $n\in \mathbb{N}$, $t_1 , \dots , t_n
\in \mathbb{R}$, and $F_1 , \dots , F_n\in \mathfrak{F}_\Lambda$, is
dense in $\mathfrak{C}_\Lambda$ in the $\sigma$-weak topology in
which the state (\ref{a4}) is continuous as a linear functional. Thus, the latter is
determined by the set of Green functions $G^\Lambda_{F_1 , \dots
F_n}$ with  $n\in \mathbb{N}$ and $F_1 , \dots , F_n\in
\mathfrak{F}_\Lambda$. The restriction of the Green functions
$G^\Lambda_{F_1 , \dots F_n}$ to the imaginary-time sets (\ref{a88})
are called {\it Matsubara functions}. For
\begin{equation} \label{a10}\tau_1 \leq \tau_2 \leq \cdots \leq
\tau_n \leq \beta,
\end{equation}
they are
\begin{equation}
\label{a9} \Gamma^\Lambda_{F_1, \dots, F_n} (\tau_1 , \dots ,
\tau_n) = G^\Lambda _{F_1, \dots, F_n} (\imath \tau_1 , \dots ,
\imath \tau_n).
\end{equation}
Since (\ref{a88}) is an inner uniqueness set, the collection of the
Matsubara functions (\ref{a9}) with all possible choices of $n\in
\mathbb{N}$ and $F_1 , \dots , F_n\in \mathfrak{F}_\Lambda$
determines the state (\ref{a4}). The extensions of the functions
(\ref{a9}) to $[0, \beta]^n$ are defined as
\[
\Gamma^\Lambda_{F_1, \dots, F_n} (\tau_1 , \dots , \tau_n) =
\Gamma^\Lambda_{F_{\sigma(1)}, \dots, F_{\sigma(n)}}
(\tau_{\sigma(1)} , \dots , \tau_{\sigma(n)})\] where $\sigma$ is
the permutation such that $\tau_{\sigma(1)}\leq \tau_{\sigma(2)}\leq
\cdots \leq \tau_{\sigma(n)}$. One can show that for every $\theta
\in [0, \beta]$,
\begin{equation} \label{a11}
\Gamma^\Lambda_{F_1, \dots, F_n} (\tau_1 +\theta , \dots ,
\tau_n+\theta) = \Gamma^\Lambda_{F_1, \dots, F_n} (\tau_1 , \dots ,
\tau_n),
\end{equation}
where addtion is modulo $\beta$.

\subsection{Path spaces}
\label{2.2ss} By (\ref{a7}), the Matsubara function (\ref{a9}) can
be written as
\begin{eqnarray}
& & \label{a12} \Gamma^\Lambda_{F_1, \dots, F_n} (\tau_1 , \dots ,
\tau_n) = \\ & & \quad  \ \ = {\rm trace}\left[F_1 e^{-(\tau_2 -
\tau_1)H_\Lambda}F_2 e^{-(\tau_3 - \tau_2)H_\Lambda} \cdot F_n
e^{-(\tau_{n+1} - \tau_n)H_\Lambda} \right] /Z_\Lambda, \nonumber
\end{eqnarray}
where $\tau_{n+1} = \beta + \tau_1$ and the arguments obey
(\ref{a10}). This expression can be rewritten
in an integral form
\begin{equation} \label{a13}
\Gamma^\Lambda_{F_1, \dots, F_n} (\tau_1 , \dots , \tau_n) =
\int_{\Omega_\Lambda} F_1 (\omega_\Lambda (\tau_1)) \cdots
 F_n (\omega_\Lambda (\tau_n)) \nu_\Lambda ({\rm d}\omega_\Lambda),
 \end{equation}
that is the main point of the Euclidean approach. Here $\nu_\Lambda$
is a probability measure on the path space $\Omega_\Lambda$ which we
introduce now. The main single-site path space is the space of
continuous periodic paths (temperature loops)
\begin{equation} \label{a14}
C_\beta = \{ \phi\in C([0, \beta]\rightarrow \mathbb{R}^\nu) \ | \
\phi(0) = \phi(\beta)\}.
\end{equation}
It is a Banach space with the usual sup-norm $\|\cdot \|_{C_\beta}$.
For an appropriate $\phi \in C_\beta$, we set
\begin{equation} \label{a15}
K_\sigma (\phi) = \beta^\sigma \cdot \sup_{\tau , \tau' \in [0,
\beta] \ \tau \neq \tau'} \frac{|\phi(\tau) - \phi(\tau')|}{|\tau -
\tau'|^\sigma_\beta}, \quad \sigma >0,
\end{equation}
where
\begin{equation} \label{a16}
|\tau - \tau'|_\beta = \min\left\{ |\tau - \tau'|; \beta - |\tau -
\tau'|\right\}
\end{equation}
 is the periodic distance on the circle $S_\beta \sim[0,
\beta]$. Then the set of H\"older-continuous periodic functions,
\begin{equation} \label{17}
C^\sigma_\beta = \{ \phi \in C_\beta \ | \ K_\sigma(\phi) <
\infty\},
\end{equation}
can be equipped with the norm
\begin{equation} \label{a18}
\|\phi \|_{C_\beta^\sigma} = |\phi(0) | + K_\sigma(\phi),
\end{equation}
which turns it into a Banach space. Along with the
spaces $C_\beta$, $C_\beta^\sigma$, we shall use the Hilbert
space $L^2_\beta = L^2 (S_\beta \rightarrow \mathbb{R}^\nu, {\rm
d}\tau)$, equipped with the inner product $(\cdot,
\cdot)_{L^2_\beta}$ and norm $\|\cdot\|_{L^2_\beta}$. By
$\mathcal{B}(C_\beta)$, $\mathcal{B}(L^2_\beta)$ we denote the
corresponding Borel $\sigma$-algebras. In a standard way, see page
21 of \cite{[Part]} and the corresponding discussion in
\cite{[KoT]}, it follows that
\begin{equation} \label{a19}
C_\beta \in \mathcal{B}(L^2_\beta) \quad \ \ {\rm and} \ \ \
\mathcal{B}(C_\beta) = \mathcal{B}(L^2_\beta) \cap C_\beta.
\end{equation}
Given $\Lambda \subseteq \mathbb{L}$, we set
\begin{eqnarray} \label{a20}
& & \Omega_\Lambda  =  \{\omega_\Lambda = (\omega_\ell)_{\ell \in
\Lambda}  \ | \ \omega_\ell \in C_\beta\},
\\ & & \Omega  =  \Omega_{\mathbb{L}} = \{\omega =
(\omega_\ell)_{\ell \in \mathbb{L}}  \ | \ \omega_\ell \in C_\beta\}.  \nonumber
\end{eqnarray}
These path spaces are equipped with the product topology and with
the Borel $\sigma$-algebras $\mathcal{B}(\Omega_\Lambda)$. Thereby,
each $\Omega_\Lambda$ is a  complete separable metric space, called {\it Polish
space}, its elements are called {\it configurations in} $\Lambda$.
For $\Lambda \subset \Lambda'$, the juxtaposition $\omega_{\Lambda'}
= \omega_\Lambda \times \omega_{\Lambda' \setminus \Lambda}$ defines
an embedding $\Omega_\Lambda \hookrightarrow \Omega_{\Lambda'}$ by
identifying $\omega_\Lambda \in \Omega_\Lambda$ with $\omega_\Lambda
\times 0_{\Lambda' \setminus \Lambda} \in \Omega_{\Lambda'}$. By
$\mathcal{P}(\Omega_\Lambda)$, $\mathcal{P}(\Omega)$ we denote the
sets of all probability measures on $(\Omega_\Lambda,
\mathcal{B}(\Omega_\Lambda))$, $(\Omega, \mathcal{B}(\Omega))$
respectively.

\subsection{Local Euclidean Gibbs measures} \label{2.3ss}

Now we construct the measure $\nu_\Lambda$ which appears in
(\ref{a13}). A single harmonic oscillator is described by the
Hamiltonian, c.f., (\ref{U2}),
\begin{equation} \label{a21}
H_\ell^{\rm har} = - \frac{1}{2m} \sum_{j=1}^\nu \left(
\frac{\partial}{\partial x_\ell^{(j)}}\right)^2 + \frac{a}{2}
|x_\ell|^2.
\end{equation}
It is a self-adjoint operator in the space $L^2(\mathbb{R}^\nu)$,
the properties of which are well-known. The operator semigroup
$\exp(-\tau H_\ell^{\rm har})$, $\tau \in S_\beta$, defines a
$\beta$-periodic Markov process, see \cite{[KLP]}. In quantum
statistical mechanics, it first appeared in R. H{\o}egh-Krohn's
paper \cite{[HK]}. The canonical realization of this process on
$(C_\beta, \mathcal{B}(C_\beta))$ is described by the path measure
which can be introduced as follows. In the space $L^2_\beta$, we
define the following self-adjoint Laplace-Beltrami type operator
\begin{equation} \label{a22}
A = \left( - m \frac{{\rm d}^2}{{\rm d}\tau^2} + a \right)\otimes
\mathbf{I},
\end{equation}
where $\mathbf{I}$ is the identity operator in $\mathbb{R}^\nu$. Its
spectrum consists of the eigenvalues
\begin{equation} \label{a23}
\lambda_l = m (2 \pi l/ \beta)^2 + a, \quad \ \ l\in \mathbb{Z}.
\end{equation}
Therefore, the inverse $A^{-1}$ is a trace-class operator on
$L^2_\beta$ and the Fourier transform
\begin{equation} \label{a24}
\int_{L^2_\beta} \exp\left[ \imath (\psi ,
\phi)_{L^2_\beta}\right]\chi({\rm d}\phi) = \exp\left\{ -
\frac{1}{2} (A^{-1} \psi, \psi)_{L^2_\beta}\right\}
\end{equation}
defines a zero mean Gaussian measure $\chi$ on $(L^2_\beta,
\mathcal{B}(L^2_\beta))$. Employing the eigenvalues (\ref{a23}) one
can show that, for any $p\in \mathbb{N}$,
\begin{eqnarray} \label{a24a}
\int_{C_\beta} \left\vert\omega (\tau) - \omega(\tau')
\right\vert^{2p} \chi({\rm d} \omega )\leq \frac{\Gamma (\nu/2 +
p)}{\Gamma (\nu/2)} \left( \frac{2}{m}\right)^p \cdot |\tau -
\tau'|_\beta^p.
\end{eqnarray}
Therefrom, by Kolmogorov's lemma (see page 43 of \cite{[Sim2]}) it
follows that
\begin{equation} \label{a25}
\chi (C_\beta^\sigma) = 1, \quad \ \ {\rm for} \ {\rm all}  \ \
\sigma \in (0, 1/2).
\end{equation}
Thereby, $\chi (C_\beta) = 1$; hence, with the help of  (\ref{a19})
we redefine $\chi$ as a measure on $(C_\beta,
\mathcal{B}(C_\beta))$, possessing the property (\ref{a25}). We
shall call it {\it H{\o}egh-Krohn's measure}. An account of the
properties of $\chi$ can be found in \cite{[RevMF]}. Here we present
the following two of them. The first  property is obtained directly
from Fernique's theorem (see Theorem 1.3.24 in \cite{[DS]}).
\begin{proposition} [Fernique]  \label{a1pn}
For every $\sigma \in (0, 1/2)$, there exists $\lambda_\sigma >0$,
which can be estimated explicitely, such that
\begin{equation} \label{a26}
\int_{L^2_\beta} \exp \left( \lambda_\sigma
\|\phi\|^2_{C^\sigma_\beta} \right) \chi({\rm d}\phi ) < \infty.
\end{equation}
\end{proposition}
The second property follows from the estimate (\ref{a24a}) by the
Garsia-Rodemich-Rumsey lemma, see \cite{[Garsia]}. For fixed $\sigma
\in (0, 1/2)$, we set
\begin{equation} \label{a26a}
\Xi_{\vartheta} (\omega ) = \sup_{\tau , \tau': \ 0< |\tau -
\tau'|_\beta < \vartheta}\left\{\frac{|\omega (\tau ) - \omega
(\tau')|^{2}}{|\tau - \tau'|_\beta^{2 \sigma } }\right\}, \  \ \quad
\vartheta \in (0, \beta/2), \quad \omega \in C_\beta^\sigma.
\end{equation}
One can show that, for each $\sigma$ and $\vartheta$, it can be extended to a measurable map $\Xi_{\vartheta} : C_\beta \rightarrow [0, +\infty]$.
\begin{proposition}[Garsia-Rodemich-Rumsey estimate] \label{grrpn}
Given $\sigma\in(0, 1/2)$, let $p\in \mathbb{N}$ be such that $(p-1)/2p >
\sigma$. Then
\begin{equation} \label{a26b}
\int_{C_\beta} \Xi^p_{\vartheta} (\omega ) \chi({\rm d}\omega) \leq
D(\sigma, p , \nu) m^{-p} \vartheta^{p(1-2\sigma)},
\end{equation}
where $m$ is the mass (\ref{In}) and
\begin{equation} \label{a26c}
D(\sigma, p , \nu) = \frac{2^{3(2p+1)}(1 + 1 / \sigma p)^{2p}}{(p -
1 - 2 \sigma p)( p - 2 \sigma p)}\cdot \frac{2^p \Gamma (\nu/2 +
1)}{\Gamma (\nu/2)}.
\end{equation}
\end{proposition}

 The H{\o}egh-Krohn measure is the local Euclidean Gibbs measure
for a single harmonic oscillator. The measure $\nu_\Lambda\in
\mathcal{P}(\Omega_\Lambda)$, which is the Euclidean Gibbs measure
corresponding to the system of interacting anharmonic oscillators
located in $\Lambda\Subset \mathbb{L}$, is defined by means of the
Feynman-Kac formula as a Gibbs modification
\begin{equation} \label{a27}
\nu_\Lambda ({\rm d}\omega_\Lambda) = \exp\left[- I_\Lambda
(\omega_\Lambda) \right]\chi_\Lambda ({\rm
d}\omega_\Lambda)/N_\Lambda
\end{equation}
of the `free measure'
\begin{equation} \label{a28}
\chi_\Lambda ({\rm d}\omega_\Lambda) = \prod_{\ell \in \Lambda}
\chi({\rm d}\omega_\ell).
\end{equation}
Here
\begin{equation} \label{a29}
I_\Lambda (\omega_\Lambda) = - \frac{1}{2} \sum_{\ell , \ell' \in
\Lambda} J_{\ell \ell'} (\omega_\ell , \omega_{\ell'})_{L^2_\beta }
+ \sum_{\ell \in \Lambda}\int_0^\beta V_\ell (\omega_\ell
(\tau)){\rm d}\tau
\end{equation}
is \emph{the energy functional} which describes the interaction of
the paths $\omega_\ell$, $\ell \in \Lambda$. The normalizing factor
\begin{equation} \label{a30}
N_\Lambda = \int_{\Omega_\Lambda} \exp\left[- I_\Lambda
(\omega_\Lambda) \right]\chi_\Lambda ({\rm d}\omega_\Lambda)
\end{equation}
is the relative partition function, whereas the Feynman-Kac
representation of the partition function (\ref{a5}) is
\begin{equation} \label{a300}
Z_\Lambda = N_\Lambda Z_\Lambda^{\rm har} ,
\end{equation}
where
\begin{eqnarray*}
Z_\Lambda^{\rm har} & \stackrel{\rm def}{=} & {\rm trace} \exp\left[
- \beta \sum_{\ell \in \Lambda}H^{\rm har}_\ell \right] \\ & = &
\left\{\frac{\exp\left[ - (\beta/2) \sqrt{a/m} \right]}{1 -\exp\left( -
\beta \sqrt{a/m} \right)} \right\}^{\nu |\Lambda|}.
\end{eqnarray*}

 Now let us summarize the connections between the description
of the subsystem located in $\Lambda \Subset \mathbb{L}$ in terms of
the states (\ref{a4}) and of the Euclidean Gibbs measures
(\ref{a27}). By the density theorem, the state $\varrho_\Lambda$ is
fully determined by the Green functions (\ref{a7}) corresponding to
all choices of $n\in \mathbb{N}$ and $F_1 , \dots , F_n\in
\mathfrak{F}_\Lambda$. Then the multiple-time analyticity theorem
leads us from the Green functions to the Matsubara functions
(\ref{a9}), which then are represented as integrals over path spaces
with respect to the local Euclidean Gibbs measures, see (\ref{a13}).
On the other hand, these integrals taken for all possible choices of
bounded continuous functions $F_1 , \dots , F_n$ fully determine the
measure $\nu_\Lambda$. Thereby, we have a one-to-one
correspondence between the local Gibbs states (\ref{a4}) and the
states on the algebras of bounded continuous functions determined by
the local Euclidean Gibbs measures (\ref{a27}). Our next aim is to
extend this approach to the global states. To this end we  make more
precise the definition of the path spaces in infinite $\Lambda$,
e.g., in
 $\Lambda = \mathbb{L}$.

\subsection{Tempered configurations} \label{2.4ss}

To describe the global thermodynamic properties we need the
conditional distributions $\pi_\Lambda ({\rm d}\omega|\xi)$,
$\Lambda \Subset \mathbb{L}$. For models with infinite-range
interactions, the construction of such distributions is a nontrivial
problem, which can be solved by imposing  a priori restrictions on
the configurations defining the corresponding conditions. In this
 and in the subsequent subsections, we present the construction of  such distributions
performed \cite{[KoT]}.

The distributions $\pi_\Lambda ({\rm d}\omega|\xi)$ are defined by
means of the energy functionals $I_\Lambda (\omega|\xi)$ describing
the interaction  of the configuration $\omega$ with the
configuration $\xi$, fixed outside of $\Lambda$. Given $\Lambda \Subset \mathbb{L}$, such a functional is
\begin{equation} \label{a31}
I_\Lambda (\omega|\xi) = I_\Lambda (\omega_\Lambda) - \sum_{\ell \in
\Lambda , \ \ell'\in \Lambda^c}J_{\ell \ell'} (\omega_\ell ,
\xi_{\ell'})_{L^2_\beta}, \quad \omega \in \Omega,
\end{equation}
where $I_\Lambda$ is given by (\ref{a29}). Recall that $\omega =
\omega_{\Lambda}\times \omega_{\Lambda^c}$; hence,
\begin{equation} \label{a32}
I_\Lambda (\omega|\xi) = I_\Lambda (\omega_\Lambda \times
0_{\Lambda^c} | 0_\Lambda \times \xi_{\Lambda^c}).
\end{equation}
The second term in (\ref{a31}) makes sense for all $\xi\in \Omega$
only if the interaction has finite range, see Definition \ref{1df}.
Otherwise, one has to impose appropriate restrictions on the
configurations $\xi$, such that, for all $\ell$ and $\omega \in
\Omega$,
\begin{equation} \label{a33}
\sum_{\ell'}J_{\ell \ell'} \cdot |(\omega_\ell ,
\xi_{\ell'})_{L^2_\beta}| < \infty.
\end{equation}
These restrictions are formulated by means of special mappings
(weights), which define the scale of growth of
$\{\|\xi_{\ell}\|_{L^2_\beta}\}_{\ell \in \mathbb{L}}$. Their choice
depends on the asymptotic properties of $J_{\ell \ell'}$, $|\ell -
\ell'|\rightarrow +\infty$, see (\ref{a1}). If for a certain $\alpha
>0$,
\begin{equation} \label{a34}
\sup_{\ell } \sum_{\ell'} J_{\ell \ell'} \exp (\alpha |\ell -
\ell'|) < \infty,
\end{equation}
then the weights $\{w_\alpha (\ell, \ell')\}_{\alpha \in
\mathcal{I}}$ are chosen as
\begin{equation} \label{a35}
w_\alpha (\ell , \ell') = \exp (-\alpha |\ell - \ell'|), \quad \ \
\mathcal{I}= (0 , \overline{\alpha}),
\end{equation}
where $\overline{\alpha}$ is the supremum of $\alpha>0$, for which
(\ref{a34}) holds. If the latter condition does not hold for any
$\alpha>0$, we assume that
\begin{equation} \label{a36}
\sup_{\ell } \sum_{\ell'} J_{\ell \ell'} \cdot ( 1+ |\ell -
\ell'|)^{\alpha d},
\end{equation}
for a certain $\alpha >1$. Then we set $\overline{\alpha}$ to be the
supremum of $\alpha>1$ obeying (\ref{a36}) and
\begin{equation} \label{a37}
w_\alpha (\ell , \ell') = ( 1+ \varepsilon |\ell - \ell'|)^{-\alpha
d},
\end{equation}
where $\varepsilon >0$ is a technical parameter. In the sequel, we restrict ourselves to these two kinds of $J_{\ell\ell'}$.
For more details on this item, we refer
the reader  to \cite{[KoT]}.

Given $\alpha \in \mathcal{I}$ and $\omega \in \Omega$, we set
\begin{equation} \label{a42}
\|\omega \|_\alpha = \left[\sum_{\ell} \|\omega_\ell
\|^2_{L^2_\beta}w_\alpha (0, \ell) \right]^{1/2},
\end{equation}
and
\begin{equation}
\label{a43} \Omega_\alpha = \{ \omega \in \Omega \ | \
\|\omega\|_\alpha < \infty\}.
\end{equation}
Thereby, we endow $\Omega_\alpha$ with the metric
\begin{equation}
\label{a44} \rho_\alpha (\omega , \omega') = \|\omega -
\omega'\|_\alpha + \sum_{\ell} 2^{-|\ell|} \frac{\|\omega_\ell -
\omega'_\ell\|_{C_\beta}}{ 1 +\|\omega_\ell -
\omega'_\ell\|_{C_\beta}},
\end{equation}
which turns it into a Polish space. The set of tempered
configurations is defined to be
\begin{equation}
\label{a45} \Omega^{\rm t} = \bigcap_{\alpha \in
\mathcal{I}}\Omega_\alpha.
\end{equation}
We endow it with the projective limit topology, which turns it into
a Polish space as well. For every $\alpha\in \mathcal{I}$, the
embeddings $\Omega^{\rm t} \hookrightarrow \Omega_\alpha
\hookrightarrow \Omega$ are continuous; hence, $\Omega_\alpha ,
\Omega^{\rm t} \in \mathcal{B}(\Omega)$ and the Borel
$\sigma$-algebras $\mathcal{B}(\Omega_\alpha)$,
$\mathcal{B}(\Omega^{\rm t})$ coincide with the ones induced on them
by $\mathcal{B}(\Omega)$.

\subsection{Local Gibbs specification}  \label{2.5.ss}
Let us turn to the functional (\ref{a31}). By standard methods, one
proves that, for every $\alpha \in \mathcal{I}$, the map
$\Omega_\alpha \times \Omega_\alpha  \mapsto I_\Lambda (\omega|\xi)$
is continuous. Furthermore, for any ball $B_\alpha (R) = \{ \omega
\in \Omega_\alpha \ | \ \rho_\alpha (0, \omega) < R\}$, $R>0$, one
has
\begin{eqnarray*}
\inf_{\omega \in \Omega , \ \ \xi \in B_\alpha (R)} I_\Lambda
(\omega |\xi) > - \infty, \quad \   \sup_{\omega , \xi \in B_\alpha
(R)}| I_\Lambda (\omega |\xi)| < + \infty.
\end{eqnarray*}
Therefore, for $\Lambda \Subset \mathbb{L}$ and $\xi \in \Omega^{\rm
t}$, the conditional relative partition function
\begin{equation}
\label{a46} N_\Lambda (\xi) = \int_{\Omega_\Lambda} \exp\left[-
I_\Lambda (\omega_\Lambda \times 0_{\Lambda^c}|\xi) \right]
\chi_\Lambda ({\rm d}\omega_\Lambda)
\end{equation}
is continuous in $\xi$. Furthermore, for any $R>0$ and $\alpha \in
\mathcal{I}$,
\[
\inf_{\xi \in B_\alpha (R) } N_\Lambda (\xi) >0.
\]
For such $\xi$ and $\Lambda$, and for $B \in \mathcal{B}(\Omega)$,
we set
\begin{equation} \label{34}
\pi_\Lambda (B|\xi) = \frac{1}{N_\Lambda(\xi)}
\int_{{\Omega}_\Lambda}\exp\left[- I_\Lambda(\omega_\Lambda
\times 0_{\Lambda^c} |\xi)  \right] \mathbb{I}_B (\omega_\Lambda
\times \xi_{\Lambda^c})\chi_\Lambda ({\rm d}\omega_\Lambda ),
\end{equation}
where $\mathbb{I}_B $ stands for the indicator of $B$. We also set
\begin{equation} \label{34a}
\pi_\Lambda (\cdot|\xi) \equiv 0, \quad {\rm for} \ \ \xi \in
{\Omega} \setminus {\Omega}^{\rm t}.
\end{equation} From
these definitions one readily derives a consistency property
\begin{equation} \label{35}
\int_{{\Omega}} \pi_\Lambda (B|\omega) \pi_{\Lambda'} ({\rm
d} \omega |\xi) = \pi_{\Lambda'} (B |\xi), \quad \Lambda \subset
\Lambda',
\end{equation}
which holds for all $B\in \mathcal{B}({\Omega})$ and $\xi \in
{\Omega}$.

 The local Gibbs specification is the family $\{\pi_\Lambda \}_{\Lambda
\Subset \mathbb{L}}$. Each $\pi_\Lambda$ is a measure kernel, which
means that, for a fixed $\xi\in \Omega$, $\pi(\cdot|\xi)$ is a
measure on $(\Omega, \mathcal{B}(\Omega))$, which is a probability
measure whenever $\xi \in \Omega^{\rm t}$. For any $B \in
\mathcal{B}(\Omega)$, $\pi_\Lambda (B|\cdot)$ is
$\mathcal{B}(\Omega)$-measurable.

By $C_{\rm b}({\Omega}_\alpha)$ (respectively, $C_{\rm
b}({\Omega}^{\rm t})$) we denote the Banach spaces of all
bounded continuous functions $f:{\Omega}_\alpha \rightarrow
\mathbb{R}$ (respectively, $f:{\Omega}^{\rm t} \rightarrow
\mathbb{R}$) equipped with the supremum norm. For every $\alpha \in
\mathcal{I}$, one has a natural embedding $C_{\rm
b}({\Omega}_\alpha ) \hookrightarrow C_{\rm
b}({\Omega}^{\rm t})$. Given $\alpha  \in \mathcal{I}$, by
$\mathcal{W}_\alpha$ we denote the usual weak topology on the set of
all probability measures $\mathcal{P}({\Omega}_\alpha)$
defined by means of $C_{\rm b}({\Omega}_\alpha)$. By
$\mathcal{W}^{\rm t}$ we denote the weak topology on
$\mathcal{P}({\Omega}^{\rm t})$. With these topologies the
sets $\mathcal{P}({\Omega}_\alpha)$ and
$\mathcal{P}({\Omega}^{\rm t})$ become Polish spaces (Theorem
6.5, page 46 of \cite{[Part]}).

By standard methods one proves the following, see Lemma 2.10 in
\cite{[KoT]},
\begin{proposition} [Feller Property] \label{2lm}
For every $\alpha \in \mathcal{I}$, $\Lambda \Subset \mathbb{L}$,
and any $f \in C_{\rm b}({\Omega}_{\alpha})$, the function
\begin{eqnarray} \label{f}
& & {\Omega}_\alpha \ni \xi \mapsto \pi_\Lambda (f | \xi) \\
& & \qquad \qquad  \stackrel{\rm def}{=} \
 \frac{1}{N_\Lambda (\xi)}
\int_{{\Omega}_\Lambda}f (\omega_\Lambda \times
\xi_{\Lambda^c}) \exp\left[- I_\Lambda (\omega_\Lambda \times
0_{\Lambda^c}|\xi) \right]\chi_\Lambda ({\rm d}\omega_\Lambda),
\nonumber
\end{eqnarray}
belongs to $C_{\rm b}({\Omega}_{\alpha})$. The linear
operator $f \mapsto \pi_\Lambda (f|\cdot)$ is a contraction on
$C_{\rm b}({\Omega}_\alpha)$.
\end{proposition}
Note that by (\ref{34}), for $\xi \in {\Omega}^{\rm t}$,
$\alpha \in \mathcal{I}$, and $f \in C_{\rm
b}({\Omega}_\alpha)$,
\begin{equation} \label{fp}
\pi_\Lambda (f|\xi) = \int_{{\Omega}}f(\omega)
\pi_\Lambda({\rm d}\omega|\xi).
\end{equation}
Recall that the particular cases of our model were specified by
Definition \ref{1df}. For $B\in \mathcal{B}({\Omega})$ and $U\in
O(\nu)$, we set
\[
  U \omega = (U
\omega_\ell)_{\ell \in \mathbb{L}} \qquad UB = \{ U \omega \  | \
\omega \in B\}.
\]
Furthermore, for a given $\ell_0$, we set
\[
t_{\ell_0} (\omega) = (\omega_{\ell - \ell_0})_{\ell \in
\mathbb{L}}, \qquad t_{\ell_0}( B) = \{t_{\ell_0}(\omega) \  | \
\omega \in B\}.
\]
Then if the model possesses the corresponding symmetry, one has
\begin{equation} \label{MA10}
\pi_\Lambda (U B |U\xi) = \pi_\Lambda (B| \xi), \qquad \pi_{\Lambda
+ \ell} (t_\ell (B)|t_\ell (\xi)) = \pi_\Lambda (B|\xi),
\end{equation}
which ought to hold for all $U$, $\ell$, $B$, and $\xi$.

\subsection{Tempered Euclidean Gibbs measures}
\label{2.6.ss}

\begin{definition} \label{3df}
A measure $\mu \in \mathcal{P}({\Omega})$ is called a
tempered Euclidean Gibbs measure if it satisfies the
Dobrushin-Lanford-Ruelle (equilibrium) equation
\begin{equation} \label{40}
\int_{{\Omega}}\pi_\Lambda (B |\omega) \mu({\rm d}\omega) =
\mu(B), \quad {\rm for} \ {\rm all} \ \ \ \Lambda \Subset \mathbb{L} \ \ {\rm
and} \ \ B \in \mathcal{B}({\Omega}).
\end{equation}
\end{definition}
\noindent By $\mathcal{G}^{\rm t}$ we denote the set of all tempered
Euclidean Gibbs measures of our model existing at a given $\beta$.
The elements of $\mathcal{G}^{\rm t}$ are supported by ${\Omega}^{\rm t}$.
Indeed, by (\ref{34}) and (\ref{34a}) $\pi_\Lambda ({\Omega}
\setminus {\Omega}^{\rm t} |\xi) = 0$ for every $\Lambda
\Subset \mathbb{L}$ and $\xi \in {\Omega}$. Then by
(\ref{40}),
\begin{equation} \label{40a}
\mu ({\Omega} \setminus {\Omega}^{\rm t})  = 0.
\end{equation}
Furthermore,
\begin{equation}
\mu \left( \left\{ \omega \in {\Omega }^{ \mathrm{t}} \ | \ \forall
\ell \in \mathbb{L}: \  \omega_\ell \in C_{\beta }^{\sigma }
\right\} \right) =1,  \label{40b}
\end{equation}%
which follows from (\ref{a25}), (\ref{a26}). If the model is
translation and/or rotation invariant, then, for every $U\in O(\nu)$
and $\ell\in \mathbb{L}$, the corresponding transformations preserve
$\mathcal{G}^{\rm t}$. That is, for any $\mu \in \mathcal{G}^{\rm
t}$,
\begin{equation} \label{MA11}
\Theta_U (\mu) \ \stackrel{\rm def}{=} \ \mu \circ U^{-1} \in
\mathcal{G}^{\rm t}, \qquad \theta_\ell (\mu) \ \stackrel{\rm
def}{=} \ \mu \circ t^{-1}_\ell \in \mathcal{G}^{\rm t}.
\end{equation}
In particular, if $\mathcal{G}^{\rm t}$ is a singleton, its unique
element should be invariant in the same sense as the model.
From Proposition \ref{2lm} one readily gets the following important
fact.
\begin{proposition} \label{3lm}
For each $\alpha \in \mathcal{I}$, every
$\mathcal{W}_\alpha$-accumulation point $\mu \in
\mathcal{P}({\Omega}^{\rm t})$ of the family $\{\pi_\Lambda
(\cdot |\xi) \ | \ \Lambda \Subset \mathbb{L}, \ \xi \in
{\Omega}^{\rm t}\}$ is an element of $\mathcal{G}^{\rm t}$.
\end{proposition}

Now let us pay some attention to the case where the model
(\ref{U1}), (\ref{U2}) is translation invariant. Recall that the
lattice $\mathbb{L} = \mathbb{Z}^d$ is considered as an additive
group. For ${\ell}_0\in \mathbb{L}$, $\Lambda \Subset \mathbb{L}$,
and $\omega \in \Omega$, we set
\begin{equation} \label{la106a}
\Lambda + \ell_0 = \{ \ell + \ell_0\ | \ \ell \in \Lambda \}; \quad
t_{\ell_0}(\omega) = (\xi^{\ell_0}_\ell)_{\ell\in \mathbb{L}}, \ \
\xi^{\ell_0}_\ell = \omega_{\ell - \ell_0}.
\end{equation}
Furthermore, for $B\in \mathcal{B}(\Omega)$, we set
\begin{equation} \label{Ula106a}
t_\ell(B) = \{ t_\ell(\omega) \ | \ \omega\in B\}.
\end{equation}
Clearly, $t_\ell (B) \in \mathcal{B}(\Omega)$ and $t_{\ell}
(\Omega^{\rm t}) = \Omega^{\rm t}$ for all $\ell$.
\begin{definition}\label{tripdf}
A probability measure $\mu \in \mathcal{P}(\Omega)$ is said to be
translation invariant if for every $\ell$ and $B \in
\mathcal{B}(\Omega)$, one has $\mu(t_\ell(B)) = \mu(B)$.
\end{definition}
As was mentioned above, the Gibbs specification $\{\pi_{
\Lambda}\}_{\Lambda \Subset \mathbb{L}}$ of the translation
invariant model  is translation invariant, that is, it has the
property (\ref{MA10}).
\begin{remark} \label{triprm}
The translation invariance of the Gibbs specification does not mean
that each probability kernel $\pi_{ \Lambda}$ as a measure is
translation invariant. Moreover, it does not mean that all the
Euclidean Gibbs measures defined by this specification are
translation invariant. One can only claim that if the set
$\mathcal{G}^{\rm t}$ consists of one element only, this element
ought to translation invariant.
\end{remark}

Set
\begin{equation} \label{ph1}
\mathcal{B}^{\rm inv} = \{ B \in \mathcal{B}(\Omega) \ | \ \forall
\ell : \ \ t_\ell(B) = B\},
\end{equation}
which is the set of all translation invariant events. By
construction, $\Omega^{\rm t} \in\mathcal{B}^{\rm inv}$. We
say that $\mu \in \mathcal{P}(\Omega)$ is trivial on
$\mathcal{B}^{\rm inv}$ if for every $B\in \mathcal{B}^{\rm inv}$,
one has $\mu(B) = 0$ or $\mu(B)=1$.  By $\mathcal{P}^{\rm
inv}(\Omega)$ we denote the set of translation invariant probability
measures on $(\Omega, \mathcal{B})$.
\begin{definition} \label{ph1df}
A probability measure $\mu \in \mathcal{P}^{\rm inv}(\Omega)$ is
said to be ergodic (with respect to the group $\mathbb{L}$) if it is
trivial on $\mathcal{B}^{\rm inv}(\Omega)$.
\end{definition}
Ergodic measures are characterized by a mixing property, which we
formulate here according to  \cite{[Simon]}, see
Theorem III.1.8 on page 244. For $L\in \mathbb{N}$, we set
\begin{equation} \label{box}
\Lambda_L = (-L, L]^d\cap \mathbb{Z}^d,
\end{equation}
which is called {\it a  box}. For a measure $\mu$ and an appropriate
function $f$, we write
\begin{equation} \label{B1}
\langle f \rangle_{\mu} = \int f {\rm d}\mu
\end{equation}
\begin{proposition}[Von Neumann Ergodic Theorem] \label{Uergpn}
Given $\mu\in \mathcal{P}^{\rm inv}(\Omega)$, the following
statements are equivalent: \vskip.1cm
\begin{tabular}{ll}
(i)   \ &$\mu$ is ergodic;\\[.2cm]
(ii)   \ &for all $f, g \in L^2(\Omega, \mu)$,
\end{tabular}
\begin{equation} \label{U9}
\lim_{L\rightarrow +\infty}  \frac{1}{|\Lambda_L|}\left\{\sum_{ \ell
\in \Lambda_L}\left(\int_{\Omega} f(\omega) g(t_\ell(\omega))
\mu({\rm d} \omega) - \langle f \rangle_\mu\cdot\langle g
\rangle_\mu \right) \right\} = 0.
\end{equation}
\end{proposition}
\begin{proposition} \label{Uergco}
If the model is  translation invariant  and $\mathcal{G}^{\rm t}$ is
a singleton, its unique element is ergodic.
\end{proposition}
Now we give a number of statements describing the properties of
$\mathcal{G}^{\rm t}$. More details can be found in \cite{[KoT]}.
\begin{proposition} \label{1tm}
For every $\beta>0$, the set of tempered Euclidean Gibbs measures
$\mathcal{G}^{\rm t}$ is non-void, convex, and $\mathcal{W}^{\rm t}$-
compact.
\end{proposition}
 Recall that the H\"{o}lder norm $\|\cdot \|_{C_{\beta
}^{\sigma }}$ was defined by (\ref{a18}).
\begin{proposition} \label{2tm}
For every $\sigma \in (0, 1/2)$ and $\varkappa >0$, there exists  a
positive constant $ C$ such that, for any $\ell $ and for all $\mu
\in \mathcal{G}^{\rm t}$,
\begin{equation} \label{43}
\int_{\mathit{\Omega}} \exp\left(\lambda_\sigma \|\omega_\ell
\|_{C^\sigma_\beta}^2 + \varkappa \|\omega_\ell \|_{L^2_\beta}^2
\right)\mu({\rm d}\omega) \leq C,
\end{equation}
where $\lambda_\sigma$ is the same as in (\ref{a26}).
\end{proposition}
In view of (\ref{43}), the one-site projections of each $\mu\in
\mathcal{G}^{\rm t}$ are sub-Gaussian. The constant $C$ does not
depend on $\ell$ and is the same for all $\mu \in \mathcal{G}^{\rm
t}$, though it may depend on $\sigma $ and $\varkappa$. The estimate
(\ref{43}) plays a crucial role in the theory of the set
$\mathcal{G}^{\rm t}$.

According to \cite{[Ge]}
certain Gibbs states correspond to the
thermodynamic phases of the underlying physical system. Thus, in our
context multiple phases exist only if
$\mathcal{G}^{\rm t}$ has more than one element for appropriate
values of $\beta$ and the model parameters. On the other hand, a
priori one cannot exclude that this set always has multiple
elements, which would make it useless for describing phase
transitions. The next statement which we present here\footnote{C.f., Theorem
3.4 in \cite{[KoT]}, Theorem 2.1 in \cite{[AKRT1]}, and Theorem 4.1 in \cite{[AKRT]}.}
clarifies the situation.    Let us decompose
\begin{equation}\label{decom}
V_{\ell} = V_{1, \ell} + V_{2, \ell},
\end{equation}
where $V_{1, \ell}\in C^2 (\mathbb{R}^\nu)$ is such that
\begin{equation} \label{dc1}
- a \leq b \ \stackrel{\rm def}{=} \ \inf_{\ell} \inf_{x, y \in
\mathbb{R}^\nu, \ y\neq 0}\left( V''_{1,\ell}(x)y, y \right)/|y|^2 <
\infty.
\end{equation}
As for the second term, we set
\begin{equation} \label{dc2}
0 \leq \delta \ \stackrel{\rm def}{=} \ \sup_{\ell} \left\{ \sup_{x
\in \mathbb{R}^\nu}V_{2, \ell}(x) - \inf_{x \in \mathbb{R}^\nu}V_{2,
\ell}(x) \right\} \leq \infty.
\end{equation}
Its role is to produce multiple minima of the potential energy
responsible for eventual phase transitions. Clearly, the
decomposition (\ref{decom}) is not unique; its optimal realizations
for certain types of $V_\ell$ are discussed in section 6 of
\cite{[AKRT]}. Recall that the interaction parameter $\hat{J}_0$ was defined in (\ref{a1}).
\begin{proposition} \label{httm}
The set $\mathcal{G}^{\rm t}$ is a singleton if
\begin{equation}
\label{dc3} e^{\beta \delta} <(a + b)/\hat{J}_0.
\end{equation}
\end{proposition}
\begin{remark} \label{apprm}
The latter condition surely holds at all $\beta$ if
\begin{equation} \label{si}
\delta = 0 \quad {\rm and} \quad \hat{J}_0 < a + b.
\end{equation}
 If the oscillators are harmonic, $\delta = b = 0$, which
yields the stability condition
\begin{equation} \label{si1}
\hat{J}_0 < a.
\end{equation}
The condition (\ref{dc3}) does not contain the particle mass $m$;
hence, the property stated holds also in the quasi-classical
limit\footnote{More details on this limit can be found in
\cite{[AKKR]}.} $m \rightarrow + \infty$.
\end{remark}
By the end of this subsection we consider the scalar case $\nu=1$. Let us introduce the following  order on
$\mathcal{G}^{\rm t}$. As the components of the configurations
$\omega\in {\Omega}$ are continuous functions $\omega_\ell
:S_\beta \rightarrow \mathbb{R}^\nu$, one can  set $\omega \leq
\tilde{\omega}$ if $\omega_\ell(\tau) \leq
\tilde{\omega}_\ell(\tau)$ for all $\ell$ and  $\tau$. Thereby,
\begin{equation} \label{MA1}
K_+ ({\Omega}^{\rm t}) \ \stackrel{\rm def}{ =} \ \{ f\in C_{\rm
b}({\Omega}^{\rm t}) \ | \ f(\omega) \leq f(\tilde{\omega}), \quad
{\rm if} \ \ \omega \leq \tilde{\omega}\},
\end{equation}
which is a cone of bounded continuous functions.
\begin{proposition} \label{MAlm}
If for given $\mu, \tilde{\mu} \in \mathcal{G}^{\rm t}$, one has
\begin{equation} \label{MA1a}
\langle f \rangle_\mu = \langle f \rangle_{\tilde{\mu}}, \qquad {\rm for} \ \ {\rm all} \ \ f \in
K_+ ({\Omega}^{\rm t}),
\end{equation}
then $\mu = \tilde{\mu}$.
\end{proposition}
This fact allows for introducing the FKG-order.
\begin{definition} \label{MAdf}
For $\mu, \tilde{\mu}\in\mathcal{G}^{\rm t}$, we say that $\mu\leq
\tilde{\mu}$, if
\begin{equation} \label{MAW}
\langle f \rangle_{{\mu}} \leq \langle f \rangle_{\tilde{\mu}}, \qquad {\rm for} \ \ {\rm all} \ \ f \in
K_+({\Omega}^{\rm t}).
\end{equation}
\end{definition}
\begin{proposition} \label{MAtm}
The set $\mathcal{G}^{\rm t}$ possesses maximal $\mu_{+}$ and
minimal $\mu_{-}$ elements in the sense of Definition \ref{MAdf}.
These elements are extreme; they also are translation invariant if
the model is translation invariant. If $V_\ell (-x ) = V_\ell (x)$
for all $\ell $, then $\mu_{+} (B) = \mu_{-} (- B)$ for all $B \in
\mathcal{B}({\Omega})$.
\end{proposition}
The proof of this statement follows from the fact that, for $f\in
K_{+}(\Omega^{\rm t})$ and any $\Lambda \Subset \mathbb{L}$,
\begin{equation}
\label{MAW1} \langle f \rangle_{\pi_\Lambda (\cdot|\xi)} \leq
\langle f \rangle_{\pi_\Lambda (\cdot|\xi')}, \quad \ \ {\rm
whenever} \ \ \xi \leq \xi',
\end{equation}
which one obtaines by the FKG inequality, see \cite{[KoT]}. By means of
this inequality, one also proves the following
\begin{proposition}
\label{MA1tm} The family $\{\pi_\Lambda (\cdot |0)\}_{\Lambda
\Subset \mathbb{L}}$ has only one $\mathcal{W}^{\rm t}$-accumulation
point, $\mu_0$, which is an element of $\mathcal{G}^{\rm t}$.
\end{proposition}

\subsection{Periodic Euclidean Gibbs measures}
\label{2.7.ss} If the model is translation invariant,  there should
exist $\phi: \mathbb{N}_0^d \rightarrow \mathbb{R}^+$ such that
\begin{equation} \label{A1}
J_{\ell \ell'} = \phi (|\ell_1 - \ell'_1|, \dots, |\ell_d -
\ell'_d|).\end{equation}  For the  box (\ref{box}), we set
\begin{equation} \label{A2}
J^\Lambda_{\ell \ell'} \ \stackrel{\rm def}{=} \ \phi (|\ell_1 -
\ell'_1|_L, \dots, |\ell_d - \ell'_d|_L),
\end{equation}
where
\begin{equation} \label{A3}
|\ell_j - \ell'_j|_L \ \stackrel{\rm def}{ =} \ \min\{ |\ell_j -
\ell'_j| \ ; \ L - |\ell_j - \ell'_j|\}, \ \ \ j= 1 , \dots , d.
\end{equation}
For $\ell , \ell'\in \Lambda$, we introduce the periodic distance
\begin{eqnarray} \label{box1}
|\ell - \ell'|_\Lambda  =  \sqrt{|\ell_1 - \ell'_1|_L^2 + \cdots +
|\ell_d - \ell'_d|_L^2}.
\end{eqnarray}
With this distance the box $\Lambda$ turns into a torus, which one
can obtained by imposing periodic conditions on its boundaries. Now
we set, c.f., (\ref{a29}),
\begin{equation} \label{A4}
I^{\rm per}_\Lambda (\omega_\Lambda) = - \frac{1}{2} \sum_{\ell ,
\ell' \in \Lambda} J^\Lambda_{\ell \ell'} (\omega_\ell ,
\omega_{\ell'})_{L^2_\beta } + \sum_{\ell \in \Lambda}\int_0^\beta
V_\ell (\omega_\ell (\tau)){\rm d}\tau ,
\end{equation}
and thereby, c.f., (\ref{a27}),
\begin{eqnarray} \label{A5}
\nu^{\rm per}_\Lambda ({\rm d}\omega_\Lambda) & = & \exp\left[-
I^{\rm per}_\Lambda (\omega_\Lambda) \right]\chi_\Lambda ({\rm
d}\omega_\Lambda)/N^{\rm per}_\Lambda, \\
N^{\rm per}_\Lambda & = & \int_{\Omega_\Lambda}\exp\left[- I^{\rm
per}_\Lambda (\omega_\Lambda) \right]\chi_\Lambda ({\rm
d}\omega_\Lambda). \nonumber
\end{eqnarray}
By means of (\ref{A2}) we introduce the periodic Hamiltonian
\begin{equation} \label{A5a}
H_\Lambda^{\rm per} = H_\Lambda = - \frac{1}{2} \sum_{\ell ,\ell'\in
\Lambda} J^\Lambda_{\ell \ell'} \cdot (q_\ell , q_{\ell'}) +
\sum_{\ell \in \Lambda}H_\ell,
\end{equation}
and the  corresponding periodic local Gibbs state
\begin{equation}
\label{A5b} \varrho^{\rm per}_\Lambda (A) = {\rm trace}[A\exp(-
\beta H^{\rm per}_\Lambda)]/ {\rm trace}[\exp(- \beta H^{\rm per}_\Lambda)], \quad A\in
\mathfrak{C}_\Lambda.
\end{equation}
The relationship between the measure $\nu_\Lambda^{\rm per}$ and this state is  the same as in the case of $\nu_\Lambda$ and $\varrho_\Lambda$.

Set, c.f., (\ref{34}),
\begin{equation}
\label{d51} \pi^{\rm per}_{ \Lambda} (B) = \frac{1}{N_{
\Lambda}^{\rm per}} \int_{\Omega_{ \Lambda}} \exp\left[ - I_{
\Lambda}^{\rm per} (\omega_{\Lambda}) \right] \mathbb{I}_B
(\omega_\Lambda \times 0_{\Lambda^c}) \chi_{ \Lambda} ({\rm
d}x_\Lambda),
\end{equation}
which is a probability measure on $\Omega^{\rm t}$. Then
\begin{equation} \label{d510}
\pi^{\rm per}_{ \Lambda}({\rm d}(\omega_\Lambda \times
\omega_{\Lambda^c})) = \nu_{ \Lambda}^{\rm per} ({\rm
d}\omega_\Lambda)\prod_{\ell'\in \Lambda^c} \delta_{0_{\ell'}}
({\rm d}x_{\ell'}),
\end{equation}
where $0_{\ell'}$ is the zero element of the Banach space $C_\beta$.
Note that  the projection of
$\pi_{\Lambda}^{\rm per}$ onto $\Omega_{\Lambda}$ is
$\nu_{\Lambda}^{\rm per}$.

Let $\mathcal{L}_{\rm box}$ be the sequence of all boxes
(\ref{box}). Arguments similar to those used in the proof of Lemma
4.4 in \cite{[KoT]} yield  the following
\begin{lemma} \label{d5lm}
For every $\alpha \in \mathcal{I}$ and $\sigma \in (0, 1/2)$, there
exists a constant $C>0$ such that, for all boxes $\Lambda$,
\begin{equation} \label{d516}
\int_{\Omega^{\rm t}} \left( \sum_{\ell} \|\omega_\ell \|^2_{
C_\beta^\sigma} w_\alpha (0,\ell) \right)^2 \pi^{\rm per}_{ \Lambda}
({\rm d}\omega) \leq C.
\end{equation}
Thereby, the family  $\{\pi_{ \Lambda}^{\rm per}\}_{\Lambda \in
\mathcal{L}_{\rm box}}$ is $\mathcal{W}^{\rm t}$-relatively compact.
\end{lemma}
Let $\mathcal{M}$ be the family of $\mathcal{W}^{\rm
t}$-accumulation points of $\{\pi_{ \Lambda}^{\rm per}\}_{\Lambda
\in \mathcal{L}_{\rm box}}$.
\begin{proposition} \label{periodtm}
It follows that $\mathcal{M}\subset \mathcal{G}^{\rm t}$. The
elements of $\mathcal{M}$, called periodic Euclidean Gibbs measures,
are translation invariant.
\end{proposition}
The proof of this statement is similar to the proof of Proposition
\ref{3lm}. It can be done by demonstrating that each $\mu \in
\mathcal{M}$ solves the DLR equation (\ref{40}). To this end, for
chosen $\Lambda \Subset \mathbb{L}$, one picks the box $\Delta$
containing this $\Lambda$, and shows that
\[
\int_\Omega \pi_\Lambda (\cdot |\xi) \pi^{\rm per}_\Delta ({\rm
d}\xi)
 \Rightarrow \mu (\cdot) , \quad \ \  {\rm if}  \ \ \pi^{\rm per}_\Delta
  \Rightarrow \mu \quad {\rm in} \ \ \mathcal{W}^{\rm t}.
\]
Here both convergence are taken along a subsequence of
$\mathcal{L}_{\rm box}$.

\subsection{The pressure}
\label{2.8.ss}

In the translation invariant case, one can introduce a thermodynamic function, which
contains important information about the
thermodynamic properties of the model. This is  the pressure, which in our case up to a factor coincides with the free energy density. As our special attention will be
given to the dependence of the pressure on the external field $h$,
c.f. (\ref{4}), we  indicate this  dependence explicitely.
For $\Lambda \Subset \mathbb{L}$, we set, see (\ref{a46}),
\begin{equation} \label{c1}
p_\Lambda (h, \xi) = \frac{1}{|\Lambda|} \log N_\Lambda (h,\xi),
\quad \xi \in {\Omega}^{\rm t}.
\end{equation}
To simplify notations we write $p_\Lambda (h) = p_\Lambda (h, 0)$.
Thereby, for $\mu \in \mathcal{G}^{\rm t}$, we set
\begin{equation} \label{c2}
p^\mu_\Lambda (h) = \int_{{\Omega}}p_\Lambda (h , \xi)
\mu({\rm d}\xi).
\end{equation}
Furthermore, we set
\begin{equation} \label{ac}
p^{\rm per}_\Lambda (h)  = \frac{1}{|\Lambda|} \log N^{\rm
per}_\Lambda (h).
\end{equation}
If, for
a cofinal sequence $\mathcal{L}$, the limit
\begin{equation} \label{c3}
p^\mu (h) \ \stackrel{\rm def}{=} \ \lim_{\mathcal{L}}p^\mu_\Lambda
(h),
\end{equation}
exists, we  call it pressure in the state $\mu$. We shall also
consider
\begin{equation} \label{CC}
p(h) \  \stackrel{\rm def}{=} \ \lim_{\mathcal{L}} p_\Lambda (h),
\quad \ \ p^{\rm per} (h) \  \stackrel{\rm def}{=} \
\lim_{\mathcal{L}_{\rm box}}  p^{\rm per}_\Lambda (h) .
\end{equation}
 Given $l = (l_1 , \dots l_d)$, $l' = (l'_1
, \dots l'_d)\in \mathbb{L}= \mathbb{Z}^d$, such that $l_j < l'_j$
for all $j=1, \dots , d$, we set
\begin{equation} \label{de1}
\Gamma = \{ \ell \in \mathbb{L} \ | \ l_j \leq  \ell_j \leq l'_j, \
\ {\rm for} \ {\rm all}\  j = 1 , \dots , d\}.
\end{equation}
For this parallelepiped, let $\mathfrak{G}(\Gamma)$ be the family of
all pair-wise disjoint translates of $\Gamma$ which cover
$\mathbb{L}$. Then for $\Lambda \Subset \mathbb{L}$, we let
$N_{-}(\Lambda|\Gamma)$ (respectively, $N_{+}(\Lambda|\Gamma)$) be the number of the elements of $\mathfrak{G}(\Gamma)$ which are
contained in $\Lambda$ (respectively, which have non-void
intersections with $\Lambda$).
\begin{definition} \label{rdf}
A cofinal sequence $\mathcal{L}$ is a van Hove sequence if for every
$\Gamma$,
\begin{equation} \label{de2}
(a) \ \  \lim_{\mathcal{L}} N_{-}(\Lambda |\Gamma) = +\infty; \quad
\quad (b) \ \ \lim_{\mathcal{L}}\left( N_{-}(\Lambda |\Gamma)/
N_{+}(\Lambda |\Gamma)\right)= 1.
\end{equation}
\end{definition}
One observes that $\mathcal{L}_{\rm box}$ is a van Hove sequence. It
is known, see Theorem 3.10 in \cite{[KoT]}, that
\begin{proposition} \label{pressuretm}
For every $h\in \mathbb{R}$ and any van Hove sequence $\mathcal{L}$,
it follows that the limits (\ref{c3}) and (\ref{CC}) exist, do not
depend on the particular choice of $\mathcal{L}$, and are equal,
that is $p(h) = p^{\rm per} (h) = p^\mu (h)$ for each $\mu\in
\mathcal{G}^{\rm t}$.
\end{proposition}
Let the model be rotation invariant, see Definition \ref{1df}. Then
the pressure depends on the norm of the vector $h\in
\mathbb{R}^\nu$. Therefore, without loss of generality one can
choose the external field to be $(h, 0, \dots , 0)$, $h \in
\mathbb{R}$. For the measure (\ref{a27}), by $\nu_\Lambda^{(0)}$ we
denote its version with $h=0$. Then
\begin{equation} \label{A10}
N_\Lambda (h) = N_\Lambda (0) \int_{\Omega_\Lambda} \exp \left( h
\sum_{\ell \in \Lambda} \int_0^\beta \omega_\ell^{(1)} (\tau) {\rm
d}\tau \right)\nu_\Lambda^{(0)} ({\rm d}\omega_\Lambda).
\end{equation}
The same representation can also be written for $N_\Lambda^{\rm
per}(h)$. One can show that the pressures $p_\Lambda (h)$ and
$p_\Lambda^{\rm per}(h)$, as functions of $h$, are analytic in a
subset of $\mathbb{C}$, which contains $\mathbb{R}$. Thus, one can
compute the derivatives and obtain
\begin{equation} \label{ZiF1}
\frac{\partial}{\partial h} p_\Lambda (h) = \beta M_\Lambda (h),
\qquad \frac{\partial}{\partial h} p^{\rm per}_\Lambda (h) = \beta
M^{\rm per}_\Lambda (h),
\end{equation}
where
\begin{equation} \label{ZiF2}
M_\Lambda (h) \ \stackrel{\rm def}{=} \ \frac{1}{|\Lambda|}
\sum_{\ell \in \Lambda} \varrho_{ \Lambda} [q^{(1)}_\ell], \quad
M^{\rm per}_\Lambda (h) \ \stackrel{\rm def}{=} \
 \varrho^{\rm per}_{ \Lambda} [q^{(1)}_\ell]
\end{equation}
are local {\it polarizations}, corresponding to the zero and
periodic boundary conditions respectively. Furthermore,
\begin{eqnarray} \label{ZiF3}
& & \frac{\partial^2}{\partial h^2} p_\Lambda (h) \\ & & \qquad =
\frac{1}{2|\Lambda|} \int_{\Omega_{ \Lambda}} \int_{\Omega_{
\Lambda}} \left[\sum_{\ell \in \Lambda} \int_0^\beta
\left(\omega^{(1)}_\ell (\tau) - \tilde{\omega}^{(1)}_\ell (\tau) \right){\rm
d}\tau \right]^2 \nu_{ \Lambda} ({\rm d}\omega_\Lambda)\nu_{
\Lambda}
({\rm d}\tilde{\omega}_\Lambda) \geq 0. \nonumber
\end{eqnarray}
The same can be said about the second derivative of $p^{\rm per}_\Lambda (h)$.
Therefore, both $p_\Lambda (h)$ and $p^{\rm per}_\Lambda (h)$ are
convex functions. For the reader convenience, we present here the
corresponding properties of convex functions following
\cite{[Simon]}, pages 34 - 37.

For a function $\varphi: \mathbb{R} \rightarrow \mathbb{R}$, by
$ \varphi'_{\pm}(t)$ we denote its one-side
derivatives at a given $t\in \mathbb{R}$.  By {\it at most countable
set} we mean the set which is void, finite, or countable.
\begin{proposition} \label{convpn}
For a convex function $\varphi: \mathbb{R} \rightarrow \mathbb{R}$,
it follows that: \vskip.1cm
\begin{tabular}{ll}
(a) \ &the derivatives $\varphi'_{\pm}(t)$ exist for every
$t\in \mathbb{R}$;\\ &the set $\{t\in \mathbb{R} \ | \ \varphi'_{+}
(t) \neq  \varphi'_{-} (t)\}$ is
at most countable;\\[.2cm]
(b) \ &for every $t\in \mathbb{R}$ and $\theta> 0$,
\end{tabular}
\vskip.1cm
\begin{equation} \label{S6z}
\varphi'_{-} (t) \leq \varphi'_{+} (t) \leq  \varphi'_{-} (t+
\theta ) \leq \varphi'_{+} (t+\theta);
\end{equation}
\vskip.1cm
\begin{tabular}{ll}
(c) \ &the point-wise limit $\varphi$ of a sequence of convex
functions $\{\varphi_n\}_{n\in \mathbb{N}}$\\ &is a convex function;
if $\varphi$ and all $\varphi_n$'s are differentiable at a\\ &given
$t$,  $\varphi'_{n}(t) \rightarrow \varphi' (t)$ as $n \rightarrow +
\infty$.
\end{tabular}
\end{proposition}
\begin{proposition} \label{ZiFtm}
The pressure $p(h)$, see Proposition \ref{pressuretm}, is a convex function of $h\in \mathbb{R}$.
Therefore, the set
\begin{equation} \label{ZiF4}
\mathcal{R} \ \stackrel{\rm def}{=} \ \{ h \in \mathbb{R} \ | \
 p'_{-} (h) <  p'_{+} (h) \}
\end{equation}
is at most countable. For any $h\in \mathcal{R}^c$ and any van Hove
sequence $\mathcal{L}$, it follows that
\begin{equation}
\label{ZiF5} \lim_{\mathcal{L}} M_\Lambda (h) =
\lim_{\mathcal{L}_{\rm box}} M^{\rm per}_\Lambda (h) = \beta^{-1}
p'(h) \  \stackrel{\rm def}{=} \ M(h) .
\end{equation}
\end{proposition}
By this statement, for any $h\in \mathcal{R}^c$, the limiting
periodic state is unique. In the scalar case, one can tell more on
this item. The following result is a consequence of Propositions
\ref{pressuretm} and \ref{MAtm}.
\begin{proposition} \label{Mco}
If $\nu = 1$ and $p(h)$ is differentiable at a given $h\in
\mathbb{R}$, then $\mathcal{G}^{\rm t}$ is a singleton at this $h$.
\end{proposition}
Returning to the general case $\nu \in \mathbb{N}$ we note that by
Proposition \ref{ZiFtm} the global polarization $M (h)$ is a
nondecreasing function of $h\in \mathcal{R}^c$; it is continuous on
each open connected component of $\mathcal{R}^c$. That is, $M(h)$ is
continuous on the intervals $(a_{-}, a_{+})\subset \mathcal{R}^c$,
where $a_{\pm}$ are two consecutive elements of $\mathcal{R}$. At
each such $a_{\pm}$, the global magnetization is discontinuous. One
observes however that the set $\mathcal{R}^c$ may have empty
interior; hence, $M(h)$ may be nowhere continuous.

In the sequel, to study phase transitions in the model with the anharmonic potentials $V$ of general type, we use the regularity of the temperature loops and Proposition \ref{grrpn}.
Let the model be just translation invariant.
i.e., the
anharmonic potential has the form (\ref{4}), where $V^0$ is
independent of $\ell$. Let us consider the following measure on
$C_\beta$:
\begin{eqnarray}
\label{ZiFF1} \lambda ({\rm d} \omega ) & = & \frac{1}{N_\beta}
\exp\left( - \int_0^\beta V^0(\omega(\tau)){\rm d}
 \tau \right) \chi ({\rm d} \omega), \\ N_\beta & = & \int_{C_\beta}
\exp\left( - \int_0^\beta V^0(\omega(\tau)){\rm d} \tau \right) \chi
({\rm d} \omega),  \nonumber
\end{eqnarray}
where $\chi$ is H{\o}egh-Krohn's measure. For a box $\Lambda$, we introduce the following functions on
$\Omega_{ \Lambda}$
\begin{eqnarray}
\label{ZiFF9}
Y_\Lambda (\omega_\Lambda) & = & \frac{1}{2} \sum_{\ell , \ell' \in \Lambda} J^\Lambda_{\ell \ell'} \sum_{j=1}^\nu \int_0^\beta \omega^{(j)}_\ell (\tau) \omega^{(j)}_{\ell'}(\tau){\rm d}\tau ,\\
X^{(j)}_\Lambda (\omega_\Lambda)& = & \sum_{\ell \in \Lambda}
\int^{\beta}_0 \omega^{(j)}_\ell (\tau) {\rm d}\tau, \quad \ \ j = 1
, \dots , \nu. \nonumber
\end{eqnarray}
Then
 from (\ref{ac}) one gets
\begin{eqnarray}
\label{ZiFF10} p^{\rm per}_{\Lambda} (h)
& = & \log N_\beta \nonumber \\ & + & \frac{1}{|\Lambda|}\log \left\{\int_{\Omega_{ \Lambda}} \exp \left[Y_\Lambda (\omega_\Lambda ) + \sum_{j=1}^\nu h^{(j)} X^{(j)}_\Lambda (\omega_\Lambda) \right]
 \prod_{\ell \in \Lambda} \lambda ({\rm d}\omega_\ell) \right\}.
\end{eqnarray}
As the measure (\ref{ZiFF1}) is a perturbation of the H{\o}egh-Krohn measure, we can study the regularity of the associated stochastic process by means of Proposition \ref{grrpn}.
Fix some $p\in
\mathbb{N}\setminus\{1\}$ and $\sigma \in (0, 1/2 - 1/2p)$. Thereby,
for $\vartheta \in (0, \beta)$, one obtains
\begin{eqnarray*}
\int_{C_\beta} \Xi_\vartheta^p (\omega) \lambda ({\rm d}\omega) \leq
e^{-\beta B_V }\cdot \langle \Xi_\vartheta^p  \rangle_{\chi}/
N_\beta ,
\end{eqnarray*}
$B_V$ being as in (\ref{a2}). By Proposition \ref{grrpn} this yields
\begin{equation}
\label{ZiFF4} \langle \Xi^p_\vartheta \rangle_{\lambda} \leq
D_V(\sigma, \nu, p) m^{-p} \vartheta^{ p(1 - 2 \sigma)},
\end{equation}
where, see (\ref{a26c}),
\[
D_V(\sigma, \nu, p) \ \stackrel{\rm def}{=} \ \frac{2^{3(2p+1)}( 1 +
1/ \sigma p)^{2p}}{(p-1 - 2 p \sigma) (p - 2 p \sigma)}\cdot \frac{2^p \exp\left( - \beta B_V\right) \Gamma
(\nu/2 + p)}{ N_\beta \Gamma (\nu/2)}.
\]
For $c>0$ and $n\in \mathbb{N}$, $n\geq 2$, we
set
\begin{equation}
\label{ZiFF5} C^{\pm} (n;c) = \{\omega\in C_\beta  \ | \ \pm
\omega^{(j)}(k\beta /n)\geq c, \ j=1 , \dots , \nu; \ k = 0, 1,
\dots n\}.
\end{equation}
For every $n\in \mathbb{N}$, $j_1, \dots, j_n \in \{1, \dots,
\nu\}$, and $\tau_1 , \dots , \tau_n \in [0,\beta]$, the joint
distribution of $\omega^{(j_1)}(\tau_1), \dots ,
\omega^{(j_n)}(\tau_n)$ induced by H{\o}egh-Krohn's measure $\chi$
is Gaussian. Therefore, $\chi (C^{\pm} (n;c)) >0$. Clearly, the same
property has the measure (\ref{ZiFF1}). Thus, we have
\begin{equation} \label{WQ}
\Sigma(n;c) \ \stackrel{\rm def}{=} \ \min\left\{\lambda \left(C^{+}(n;c) \right);\lambda
\left(C^{-}(n;c) \right) \right\}>0.
\end{equation}
For $\varepsilon \in (0, c)$, we set
\begin{eqnarray}
\label{ZiFF6}
A(c;\varepsilon) & = & \{\omega\in C_\beta  \ | \ \Xi_{\beta/n}
(\omega) \leq (c - \varepsilon)^{2}(\beta/n)^{-2 \sigma } \},\\
B^{\pm}(\varepsilon,c)& = & A(c;\varepsilon)\bigcap  C^{\pm} (n;c).
\nonumber
\end{eqnarray}
Then for any $\tau \in [0, \beta]$, one finds $k \in \mathbb{N}$
such that $|\tau - k \beta /n| \leq \beta/n$, and hence, for any
$j=1, \dots , \nu$,
\[
|\omega^{(j)} (\tau) - \omega^{(j)} (k\beta/n)| \leq
\left[\Xi_{\beta/n} (\omega)\right]^{1/2}(\beta/ n)^{\sigma},
\]
which yields $\pm \omega^{(j)} (\tau) \geq \varepsilon$ if
$\omega\in B^{\pm}(\varepsilon,c)$. Let us estimate $\lambda [
B^{\pm}(\varepsilon,c)]$. By (\ref{ZiFF4}) and Chebyshev's
inequality, one gets
\begin{eqnarray*}
\lambda \left(C_\beta \setminus A(c; \varepsilon) \right)
&\leq & \frac{\beta^{ 2 \sigma p}}{n^{ 2 \sigma p}(c-\varepsilon)^{2p}} \langle \Xi^p_{\beta/n} \rangle_{\lambda}\\
 & \leq & \frac{\beta^p D_V (\sigma, \nu, p)}{[m n (c-\varepsilon)^2]^p} .
\end{eqnarray*}
Thereby,
\begin{eqnarray}
\label{ZiFF7} \lambda \left[B^{\pm}(\varepsilon,c) \right] & = &
\lambda\left[ C^{\pm}(n;c) \setminus \left(C_\beta \setminus
A(c;\varepsilon)  \right)\right] \\ &\geq&
\Sigma(n;c) - \lambda  \left(C_\beta \setminus A(c;\varepsilon)  \right) \nonumber \\
& \geq & \Sigma(n;c) - \frac{\beta^p D_V (\sigma , \nu, p)}{\left[m n (c-\varepsilon)^2\right]^p} \nonumber\\
&  \stackrel{\rm def}{=} & \gamma (m), \nonumber
\end{eqnarray}
which is positive, see (\ref{WQ}), for all
\begin{equation}
\label{ZiFF8} m \geq m_* \ \stackrel{\rm def}{=} \ \frac{\beta}{n
(c-\varepsilon)^2} \cdot\left( \frac{D_V (\sigma , \nu,
p)}{\Sigma(n;c)}\right)^{1/p}.
\end{equation}
This result will be used for estimating the integrals in (\ref{ZiFF10}).

\section{Phase Transitions}
\label{3s}

There exist several approaches to describe phase transitions. Their
common point is that the macroscopic equilibrium properties of a
statistical mechanical model can be different at the same values of
the model parameters.
That is, one speaks about the possibility for the multiple states to
exist rather than the transition (as a process) between these states
or between their uniqueness and multiplicity.

\subsection{Phase transitions and order parameters}
\label{3.1.ss}

We begin by introducing the main notion of this section.
\begin{definition} \label{phdef}
The model described by the Hamiltonians (\ref{U1}), (\ref{U2}) has a
phase transition if $|\mathcal{G}^{\rm t}|>1$ at certain values of
$\beta$ and the model parameters.
\end{definition}
Note that here we demand the existence of multiple \emph{tempered}
Euclidean Gibbs measures. For models with finite range interactions,
there may exist Euclidean Gibbs measures, which are not tempered.
Such measures should not be taken into account. Another observation
is that in Definition \ref{phdef} we do not assume any symmetry of
the model, the translation invariance including. If the model is
rotation invariant (symmetric for $\nu=1$, see Definition
\ref{1df}), the unique element of $\mathcal{G}^{\rm t}$ should have
the same symmetry. If $|\mathcal{G}^{\rm t}|>1$, the symmetry can be
`distributed' among the elements of $\mathcal{G}^{\rm t}$. In this
case, the phase transition is connected with {\it a symmetry
breaking}. In the sequel, we consider mostly phase transitions of
this type. However, in subsection \ref{6.3.3.ss} we study the case where the anharmonic potentials $V_\ell$ have no symmetry and hence there is no symmetry breaking connected with the phase transition.

If the model is translation invariant, the multiplicity of its
Euclidean Gibbs states is equivalent to the existence of non-ergodic
elements of $\mathcal{G}^{\rm t}$, see Corollary \ref{Uergco}. Thus,
to prove that the model has a phase transition it is enough to show
that there exists an element of $\mathcal{G}^{\rm t}$, which fails
to obey (\ref{U9}). In the case where the model is not translation invariant, we employ a comparison
method, based on correlation inequalities. Its main idea is that the
model has a phase transition if the translation invariant model with
which we compare it has a phase transition.

Let us consider first the translation and rotation invariant case. Given
$\ell$ and $j = 1, \dots , \nu$, we set
\begin{equation} \label{nrp10}
D^\Lambda_{\ell \ell'} =  \beta \int_0^\beta  \big{\langle}
\left(\omega_\ell (\tau), \omega_{\ell'} (\tau')\right)
\big{\rangle}_{\nu_{\Lambda}^{\rm per}} {\rm d}\tau'.
\end{equation}
The right-hand side in (\ref{nrp10}) does not depend on $\tau$ due
to the property (\ref{a11}). To introduce the Fourier transformation
in the box $\Lambda$  we employ the conjugate set $\Lambda_*$
(Brillouin zone), consisting of the vectors $p = (p_1 , \dots ,
p_d)$, such that
\begin{equation} \label{rp39}
 \ p_j = - \pi + \frac{
\pi}{L} s_j, \ s_j = 1 , \dots , 2L, \ j = 1, \dots , d.
\end{equation}
Then the Fourier transformation is
\begin{eqnarray} \label{rp40}
{\omega}^{(j)}_{\ell} (\tau) & = & \frac{1}{|\Lambda|^{1/2}} \sum_{p
\in \Lambda_*} \hat{\omega}^{(j)}_p (\tau) e^{\imath (p,\ell)}, \\
\hat{\omega}^{(j)}_p (\tau) & = & \frac{1}{|\Lambda|^{1/2}}
\sum_{\ell \in \Lambda} {\omega}^{(j)}_\ell (\tau) e^{-\imath
(p,\ell)}. \nonumber
\end{eqnarray}
In order that ${\omega}^{(j)}_\ell (\tau)$ be real, the Fourier
coefficients should satisfy
\[
\overline{ \hat{\omega}^{(j)}_p (\tau)} =  \hat{\omega}^{(j)}_{ -p}
(\tau).
\]
By the rotation invariance of the state $\langle \cdot
\rangle_{\nu_{ \Lambda}^{\rm per}}$, as well as by its invariance
with respect to the translations of the torus $\Lambda$, it follows
that
\begin{equation} \label{rp40k}
\langle \hat{\omega}^{(j)}_p (\tau) \hat{\omega}^{(j')}_{p'} (\tau')
\rangle_{\nu_{ \Lambda}^{\rm per}}  =  \delta_{jj'} \delta (p + p')
\sum_{\ell'\in \Lambda} \langle {\omega}_\ell^{(j)} (\tau)
{\omega}^{(j)}_{\ell'} (\tau') \rangle_{\nu_{ \Lambda}^{\rm per}}
e^{\imath (p, \ell' - \ell)}.
\end{equation}
Thus,  we set
\begin{eqnarray} \label{rp40z}
\widehat{D}^\Lambda_p & = & \sum_{\ell' \in \Lambda} D^\Lambda_{\ell
\ell'}e^{\imath (p, \ell' - \ell)}
,\\
D^\Lambda_{\ell \ell'} & = & \frac{1}{|\Lambda|} \sum_{p \in
\Lambda_*}\widehat{D}^\Lambda_p e^{\imath (p, \ell - \ell')}.
\nonumber
\end{eqnarray}
One observes that $\widehat{D}^\Lambda_p$ can be extended to
all $p \in (-\pi, \pi]^d$. Furthermore,
\begin{equation}
\label{RP} \widehat{D}^\Lambda_p = \widehat{D}^\Lambda_{-p} =
\sum_{\ell'\in \Lambda} D^\Lambda_{\ell \ell'}\cos (p, \ell' - \ell)
,
\end{equation}
and
\begin{equation}
\label{RP1} D^\Lambda_{\ell \ell'} = \frac{1}{|\Lambda|} \sum_{p \in
\Lambda_*} \widehat{D}^\Lambda_p e^{\imath (p, \ell - \ell')} =
\frac{1}{|\Lambda|} \sum_{p \in \Lambda_*} \widehat{D}^\Lambda_p
\cos (p, \ell - \ell').
\end{equation}
For $u_\Lambda = (u_\ell)_{\ell \in \Lambda}$, $u_\ell \in
\mathbb{R}$,
\begin{eqnarray} \label{MaR1}
\left(u_\Lambda, D^\Lambda u_\Lambda \right)_{l^2(\Lambda)} &
\stackrel{\rm def}{=} & \sum_{\ell, \ell'\in \Lambda}
D^\Lambda_{\ell \ell'} u_{\ell} u_{\ell'} \\ & = & \sum_{j=1}^\nu
\bigg{\langle}
 \left[\sum_{\ell \in \Lambda} u_\ell \int_0^\beta
\omega_\ell^{(j)}(\tau) {\rm d}\tau
\right]^2\bigg{\rangle}_{\nu_{\Lambda}^{\rm per}}\geq 0. \nonumber
\end{eqnarray}
Thereby, the operator $D^\Lambda : l^2(\Lambda) \rightarrow
l^2(\Lambda)$ is strictly positive; hence, all its eigenvalues
$\widehat{D}^\Lambda_p$ are also strictly positive.

Suppose now that we are given a continuous function $\widehat{B}:
(-\pi, \pi]^d \rightarrow (0, +\infty]$ with the following
properties:
\begin{eqnarray} \label{rp40x}
& & {\rm (i)} \qquad \int_{(-\pi, \pi]^d} \widehat{B} (p) {\rm d}p < \infty,\\
& & {\rm (ii)} \qquad  \widehat{D}^\Lambda_p \leq \widehat{B} (p),
\quad {\rm for} \ {\rm all} \ p \in \Lambda_*\setminus \{0\},
\nonumber
\end{eqnarray}
holding for all boxes $\Lambda$.  Then we set
\begin{equation}
\label{RP2} B_{\ell \ell'} = \frac{1}{(2\pi)^d} \int_{(-\pi, \pi]^d}
\widehat{B} (p)
 \cos (p , \ell -\ell'){\rm d}p, \quad \ell , \ell' \in \mathbb{L},
\end{equation}
and
\begin{equation}
\label{RP3} B_{\ell \ell'}^\Lambda  = \frac{1}{|\Lambda|} \sum_{p
\in \Lambda_* \setminus\{0\}} \widehat{B} (p) \cos (p , \ell
-\ell'), \quad \ell, \ell' \in  \Lambda.
\end{equation}
We also set $B_{\ell \ell'}^\Lambda = 0$ if either of $\ell , \ell'$
belongs to $\Lambda^c$.
\begin{proposition}
\label{RPpn} For every $\ell , \ell'$, it follows that
$B^{\Lambda_L}_{\ell \ell'} \rightarrow B_{\ell \ell'}$ as $L
\rightarrow +\infty.$
\end{proposition}
{\it Proof:} By (\ref{rp40x}), $\widehat{B} (p) \cos (p , \ell -\ell')$ is an
absolutely integrable function in the sense of improper Riemann
integral. The right-hand side of (\ref{RP3}) is its integral sum;
thereby, the convergence stated is obtained in a standard way.
$\square$

From claim (i) of (\ref{rp40x}) by the Riemann-Lebesgue lemma, see
page 116 in \cite{[LiebL]}, one obtains
\begin{equation} \label{RL}
\lim_{|\ell - \ell'|\rightarrow +\infty} B_{\ell \ell'} = 0.
\end{equation}

\begin{lemma}
\label{RPlm} For every box $\Lambda$ and any $\ell , \ell'\in
\Lambda$, it follows that
\begin{equation}
\label{RP5} D^\Lambda_{\ell \ell'} \geq \left(D^\Lambda_{\ell \ell}
- B^\Lambda_{\ell \ell} \right) + B^\Lambda_{\ell \ell'}.
\end{equation}
\end{lemma}
{\it Proof:}
By (\ref{RP1}), (\ref{RP3}), and claim (ii) of (\ref{rp40x}), one
has
\begin{eqnarray*}
D^\Lambda_{\ell \ell} - D^\Lambda_{\ell \ell'} & = &
\frac{2}{|\Lambda|} \sum_{p\in \Lambda_* \setminus
 \{0\}} \widehat{D}^\Lambda_p \sin^2 (p , \ell - \ell') \\ & \leq & \frac{2}{|\Lambda|} \sum_{p\in \Lambda_* \setminus \{0\}}
  \widehat{B}(p) \sin^2 (p , \ell - \ell')  \\ & = & B^\Lambda_{\ell \ell} - B^\Lambda_{\ell \ell'},
\end{eqnarray*}
which yields (\ref{RP5}).
$\square$

For $\mu \in \mathcal{G}^{\rm t}$, we set, c.f., (\ref{nrp10}),
\begin{equation} \label{nrp90}
D^\mu_{\ell\ell'} = \beta \int_0^\beta \langle (\omega_\ell (\tau), \omega_{\ell'} (\tau') )\rangle_{\mu} {\rm d}\tau'.
\end{equation}
\begin{corollary}
\label{RPco} For every periodic $\mu\in \mathcal{G}^{\rm t}$, it
follows that
\begin{equation}
\label{RP6} D^\mu_{\ell \ell'} \geq \left(D^\mu_{\ell \ell} -
B_{\ell \ell} \right) + B_{\ell \ell'},
\end{equation}
holding for any $\ell , \ell'$.
\end{corollary}
{\it Proof:}
For periodic $\mu\in \mathcal{G}^{\rm t}$, one finds the sequence
$\{L_n\}_{n\in \mathbb{N}}\subset \mathbb{N}$, such that
$\pi^{\rm per}_{\Lambda_{L_n}} \Rightarrow \mu$ as $n \rightarrow +\infty$, see Proposition \ref{periodtm}.
This fact alone does not yet mean that $D^{\Lambda_{L_n}}_{\ell \ell'} \rightarrow D^{\mu}_{\ell \ell'}$, what we would like to get. To prove the latter convergence one employs Lemma \ref{d5lm} and
proceeds as in the proof of claim (b) of Lemma 5.2 in \cite{[KoT]}.
 Then (\ref{RP6})
 follows from (\ref{RP5}) and Proposition \ref{RPpn}.
$\square$

One observes that the first summand in (\ref{RP6}) is independent of
$\ell$, whereas the second one may be neither positive nor summable.
Suppose now that there exists a positive $\vartheta$ such that, for
any box $\Lambda$,
\begin{equation}\label{rp40w}
D^\Lambda_{\ell \ell} \geq \vartheta.
\end{equation}
Then, in view of (\ref{RL}), the phase transition occurs if
\begin{equation} \label{RP4}
 \quad \vartheta > B_{\ell \ell}.
\end{equation}
For certain versions of our model, we find the function
$\widehat{B}$ obeying the conditions (\ref{rp40x}) and the bound
(\ref{RP4}). Note that under (\ref{rp40w}) and (\ref{RP4}) by
(\ref{RP6}) it follows that
\begin{equation} \label{nrp1}
\lim_{L\rightarrow +\infty} \frac{1}{|\Lambda_L|} \sum_{\ell'\in
\Lambda_L}  D^\mu_{\ell \ell'} = \lim_{L\rightarrow +\infty}
\frac{1}{|\Lambda_L|^2} \sum_{\ell, \ell'\in \Lambda_L} D^\mu_{\ell
\ell'} > 0.
\end{equation}
Let us consider now another possibilities to define phase
transitions in translation invariant versions of our model. For a
box $\Lambda$, see (\ref{box}), we introduce
\begin{eqnarray} \label{rpp}
P_\Lambda & = & \frac{1}{(\beta |\Lambda|)^2} \sum_{\ell ,\ell'\in
\Lambda} D^\Lambda_{\ell \ell'} \\ & = & \int_{\Omega_{ \Lambda}}
\left\vert\frac{1}{\beta |\Lambda|}\sum_{\ell \in
\Lambda}\int_0^\beta \omega_\ell (\tau) {\rm d}\tau \right\vert^2
\nu^{\rm per}_{ \Lambda}({\rm d }\omega_\Lambda), \nonumber
\end{eqnarray}
and set
\begin{equation} \label{rpp1}
 P \ \stackrel{\rm
def}{=} \ \limsup_{L \rightarrow +\infty } P_{\Lambda_L}.
\end{equation}
\begin{definition} \label{rppdf}
The above $P$ is called the order parameter. If $P>0$ for given
values of $\beta$ and the model parameters, then there exists a long
range order.
\end{definition}
By standard arguments one proves the following
\begin{proposition} \label{rpppn}
If (\ref{rp40w}) and (\ref{RP4}) hold, then  $P>0$.
\end{proposition}
The appearance of the long range order, which in a more `physical'
context is identified with a phase transition, does not imply the
phase transition in the sense of Definition \ref{phdef}. At the same
time, Definition \ref{phdef}  describes models without translation
invariance. On the other hand, Definition \ref{rppdf} is based upon
the local states only and hence can be formulated without employing
$\mathcal{G}^{\rm t}$.
 Yet another `physical'
approach to phase transitions in translation invariant models like
(\ref{U1}), (\ref{U2}) is based on the properties of the pressure
$p(h)$,  which by Proposition \ref{pressuretm} exists and is the
same in every state. It  does not employ  the set $\mathcal{G}^{\rm
t}$ and is based on the continuity of the global polarization
(\ref{ZiF5}), that is, on the differentiability of $p(h)$.
\begin{definition} [Landau Classification] \label{landau}
The model has a first order phase transition  if $p'(h)$
 is discontinuous at a certain $h_*$. The model has a second order phase transition
 if there exists $h_*\in \mathbb{R}^\nu$ such that $p'(h)$ is continuous but  $p''(h)$
 is discontinuous at
 $h=h_*$.
\end{definition}
\begin{remark} \label{landaurk}
Like in Definition \ref{phdef}, here we do not assume any symmetry
of the model (except for the translation invariance). As $p(h)$ is convex, $p'(h)$ is increasing; hence, $p''(h) \geq 0$. The discontinuity of the latter mentioned in Definition \ref{landau} includes the case $p''(h_*) = +\infty$, where the polarization $M(h)$ at $h=h_*$ grows infinitely fast, but still is continuous.
\end{remark}
The
relationship between the first order phase transition and  the long
range order is established with the help of the following result,
the proof of which can be done by a slight modification of the
arguments used in \cite{[DLS]}, see Theorem 1.1 and its corollaries.
Let $\{\mu_n\}_{N\in \mathbb{N}}$ (respectively, $\{M_n\}_{n\in
\mathbb{N}}$)  be a sequence of probability measures on $\mathbb{R}$
(respectively, positive real numbers, such
that $\lim M_n = + \infty$). We also suppose that,  for any $y\in\mathbb{R}$,
\begin{equation} \label{ggri}
f(y) = \lim_{n \rightarrow + \infty} \frac{1}{M_n} \log \int e^{y
u}\mu_n({\rm d}u)
\end{equation}
exists and is finite. As the function $f$ is convex, it has
one-sided derivatives $f'_{\pm}(0)$, see Proposition \ref{convpn}.
\begin{proposition}[Griffiths] \label{Grpn}
Let the sequence of measures $\{\mu_n\}_{N\in \mathbb{N}}$ be as
above. If $f'_+ (0) = f'_- (0) = \phi$ (i.e., $f$ is differentiable
at $y=0$), then
\begin{equation} \label{gri}
\lim_{n\rightarrow +\infty}  \int g(u/M_n) \mu_n({\rm d}u) =
g(\phi),
\end{equation}
for any continuous $g:\mathbb{R}\rightarrow \mathbb{R}$, such that
$|g(u)| \leq \lambda e^{\varkappa |u|}$ with certain $\lambda,
\varkappa >0$. Furthermore, for each such a function $g$,
\begin{equation} \label{gri1}
\limsup_{n\rightarrow +\infty}\int g(u/M_n) \mu_n({\rm d}u) \leq
\max_{z \in [f'_- (0), f'_+ (0)]} g (z).
\end{equation}
In particular, if $f'_- (0) = - f'_+ (0)$, then  for any $k \in
\mathbb{N}$,
\begin{equation} \label{gri2}
f'_+ (0) \geq \limsup_{n\rightarrow +\infty}\left(\int (u/M_n)^{2k}
\mu_n({\rm d}u)\right)^{1/2k}.
\end{equation}
\end{proposition}
Write, c.f., (\ref{A10}),
\begin{equation} \label{gri3}
N_{ \Lambda}^{\rm per} (h) = N_{ \Lambda}^{\rm per} (0)
\int_{\Omega_{ \Lambda}} \exp\left( h \sum_{\ell \in \Lambda}
\int_0^\beta \omega^{(1)}_\ell (\tau){\rm d}\tau \right) \nu^{0,\rm
per}_{\Lambda}({\rm d}\omega_\Lambda),
\end{equation}
where $\nu^{0,\rm per}_{ \Lambda}$ is the local periodic Euclidean
Gibbs measure with $h=0$. Now let $\{L_n\}_{n\in \mathbb{N}}\subset
\mathbb{N}$ be the sequence such that the sequences of local
measures $\{\nu^{0,\rm per}_{ \Lambda_{L_n}}\}$ and $\{\nu^{\rm
per}_{ \Lambda_{L_n}}\}$ converge to the corresponding periodic
Euclidean Gibbs measures $\mu^{0}$ and $\mu$ respectively. Set
\begin{equation} \label{gri4}
\mathcal{X}_n = \left\{ \omega_{\Lambda_{L_n}} \in
\Omega_{\Lambda_{L_n}} \ \left\vert \ \exists u \in \mathbb{R}:  \ \
\sum_{\ell \in \Lambda_{L_n}} \int_0^\beta \omega^{(1)}_\ell (\tau)
{\rm d}\tau = u \right. \right\}.\end{equation} Clearly, each such
$\mathcal{X}_n$ is measurable and isomorphic to $\mathbb{R}$. Let
$\mu_n$, $n \in \mathbb{N}$, be the projection of $\{\nu^{0,\rm
per}_{ \Lambda_{L_n}}\}$ onto this $\mathcal{X}_n$. Then
\begin{equation} \label{gri5}
p(h) = p(0) + f(h),
\end{equation}
where $f$ is given by (\ref{ggri}) with such $\mu_n$ and $M_n =
|\Lambda_{L_n}|= (2L_n)^{d}$. Thereby, we apply (\ref{gri2}) with $k=2$ and obtain
\[
p'_{+} (0) \geq \beta \limsup_{n\rightarrow +\infty}\sqrt{ P_{\Lambda_{L_n}}}.
\]
Thus, in the case where the model
is just rotation and translation invariant,
the existence of the long range order implies the first order phase
transition.

Consider now the second order phase transitions in the rotation invariant
case. For $\alpha \in [0, 1]$, we set, c.f., (\ref{rpp}),
\begin{equation} \label{gri6}
P_\Lambda^{(\alpha)} =
\frac{\beta^{-2}}{|\Lambda|^{1+\alpha}}\int_{\Omega_{
\Lambda}} \left\vert\sum_{\ell \in \Lambda}\int_0^\beta \omega_\ell
(\tau) {\rm d}\tau \right\vert^2 \nu^{\rm per}_{\beta, \Lambda}({\rm
d }\omega_\Lambda),
\end{equation}
where $\Lambda$ is a box. Then $P_\Lambda^{(1)} =P_\Lambda$ and, as we just have shown, the
existence of a positive limit (\ref{rpp1}) yields a first order
phase transition.
\begin{proposition} \label{sophpn}
If there exists $\alpha \in (0,1)$, such that for a sequence
$\{L_n\}$, there exists a finite limit
\begin{equation} \label{gri7}
\lim_{n\rightarrow +\infty} P_{\Lambda_{L_n}}^{(\alpha)} \ \stackrel{\rm
def}{=} \ P^{(\alpha)} >0.
\end{equation}
Then the model has at $h=0$ a second order phase transition.
\end{proposition}
{\it Proof:}
We observe that
\[
P_{\Lambda}^{(\alpha)} = \nu p_\Lambda ''(0)/\beta^2 |\Lambda|^\alpha.
\]
Then there exists $c>0$, such that
\[
p''_{\Lambda_{L_n}} (0) \geq c |\Lambda_{L_n}|^\alpha, \quad \ \  {\rm for} \ \  {\rm all} \ \  n \in \mathbb{N}.
\]
As each $p''_\Lambda$ is continuous, one finds the sequence $\{\delta_n\}_{n\in \mathbb{N}}$ such that $\delta_n \downarrow 0$ and
\begin{equation} \label{gri8}
p''_{\Lambda_{L_n}} (h) \geq \frac{1}{2} c |\Lambda_{L_n}|^\alpha, \quad \ \  {\rm for} \ \  {\rm all} \ \  h \in [0, \delta_n] \ \ \  {\rm and} \ \ \ n \in \mathbb{N}.
\end{equation}
If $p''(0)$ were finite, see Remark \ref{landaurk}, one would get
\[
p''(0) = \lim_{n\rightarrow +\infty} \left[ p'_{\Lambda_{L_n}}(\delta_n) - p'_{\Lambda_{L_n}} (0)
\right]/\delta_n,
\]
which contradicts (\ref{gri8}).
$\square$
\vskip.1cm Proposition \ref{sophpn}
remains true if one replaces in (\ref{gri6}) the periodic local
measure $\nu_{ \Lambda}^{\rm per}$ by the one corresponding to the
zero boundary condition, i.e., by $\nu_{ \Lambda}$. Then the limit
in (\ref{gri7}) can be taken along any van Hove sequence
$\mathcal{L}$. We remind that Proposition \ref{sophpn} describes the
rotation invariant case. The existence of a positive $P^{(\alpha)}$
with $\alpha>0$ may be interpreted as follows. According to the
central limit theorem for independent identically distributed random
variables, for our model with $J_{\ell \ell'} =0$ and $V_\ell = V$,
the only possibility to have a finite positive limit in (\ref{gri7})
is to set $\alpha =0$. If $P^{(0)}< \infty$ for nonzero interaction, one
can say that the dependence between the temperature loops is weak;
it holds for small $\hat{J}_0$. Of course, in this case
$P^{(\alpha)} = 0$ for any $\alpha >0$. If $P^{(\alpha)}$ gets
positive for a certain $\alpha \in (0,1)$, one says that a strong
dependence between the loops appears. In this case, the central
limit theorem holds with an abnormal normalization. However, this
dependence is not so strong to make $p'$ discontinuous, which occurs
for $\alpha =1$, where a new law of large numbers comes to power. In
statistical physics, the point at which $P^{(\alpha)} > 0$ for
$\alpha \in(0,1)$ is called {\it a critical point}. The quantity
$P^{(0)}$ is called {\it susceptibility,} it gets discontinuous at
the critical point. Its singularity at this point is connected with
the value of $\alpha$ for which $P^{(\alpha)} > 0$. The above
analysis opens the possibility to extend the notion of the critical
point to the models which are not translation invariant.
\begin{definition} \label{Weberdf}
The rotation invariant model has a critical point if there exist a
van Hove sequence $\mathcal{L}$ and   $\alpha \in (0,1)$ such that
\begin{equation} \label{gri9}
\lim_{\mathcal{L}}\frac{1}{|\Lambda|^{1+\alpha}}\int_{\Omega_{
\Lambda}} \left\vert\sum_{\ell \in \Lambda}\int_0^\beta \omega_\ell
(\tau) {\rm d}\tau \right\vert^2 \nu_{ \Lambda}({\rm d
}\omega_\Lambda) >0
\end{equation}
at certain values of the model parameters, including $h$ and
$\beta$.
\end{definition}
Note that by Proposition \ref{sophpn}, it follow that in the translation invariant case the notions of the critical point
and of the second order phase transition coincide.

\subsection{Infrared bound}
\label{3.2.ss} Here, for  the
translation and rotation invariant version of our model, we find the function
$\widehat{B}$ obeying (\ref{rp40x}).

For a box $\Lambda$, let $E$ be the set of all unordered pairs
$\langle \ell , \ell'\rangle$, $\ell , \ell'\in \Lambda$, such that
$|\ell - \ell'|_\Lambda=1$, see (\ref{box1}). Suppose also that the interaction
intensities (\ref{A2}) are such that $J^\Lambda_{\ell \ell'} = J>0$
if and only if $\langle \ell , \ell'\rangle \in E$ and hence the measure
(\ref{A5}) can be written
\begin{equation} \label{rp16}
\nu^{\rm per}_{ \Lambda} ({\rm d}\omega_\Lambda) =
\frac{1}{Y_\Lambda (0) } \exp\left( - \frac{J}{2}\sum_{\langle \ell,
\ell'\rangle \in E} \|\omega_\ell - \omega_{\ell'}\|^2_{L^2_\beta}
\right) \sigma_{ \Lambda} ({\rm d}\omega_\Lambda),
\end{equation}
where
\begin{eqnarray} \label{rp17}
& & \sigma_{ \Lambda} ({\rm d}\omega_\Lambda) \\ & & \qquad  =
 \exp\left( Jd \sum_{\ell \in
\Lambda} \|\omega_\ell\|^2_{L^2_\beta} - \sum_{\ell \in \Lambda}
\int_0^\beta V(\omega_\ell (\tau)) {\rm d}\tau  \right)\chi_{
\Lambda}({\rm d}\omega_\Lambda), \nonumber
\end{eqnarray}
and
\begin{equation} \label{rp18}
Y_\Lambda (0) = \int_{\Omega_{ \Lambda}} \exp\left( -
\frac{J}{2}\sum_{\langle \ell, \ell'\rangle \in E} \|\omega_\ell -
\omega_{\ell'}\|^2_{L^2_\beta} \right) \sigma_{ \Lambda} ({\rm
d}\omega_\Lambda).
\end{equation}
 With every edge $\langle\ell , \ell' \rangle \in E$ we associate
$b_{\ell \ell'} \in L^2_\beta$ and consider
\begin{equation}\label{rp19}
Y_\Lambda (b) = \int_{\Omega_{ \Lambda}} \exp\left( -
\frac{J}{2}\sum_{\langle \ell, \ell'\rangle \in E} \|\omega_\ell -
\omega_{\ell'} - b_{\ell \ell'}\|^2_{L^2_\beta} \right) \sigma_{
\Lambda} ({\rm d}\omega_\Lambda).
\end{equation}
By standard arguments, see \cite{[KKE]} and the references therein,
one proves the following
\begin{lemma} [Gaussian Domination] \label{irelm}
For every $b = (b_{\ell \ell'})_{\langle \ell , \ell'\rangle \in
E}$, $b_{\ell \ell'} \in L^2_\beta$, it follows that
\begin{equation}
\label{rp20} Y_\Lambda (b) \leq Y_\Lambda (0).
\end{equation}
\end{lemma}
Let $\mathcal{X}_E$ be the real Hilbert space
\begin{equation} \label{rp33}
\mathcal{X}_E = \{ b = (b_{\ell\ell'})_{\langle \ell , \ell'\rangle
\in E} \ | \ b_{\ell \ell'} \in L^2_\beta\},
\end{equation}
with scalar product
\begin{equation} \label{rp34}
(b,c)_{\mathcal{X}_E} = \sum_{\langle \ell , \ell'\rangle \in E}
(b_{\ell \ell'}, c_{\ell \ell'})_{L^2_\beta}.
\end{equation}
To simplify notations we write $e = \langle \ell , \ell' \rangle$. A
bounded linear operator $Q: \mathcal{X}_E \rightarrow \mathcal{X}_E
$ may be defined by means of its kernel $Q^{jj'}_{ee'} (\tau,
\tau')$, $j,j' = 1, \dots , \nu$, $e,e' \in E$, and $\tau , \tau'
\in [0,\beta]$. That is
\begin{equation} \label{rp35}
\left( Q b  \right)^{(j)}_e (\tau) = \sum_{j'=1}^d \sum_{e' \in E}
\int_0^\beta Q^{jj'}_{ee'} (\tau, \tau') b_{e'}^{(j')} (\tau') {\rm
d}\tau'.
\end{equation}
Let us study the operator with the following
kernel
\begin{equation} \label{rp36}
Q^{jj'}_{\langle \ell_1 , \ell_1' \rangle \langle \ell_2 , \ell_2'
\rangle} (\tau, \tau') = \bigg{ \langle} \left[
\omega^{(j)}_{\ell_1} (\tau) - \omega^{(j)}_{\ell_1'} (\tau)\right]
\cdot\left[ \omega^{(j')}_{\ell_2} (\tau') - \omega^{(j')}_{\ell_2'}
(\tau')\right] \bigg{\rangle}_{\nu_{ \Lambda}^{\rm per}},
\end{equation}
where the expectation is taken with respect to the measure
(\ref{rp16}). This operator in positive. Indeed,
\begin{eqnarray*}
(b , Q b)_{\mathcal{X}_E} = \bigg{ \langle} \left[ \sum_{\langle
\ell , \ell'\rangle \in E} (\omega_\ell - \omega_{\ell'}, b_{\ell
\ell'})_{L^2_\beta}\right]^2 \bigg{\rangle}_{\nu_{ \Lambda}^{\rm
per}} \geq 0.
\end{eqnarray*}
The kernel (\ref{rp36}) can be expressed in terms of the Matsubara
functions; thus, as a function of $\tau, \tau'$, it has the property
(\ref{a11}). We employ the latter by introducing yet another Fourier
transformation. Set
\begin{equation} \label{FY}
\mathcal{K} = \{ k = (2\pi/\beta) \kappa \  | \ \kappa \in
\mathbb{Z}\},
\end{equation}
\begin{equation} \label{FY1}
e_k (\tau) = \left\{ \begin{array}{ll} \beta^{-1/2} \cos k \tau ,
\quad &{\rm if} \ k>0; \\ - \beta^{-1/2} \sin k \tau , \quad &{\rm
if} \ k<0; \\ \sqrt{2/\beta}, \quad &{\rm if} \ k=0. \end{array}
\right.
\end{equation}
The transformation we need is
\begin{eqnarray} \label{FY2}
\hat{\omega}_\ell^{(j)} (k) & =  & \int_0^\beta  {\omega}_\ell^{(j)}
(\tau) e_k (\tau) {\rm d}\tau, \\ {\omega}_\ell^{(j)} (\tau) & = &
\sum_{k \in \mathcal{K}} \hat{\omega}_\ell^{(j)} (k) e_k (\tau).
\end{eqnarray}
Then the property (\ref{a11}) yields, c.f., (\ref{rp40k})
\[
\langle \hat{\omega}^{(j)}_\ell (k) \hat{\omega}^{(j')}_{\ell'} (k'
) \rangle_{\nu_{\Lambda}^{\rm per}} = 0 \quad {\rm if} \ \ k \neq
k', \ \ {\rm or} \ \ j \neq j'.
\]
Taking this into account we employ in (\ref{rp36}) the
transformation (\ref{FY2}) and obtain
\begin{equation} \label{rp37}
Q^{jj'}_{\langle \ell_1 , \ell_1' \rangle \langle \ell_2 , \ell_2'
\rangle} (\tau, \tau') = \delta_{jj'} \sum_{k \in \mathcal{K}}
\widehat{Q}_{\langle \ell_1 , \ell_1' \rangle \langle \ell_2 , \ell_2'
\rangle} (k) e_k (\tau) e_k (\tau'),
\end{equation}
with
\begin{equation} \label{rp38}
\widehat{Q}_{\langle \ell_1 , \ell_1' \rangle \langle \ell_2 , \ell_2'
\rangle} (k) = \bigg{ \langle }\left[\hat{\omega}^{(j)}_{\ell_1} (k)
- \hat{\omega}^{(j)}_{\ell'_1} (k) \right] \cdot
\left[\hat{\omega}^{(j)}_{\ell_2} (k) - \hat{\omega}^{(j)}_{\ell'_2}
(k) \right]\bigg{\rangle}_{\nu_{\Lambda}^{\rm per}}.
\end{equation}
In view of the periodic conditions imposed on the boundaries of the
box $\Lambda$ the latter kernel, as well as the one given by
(\ref{rp36}), are invariant with respect to the translations of the
corresponding torus. This allows us to `diagonalize' the kernel
(\ref{rp38}) by means of a spatial Fourier transformation
(\ref{rp39}), (\ref{rp40}). Then the spacial periodicity of the
state $\langle \cdot \rangle_{\nu_{ \Lambda}^{\rm per}}$ yields
\begin{equation} \label{rp40a}
\langle \hat{\omega}^{(j)} (p,k) \hat{\omega}^{(j)} (p',k)
\rangle_{\nu_{ \Lambda}^{\rm per}} = 0 \quad {\rm if} \ \ p + p'
\neq 0.
\end{equation}
Taking this into account we obtain
\begin{eqnarray} \label{rp40b}
\widehat{Q}_{\langle \ell_1 , \ell_1' \rangle \langle \ell_2 , \ell_2'
\rangle} (k) & = & \sum_{p \in \Lambda_*} \langle \hat{\omega}^{(j)}
(p,k) \hat{\omega}^{(j)} (-p,k) \rangle_{\nu_{\beta , \Lambda}^{\rm
per}}\\ & \times &\left(e^{\imath (p,\ell_1)} - e^{\imath
(p,\ell'_1)} \right)/|\Lambda|^{1/2} \nonumber \\ & \times &
\left(e^{-\imath (p,\ell_2)} - e^{\imath (-p,\ell'_2)}
\right)/|\Lambda|^{1/2}. \nonumber
\end{eqnarray}
Since the summand corresponding to $p=0$ equals zero,  the sum can
be restricted to $\Lambda_* \setminus \{0\}$. This representation
however cannot serve as a spectral decomposition similar to
(\ref{rp37}) because the eigenfunctions here are not normalized.
Indeed,
\begin{eqnarray*}
\sum_{\langle \ell , \ell' \rangle \in E} \left(e^{\imath (p,\ell)}
- e^{\imath (p,\ell')} \right)/|\Lambda|^{1/2} \times
\left(e^{-\imath (p,\ell)} - e^{-\imath (p,\ell')}
\right)/|\Lambda|^{1/2} = 2 \mathcal{E}(p)
\end{eqnarray*}
where
\begin{equation}
 \label{rp40c}
\mathcal{E}(p) \ \stackrel{\rm def}{=} \ \sum_{j=1}^d [ 1 - \cos p_j
].
\end{equation}
Then we set
\begin{equation} \label{rp43}
\sigma_{\ell \ell'} (p) = \left(e^{\imath (p,\ell)} - e^{\imath
(p,\ell')} \right)/\sqrt{2|\Lambda| \mathcal{E}(p)}, \quad p \in
\Lambda_* \setminus \{0\},
\end{equation}
and
\begin{equation} \label{rp42}
\widehat{Q} (p,k) =  2 \mathcal{E}(p) \langle \hat{\omega}^{(j)} (p,k)
\hat{\omega}^{(j)} (-p,k) \rangle_{\nu_{ \Lambda}^{\rm per}}, \quad
p \in \Lambda_* \setminus \{0\}.
\end{equation}
Thereby,
\begin{eqnarray}  \label{rp41}
 & & Q_{\langle \ell_1 , \ell_1' \rangle \langle \ell_2 , \ell_2'
\rangle} (\tau, \tau') = \\ & & \quad  = \sum_{p \in \Lambda_*
\setminus\{0\}} \sum_{k \in \mathcal{K}}\widehat{Q} (p,k) \sigma_{\ell_1
\ell_1'} (p) \sigma_{\ell_2 \ell_2'} (-p) e_k(\tau) e_k (\tau'),
\nonumber
\end{eqnarray}
which is the spectral decomposition of the operator (\ref{rp35}).
Now we show that the eigenvalues (\ref{rp42}) have a specific upper
bound\footnote{Their natural lower bound is zero as the operator
(\ref{rp35}) is positive}.
\begin{lemma}  \label{rp2tm}
For every $p \in \Lambda_* \setminus\{0\}$ and $k \in \mathcal{K}$,
the eigenvalues (\ref{rp42}) obey the estimate
\begin{equation} \label{rp44a}
\widehat{Q} (p,k) \leq 1 /J,
\end{equation}
where $J$ is the same as in (\ref{rp16}). From this estimate one
gets
\begin{equation} \label{rp44}
\langle \hat{\omega}^{(j)} (p,k) \hat{\omega}^{(j)} (-p,k)
\rangle_{\nu_{\Lambda}^{\rm per}} \leq \frac{1}{ 2 J \mathcal{E}(p)}
, \quad p \in \Lambda_* \setminus \{0\}.
\end{equation}
\end{lemma}
{\it Proof:}
The estimate in question will be obtained from the Gaussian
domination (\ref{rp20}). For $t \in \mathbb{R}$ and a given $b \in
\mathcal{X}_E$, we consider the function $\phi(t) = Y_\Lambda (t
b)$. By Lemma \ref{irelm},  $\phi'' (0) \leq 0$. Computing the
derivative from (\ref{rp19}) we get
\[
\phi'' (0) = J (b, Q b)_{\mathcal{X}_E} - \|b\|^2_{\mathcal{X}_E},
\]
where the operator $Q$ is defined by its kernel (\ref{rp36}). Then
the estimate (\ref{rp44a}) is immediate.
$\square$
\vskip.1cm By (\ref{rp40}), (\ref{rp37}), and (\ref{rp42}), we
readily obtain
\[
\langle (\hat{\omega}_p (\tau) , \hat{\omega}_{-p} (\tau')
)\rangle_{\nu_{ \Lambda}^{\rm per}} = \frac{\nu}{2 \beta
\mathcal{E}(p)} \sum_{k \in \mathcal{K}} \widehat{Q}(p,k) \cos[k(\tau -
\tau')], \quad p\neq 0,
\]
which yields, see (\ref{rp40z}) and (\ref{rp44a}),
\begin{equation} \label{rp45}
\widehat{D}^\Lambda_p = \frac{ \beta \nu}{2 \mathcal{E}(p)}
\widehat{Q}(p, 0) \leq \frac{ \beta \nu}{2 J \mathcal{E}(p)}, \quad p
\neq 0.
\end{equation}
Comparing this estimate with (\ref{rp40x}) we have the following
\begin{corollary} \label{Irco}
If the model is translation and rotation invariant with the nearest
neighbor interaction, then the infrared estimate (\ref{rp40x}) holds
with
\begin{equation} \label{rp45z}
\widehat{B}(p) = \frac{ \beta \nu}{2 J \mathcal{E}(p)}, \quad p\in
(-\pi , \pi]^d \setminus \{0\}, \qquad \widehat{B}(0) = +\infty.
\end{equation}
\end{corollary}

\subsection{Phase transition in the translation and rotation invariant model}
\label{3.3.ss}

In this subsection, we consider the model described by Corollary
\ref{Irco}. First we obtain the lower bounds for
\[
\langle (\omega_{\ell} (\tau),
\omega_\ell(\tau))\rangle_{\nu_\Lambda^{\rm per}},
\]
from which we then obtain the bounds (\ref{rp40w}). In the case
where the anharmonic potential has the form
\begin{equation} \label{rp46}
V(u) = - b |u|^2 + b_2|u|^4, \quad b >  a/2, \ \ b_2>0,
\end{equation}
$a$ being the same as in (\ref{U1}), the bound (\ref{rp40w}) can be
found explicitly. We begin by considering this special case.
\begin{lemma} \label{ph1lm}
Let $V$ be as in (\ref{rp46}). Then, for every $\Lambda\Subset
\mathbb{L}$,
\begin{equation} \label{rp47a}
\langle (\omega_{\ell} (\tau),
\omega_\ell(\tau))\rangle_{\nu_\Lambda^{\rm per}} \geq \frac{(2b -
a) \nu}{ 4 b_2 (\nu + 2)}\ \stackrel{\rm def}{=} \vartheta_{*}.
\end{equation}
\end{lemma}
{\it Proof:}
Let $A$ be a self-adjoint operator, such that the expressions below
make sense. Then
\begin{eqnarray} \label{rp47b}
& & \varrho_{ \Lambda}^{\rm per} \left([A, [H^{\rm
per}_\Lambda,A]]\right) \\ & & \quad  =   \varrho_{\beta ,
\Lambda}^{\rm per} \left( A H^{\rm per}_\Lambda A + A H^{\rm
per}_\Lambda A - A A H^{\rm per}_\Lambda
 -  H^{\rm per}_\Lambda A A \right) \nonumber \\
& &  \quad =  \frac{1}{Z^{\rm per}_{\beta ,\Lambda}} \sum_{s, s' \in
\mathbb{N}} \left\vert A_{ss'}\right\vert^2 \left(E^{\rm per}_{s'} -
E^{\rm per}_{s} \right)\left\{ \exp\left[ - \beta E^{\rm per}_{s}
\right] - \exp\left[ - \beta E^{\rm per}_{s'} \right]\right\}
\nonumber \\ & & \quad \geq 0.\nonumber
\end{eqnarray}
Here $E^{\rm per}_s$, $s\in \mathbb{N}$ are the eigenvalues of the
periodic Hamiltonian (\ref{A5a}), $A_{ss'}$ are the corresponding
matrix elements of $A$, and $\varrho_{\Lambda}^{\rm per}$ is the periodic local
Gibbs state (\ref{A5b}). By the Euclidean representation,
\[
\langle (\omega_{\ell} (\tau),
\omega_\ell(\tau))\rangle_{\nu_\Lambda^{\rm per}}
 = \sum_{j=1}^\nu\bigg{\langle}
\left( \omega_\ell^{(j)} (0) \right)^2 \bigg{\rangle}_{\nu_{\beta,
\Lambda}^{\rm per}} = \sum_{j=1}^\nu \varrho^{\rm per}_{
\Lambda}\left[ \left( q_\ell^{(j)} \right)^2 \right].
\]
Then we take in (\ref{rp47b}) $A =  p_\ell^{(j)}$,  $j = 1, \dots ,
\nu$, make use of the commutation relation (\ref{cr}), take into
account
 the rotation invariance, and arrive at
\begin{eqnarray} \label{rp47c}
\varrho_{ \Lambda}^{\rm per} \left([A, [H^{\rm
per}_\Lambda,A]]\right) &  = & \varrho_{\beta , \Lambda}^{\rm per}
\left( - 2 b + a + 2 b_2 |q_\ell|^2 + 4 b_2 (q_\ell^{(j)})^2 \right)
 \\
& = &
 - 2b + a + 4 b_2 (\nu + 2) \bigg{\langle} \left[ \omega_\ell^{(j)} (0)
\right]^2\bigg{\rangle}_{\nu_{\beta, \Lambda}^{\rm per}} \nonumber
\\ & \geq & 0, \nonumber
\end{eqnarray}
which yields  (\ref{rp47a}).
$\square$

Now we consider the case where $V$ is more general as to compare with (\ref{rp46}).
\begin{lemma} \label{FUlm}
Let the model be translation and rotation invariant, with nearest
neighbor interaction. Then, for every $\theta >0$, there exist
positive $m_*$ and $J_*$, which may depend on $\beta$, $\theta$, and
on  the potential $V$, such that, for $m>m_*$ and $J>J_*$,
\begin{equation}
\label{FU2} \langle (\omega_{\ell} (\tau),
\omega_\ell(\tau))\rangle_{\nu_\Lambda^{\rm per}} \geq \theta.
\end{equation}
 \end{lemma}
{\it Proof:}
Let us rewrite (\ref{ZiFF10})
\begin{eqnarray}
\label{KK}
p^{\rm per}_{\Lambda} (J)
& = & \log N_\beta \nonumber \\ & + & \frac{1}{|\Lambda|}\log \left\{\int_{\Omega_{ \Lambda}} \exp \left[Y_\Lambda (\omega_\Lambda )  \right]
 \prod_{\ell \in \Lambda} \lambda ({\rm d}\omega_\ell) \right\},
\end{eqnarray}
where we indicate the dependence of the pressure on the interaction
intensity and have set $h=0$ since the potential $V$ should be
rotation invariant. Clearly, $p_\Lambda^{\rm per}(J)$ is convex; its
derivative can be computed from (\ref{KK}). Then  we get
\begin{eqnarray} \label{FU1}
\frac{J}{|\Lambda|} \sum_{\langle \ell , \ell' \rangle \in E}
\big{\langle} (\omega_{\ell},
\omega_{\ell'})_{L^2_\beta}\big{\rangle}_{\nu^{\rm per}_{
\Lambda}}
& = & J \frac{\partial}{\partial J}p_\Lambda^{\rm per}(J) \\
& \geq & p_\Lambda^{\rm per}(J) - p_\Lambda^{\rm per}(0) \nonumber
\\ & = & \frac{1}{|\Lambda|}\log \left\{\int_{\Omega_{ \Lambda}}
\exp \left[Y_\Lambda (\omega_\Lambda )  \right]
 \prod_{\ell \in \Lambda} \lambda ({\rm d}\omega_\ell) \right\}, \nonumber
\end{eqnarray}
where $E$ is the same as in (\ref{rp33}). By  the translation
invariance and (\ref{a11}), one gets
\begin{eqnarray*}
\big{\langle} (\omega_{\ell},
\omega_{\ell'})_{L^2_\beta}\big{\rangle}_{\nu^{\rm per}_{
\Lambda}} & \leq & \left( \big{\langle} (\omega_{\ell},
\omega_{\ell})_{L^2_\beta}\big{\rangle}_{\nu^{\rm per}_{\Lambda}} + \big{\langle} (\omega_{\ell'},
\omega_{\ell'})_{L^2_\beta}\big{\rangle}_{\nu^{\rm per}_{\Lambda}}\right)/2 \\ & = & \big{\langle} (\omega_{\ell},
\omega_{\ell})_{L^2_\beta}\big{\rangle}_{\nu^{\rm per}_{\Lambda}} = \beta \big{\langle} (\omega_{\ell} (\tau),
\omega_\ell(\tau))\big{\rangle}_{\nu_\Lambda^{\rm per}}.
\end{eqnarray*}
Then we choose $\varepsilon$, $c$, and $n$ as in (\ref{ZiFF7}),
apply this estimate in (\ref{FU1}), and obtain
\begin{eqnarray}
\label{KK1} \beta J d \big{\langle} (\omega_{\ell} (\tau),
\omega_\ell(\tau))\big{\rangle}_{\nu_\Lambda^{\rm per}} & \geq &
\frac{1}{|\Lambda|}\log \left\{\int_{\left[B^+(\varepsilon;c)
\right]^{\nu|\Lambda|}} \exp \left[Y_\Lambda (\omega_\Lambda )
\right]
 \prod_{\ell \in \Lambda} \lambda ({\rm d}\omega_\ell) \right\} \nonumber \\
  & \geq & \beta J \nu d \varepsilon^2 + \nu \log \gamma (m).
\end{eqnarray}
For $m>m_*$ given by (\ref{ZiFF8}), $\gamma (m)>0$ and the latter
estimate makes sense. Given $\theta>0$, one picks $\varepsilon
>\sqrt{\theta/\nu}$ and then finds $J_*$ such that the right-hand
side of the latter estimate equals $\theta$ for $J=J_*$.
$\square$

To convert (\ref{rp47a}) and (\ref{FU2}) into the bound
(\ref{rp40w}) we need the function $f: [0, +\infty) \rightarrow
[0,1)$ defined implicitly by
\begin{equation} \label{rp48}
f( u \tanh u) = u^{-1} \tanh u, \quad {\rm for} \ \ u>0; \quad {\rm
and} \ \ f(0)=1.
\end{equation}
It is differentiable, convex, monotone decreasing on $(0, +\infty)$,
such that  $t f(t) \rightarrow 1$. For $t\geq 6$, $f(t) \approx 1/t$
to five-place accuracy, see Theorem A.2 in \cite{[DLS]}. By direct
calculation,
\begin{equation} \label{RRP}
\frac{f' (u \tau) }{f(u \tau)} = - \frac{1}{ u \tau} \cdot \frac{
\tau - u (1 - \tau^2)}{\tau + u (1 - \tau^2)}, \qquad \tau = \tanh
u.
\end{equation}
\begin{proposition} \label{rp48pn}
For every fixed $\alpha >0$, the function
\begin{equation} \label{rp48Z}
\phi (t) = t \alpha f (t/\alpha), \quad t >0
\end{equation}
is differentiable and monotone increasing to $\alpha^2$ as $t
\rightarrow +\infty$.
\end{proposition}
{\it Proof:}
By (\ref{RRP}),
\[
\phi ' (t) =  \frac{ 2 \alpha \tau  (1 -\tau^2)} {\tau + u (1 -
\tau^2)} >0, \qquad  u \tau = u \tanh u = t /\alpha.
\]
The limit $\alpha^2$ is obtained from the corresponding asymptotic
property of $f$.
$\square$
\vskip.1cm Next, we need the following fact, known as Inequality of
Bruch and Falk, see Theorem IV.7.5 on page 392 of \cite{[Simon]} or
Theorem 3.1 in \cite{[DLS]}.
\begin{proposition} \label{BFpn}
Let $A$  be as in (\ref{rp47b}). Let also
\begin{eqnarray*}
& &  b(A) = \beta^{-1} \int_0^\beta \varrho^{\rm per}_{ \Lambda}
\left\{ A \exp[- \tau H^{\rm per}_\Lambda] A \exp[ \tau H^{\rm
per}_\Lambda] \right\}{\rm d}\tau, \\ & & g(A) =  \varrho^{\rm per}_{
\Lambda} \left( A^2 \right); \quad  \ \ c(A) = \varrho^{\rm per}_{
\Lambda}\left\{[A,[\beta H^{\rm per}_\Lambda, A]] \right\},
\end{eqnarray*}
Then
\begin{equation} \label{rp49}
b(A) \geq g(A) f\left(\frac{c(A)}{4 g(A)} \right),
\end{equation}
where $f$ is the same as in (\ref{rp48}).
\end{proposition}

 Set
\begin{equation} \label{rp50}
\mathcal{J}(d) = \frac{1}{(2\pi)^d}\int_{(-\pi, \pi]^d} \frac{{\rm
d} p}{\mathcal{E}(p)},
\end{equation}
where $\mathcal{E}(p)$ is given by (\ref{rp40c}).  The exact value of $\mathcal{J}(3)$ can
be expressed in terms of complete elliptic integrals, see
\cite{[Watson]} and also \cite{[Joyce]} for more recent
developments. For our aims, it is enough to have the following
property, see Theorem 5.1 in \cite{[DLP]}.
\begin{proposition} \label{DLPpn}
For $d\geq 4$, one has
\begin{equation} \label{rp500}
\frac{1}{d - 1/2} < \mathcal{J}(d) < \frac{1}{d - \alpha (d)}<
\frac{1}{d - 1},
\end{equation}
where $\alpha (d) \rightarrow 1/2$ as $d\rightarrow +\infty$.
\end{proposition}

Recall that $m$ is the reduced particle mass (\ref{In}).
\begin{theorem} \label{phrottm}
Let $d\geq 3$, the interaction be of nearest neighbor type, and the
anharmonic potential be of the form (\ref{rp46}), which defines the
parameter $\vartheta_*$. Let also
 the following condition be satisfied
\begin{equation} \label{rp52}
8 m \vartheta_*^2 J >  \mathcal{J}(d) .
\end{equation}
Then for every $\beta> \beta_*$, where the latter is the unique
solution of the equation
\begin{equation} \label{rp52a}
2 \beta J \vartheta_* f (\beta/ 4 m \vartheta_*) = \mathcal{J}(d),
\end{equation}
the model has a phase transition in the sense of Definition
\ref{phdef}.
\end{theorem}
{\it Proof:}
One observes that
\begin{equation} \label{rp53}
[ q^{(j)}_\ell, [ H_\Lambda^{\rm per}, q_\ell^{(j)}]] = 1/m, \quad
\ell \in \Lambda.
\end{equation}
Then we take in (\ref{rp49}) $A= q_{\ell}^{(j)}$ and obtain
 \[
b(A)  \geq  \big{\langle} \left( \omega_\ell^{(j)}(0) \right)^2 \big{\rangle}_{\nu_{
\Lambda}^{\rm per}} f\left(\frac{\beta}{ 4 m \big{\langle}
\left( \omega_\ell^{(j)}(0) \right)^2 \big{\rangle}_{\nu_{
 \Lambda}^{\rm per}}} \right).
\]
 By
Proposition \ref{rp48pn}, $\vartheta f(\beta / 4m \vartheta)$ is an
increasing function of $\vartheta$. Thus, by (\ref{rp47a}) and
(\ref{nrp10}),
\begin{equation}
\label{rp54}  D^\Lambda_{\ell \ell} \geq \beta^{2} \nu\vartheta_*
f(\beta/ 4m \vartheta_* ),
\end{equation}
which  yields the bound (\ref{rp40w}). Thereby, the condition (i) in
(\ref{RP4}) takes the form
 \begin{equation} \label{rp51} \vartheta_* f \left( \beta / 4 m \vartheta_* \right) >
\mathcal{J}(d) / 2 \beta J.
\end{equation}
By Proposition \ref{rp48pn}, the function
\[
\phi (\beta) = 2 \beta J \vartheta_* f (\beta/ 4 m \vartheta_*)\] is
monotone increasing and hits the level $\mathcal{J}(d)$ at certain
$\beta_*$. For $\beta > \beta_*$, the estimate (\ref{rp51}) holds,
which yields $|\mathcal{G}_\beta^{\rm t}|>1$.
$\square$
\vskip.1cm One observes that $f(\beta/4 m \vartheta_*) \rightarrow
1$ as $m \rightarrow +\infty$. In this limit, the condition
(\ref{rp52}) turns into the corresponding condition for a classical
model of $\phi^4$ anharmonic oscillators, Now let us turn to the
general case.
\begin{theorem} \label{phrot1tm}
Let $d\geq 3$, the interaction be of nearest neighbor type, and the
anharmonic potential be rotation invariant. Then, for every
$\beta>0$, there exist $m_*$ and $J_*>0$, which may depend on
$\beta$ and on the anharmonic potential, such that
$|\mathcal{G}^{\rm t} | >1$ for $m>m_*$ and $J>J_*$.
\end{theorem}
{\it Proof:}
Given positive $\beta$ and $\theta$, the estimate (\ref{FU2}) for big
enough $m$ and $J$. Then one applies Proposition \ref{BFpn}, which
yields that the condition (i) in (\ref{RP4}) is satisfied if
\[
\theta f(\beta /4m \theta) > \mathcal{J} (d) / 2 \beta J.
\]
 Then one sets $m_*$ to be as in (\ref{ZiFF8}) and $J_*$ to be the smallest value of $J$ for which both (\ref{FU2}) and the latter inequality hold.
$\square$

\subsection{Phase transition in the symmetric scalar models}
\label{6.3.2.ss}

In the case $\nu=1$,  we can extend the above results to the models
without translation invariance and with much more general $J_{\ell
\ell'}$ and $V_\ell$. However, certain assumptions beyond (\ref{a1})
and  (\ref{a2}) should be made.
 Suppose
also that the interaction between the nearest neighbors is uniformly
nonzero, i.e.,
\begin{equation} \label{Ip1}
\inf_{|\ell -\ell'|=1} J_{\ell \ell'} \ \stackrel{\rm def}{=} \ J
>0.
\end{equation}
Next we suppose that all $V_\ell$'s are even continuous functions
and the upper bound in (\ref{A3}) can be chosen to obey the
following conditions: \vskip.1cm
\begin{tabular}{ll}
(a) \  &for every  $\ell$,
\end{tabular}
\vskip.1cm
\begin{equation} \label{Iub1}
V(u_\ell ) - V_\ell (u_\ell) \leq V(\tilde{u}_\ell )  - V_\ell
(\tilde{u}_\ell) , \quad {\rm whenever} \  \ u_\ell^2 \leq
\tilde{u}_\ell^2;
\end{equation}
\vskip.1cm
\begin{tabular}{ll}
(b) \ &the function $V$ has the form
\end{tabular}
\vskip.1cm
\begin{equation} \label{Iub}
V(u_\ell ) = \sum_{s=1}^r b^{(s)} u_\ell^{2s}; \quad 2 b^{(1)} < - a
; \ \ \ b^{(s)}\geq 0, \  s\geq 2,
\end{equation}
\vskip.1cm
\begin{tabular}{ll}
 &where $a$ is as in (\ref{U1}) and $r\geq 2$ is either positive
integer or infinite;\\[.1cm] (c) &if $r=+\infty$, the series
\end{tabular}
\vskip.1cm
\begin{equation} \label{Ip3}
\mathit{\Phi}(\vartheta) = \sum_{s=2}^{+\infty}
\frac{(2s)!}{2^{s-1}(s-1)!}{b}^{(s)} \vartheta^{s-1},
\end{equation}
\vskip.1cm
\begin{tabular}{ll}
&converges at some $\vartheta>0$.
\end{tabular}
\vskip.1cm \noindent Since $2b^{(1)} + a <0$, the equation
\begin{equation} \label{Ip4}
a + 2b^{(1)} + \mathit{\Phi}(\vartheta) = 0,
\end{equation} has a unique solution $\vartheta_* >0$.
 By the above assumptions, all $V_\ell$ are
`uniformly double-welled'. If $V_\ell (u_\ell)= v_\ell (u_\ell^2)$
and $v_\ell$ are differentiable, the condition (\ref{Iub1}) can be
formulated as an upper bound for $v_\ell'$. Note that the pressure
as a unified  characteristics of all Euclidean Gibbs states makes
senses for translation invariant models only. Thus, the notions
mentioned in Definition \ref{landau} are not applicable to the
versions of the model which do not possess this property.

The main result of this subsection is contained in the following
statement.
\begin{theorem} \label{phsctm}
Let the model be as just described. Let also the condition
(\ref{rp52}) with $\vartheta_*$ defined by the equation (\ref{Iub1})
and $J$ defined by (\ref{Ip1}) be satisfied. Then for every $\beta
>\beta_*$, where $\beta_*$ is defined by the equation (\ref{rp52}),
 the model has a phase transition in the sense of Definition
\ref{phdef}. If the model is translation invariant, the long range
order and the first order phase transition take place at such
$\beta$.
\end{theorem}
{\it Proof:}
The proof is made by comparing the model under consideration with a
reference model, which is
 the scalar model with the nearest
neighbor interaction of intensity (\ref{Ip1}) and with the
anharmonic potential (\ref{Iub}). Thanks to the condition  (\ref{Iub1}), the
reference model is more stable; hence, the phase transition in this
model implies the same for the model considered. The comparison is conducted by means of the correlation inequalities.

The reference model
is translation invariant and hence can be defined  by its local periodic Hamiltonians
\begin{equation} \label{Irf1}
H^{\rm low}_{\Lambda} = \sum_{\ell \in \Lambda}\left[ H_\ell^{\rm
har} + V(q_\ell)\right] -  J \sum_{\langle \ell, \ell' \rangle\in E}
 q_\ell q_{\ell'},
\end{equation}
where for a box $\Lambda$, $E$ is the same as in (\ref{rp16});
$H_\ell^{\rm har}$ is  as in (\ref{U1}).  For this model, we have
the infrared estimate (\ref{rp45}) with $\nu=1$. Let us obtain the
lower bound, see (\ref{rp47a}). To this end we use the inequalities
(\ref{rp47b}), (\ref{rp47c}) and obtain
\begin{eqnarray} \label{rp55}
\qquad 0 &\leq  & a + 2 b^{(1)} + \sum_{s=2}^r 2 s (2 s-1)b^{(s)}
\big{\langle }\left[
\omega_\ell (0) \right]^{2(s-1)} \big{\rangle}_{\nu^{\rm low}_{ \Lambda}} \\
&  \leq &  a + 2 b^{(1)} + \sum_{s=2}^r 2 s (2
s-1)\frac{(2s-2)!}{2^{s-1} (s-1)!} \cdot b^{(s)} \left[\big{\langle}
\left(\omega_\ell (0)\right)^2 \big{\rangle}_{\nu^{\rm low}_{
\Lambda}}\right]^{s-1}. \nonumber
\end{eqnarray}
Here $\nu^{\rm low}_{\Lambda}$ is the periodic Gibbs measure for the
model (\ref{Irf1}). To get the second line we used the Gaussian
upper bound inequality, see page 1031 in \cite{[KoT]} and page 1372
in \cite{[RevMF]}, which is possible since all $b^{(s)}$, $s\geq 2$
are nonnegative. The solution of the latter inequality is
\begin{equation} \label{rp55A}
\big{\langle} \left( \omega_\ell(0) \right)^2 \big{\rangle}_{\nu_{
 \Lambda}^{\rm low}} \geq \vartheta_*.
\end{equation}
Then the proof of the phase transitions in the model (\ref{Irf1})
goes along the line of arguments used in proving Theorem
\ref{phrottm}. Thus, for $\beta>\beta_*$, $\langle \omega_\ell (0)
\rangle_{\mu^{\rm low}_+} >0$, where $\mu^{\rm low}_+$ is the
corresponding maximal Euclidean Gibbs measure, see Proposition
\ref{MAtm}. But,
\begin{equation} \label{kot}
\langle \omega_\ell (0) \rangle_{\mu_+} > \langle \omega_\ell (0)
\rangle_{\mu^{\rm low}_+},
\end{equation}
see Lemma 7.7 in \cite{[KoT]}. At the same time $\langle \omega_\ell
(0) \rangle_{\mu} =0$ for any periodic $\mu \in \mathcal{G}^{\rm
t}$, which yields the result to be proven.
$\square$

\subsection{Phase transition in the scalar model with asymmetric potential}
\label{6.3.3.ss}

The phase transitions proven so far have a common feature -- the
spontaneous symmetry breaking. This means that the symmetry, e.g.,
rotation invariance, possessed by the model and hence by the unique
element of $\mathcal{G}^{\rm t}$ is no longer possessed by the
multiple Gibbs measures appearing as its result. In this subsection,
we show that the translation invariant scalar version o the model
(\ref{U1}), (\ref{U2}) has a phase transition without symmetry
breaking. However, we restrict ourselves to the case of first order
phase transitions, see Definition \ref{landau}. The reason for this
can be explained as follows. The fact that $D_{\ell \ell'}^\mu$ does
not decay to zero as $|\ell - \ell'| \rightarrow +\infty$, see
(\ref{nrp1}), implies that $\mu$ is non-ergodic only if $\mu$ is
symmetric. Otherwise, to show that $\mu$ is non-ergodic one should
prove that the difference $D_{\ell \ell'}^\mu - \langle f_\ell
\rangle_\mu \cdot \langle f_{\ell'} \rangle_\mu$ does not decay to
zero, which cannot be done by means of our methods based on the
infrared estimate.

In what follows, we consider the translation invariant scalar
version of the model (\ref{U1}), (\ref{U2}) with the nearest
neighbor interaction. The only condition imposed on the anharmonic
potential is (\ref{a2}). Obviously, we have to include the external
field, that is the anharmonic potential is now $V(u) - h u$. Since
we are not going to impose any conditions on the odd part of $V$, we
cannot apply the GKS inequalities, see \cite{[RevMF],[KoT]}, the
comparison methods are based on, see (\ref{kot}). In view of this
fact we suppose that the interaction is of nearest neighbor type.
Thus, for a box $\Lambda$, the periodic local Hamiltonian of the
model has the form (\ref{Irf1}).

In accordance with Definition \ref{landau}, our goal is to show that
the model parameters (except for $h$) and the inverse temperature
$\beta$ can be chosen in such a way that the set $\mathcal{R}$,
defined by (\ref{ZiF4}), is non-void. The main idea on how to do
this can be explained as follows. First we find a condition,
independent of $h$, under which $D^\mu_{\ell \ell'}$ does not decay
to zero for a certain periodic $\mu$. Next we prove the following
\begin{lemma} \label{ZiFFtm}
There exist $h_{\pm}$, $h_{-} < h_{+}$, which may depend on the
model parameters and $\beta$, such that the magnetization
(\ref{ZiF5}) has the property:
\[
M(h) < 0, \ \ \ {\rm for} \ \ h \in \mathcal{R}^{c} \cap (- \infty,
h_{-}); \quad \ M(h) > 0, \ \ \ {\rm for} \ \ h \in \mathcal{R}^{c}
\cap (h_{+} + \infty).
\]
\end{lemma}
 Thereby, if $\mathcal{R}$ were void, one would find $h_* \in
(h_{-}, h_{+})$ such that $M(h_*)= 0$. At such $h_*$, the
aforementioned property of $D^\mu$ would yield the non-ergodicity of
$\mu$ and hence the first order phase transition, see Theorem
\ref{phsctm}.

In view of Corollary \ref{RPco}, $D^\mu_{\ell \ell'}$ does not decay
to zero if (\ref{rp40w}) holds with  big enough $\vartheta$. By
Proposition \ref{BFpn}, the lower bound (\ref{rp40w}) can be
obtained from the estimate (\ref{FU2}). The only problem with the
latter estimate is that it holds for $h=0$.
\begin{lemma} \label{Ziplm}
For every $\beta>0$ and $\theta$, there exist positive $m_*$ and
$J_*$, which may depend on $\beta>0$ and $\theta$ but are
independent of $h$,  such that, for any box $\Lambda$ and any $h\in
\mathbb{R}$,
\begin{equation} \label{zifc}
\big{\langle} \left[ \omega_\ell(0) \right]^2
\big{\rangle}_{\nu_{\Lambda}^{\rm per}} \geq \theta, \quad  \ {\rm
if} \ \ {J}
> J_* \ \ {\rm and}  \ \ m>m_* .
\end{equation}
\end{lemma}
{\it Proof:}
For $h\in \mathbb{R}$, we set
\begin{eqnarray} \label{zifa}
\lambda^h ({\rm d} \omega) & = & \frac{1}{N_\beta^h} \exp\left( h
\int_0^\beta \omega(\tau){\rm d}\tau\right)\lambda ({\rm d}\omega), \\
{N_\beta^h} & = & \int_{C_\beta} \exp\left( h \int_0^\beta
\omega(\tau){\rm d}\tau\right)\lambda ({\rm d}\omega), \nonumber
\end{eqnarray}
where $\lambda$ is as in (\ref{ZiFF1}). Then for $\pm h>0$, we get
the estimate (\ref{KK1}) in the following form
\begin{equation}
\label{ziffa} \beta J d  \big{\langle} \left[ \omega_\ell(0)
\right]^2 \big{\rangle}_{\nu_{\Lambda}^{\rm per}} \geq \beta J d
\varepsilon^2 + \log \lambda^h \left[ B^{\pm}(\varepsilon, c)\right]
,
\end{equation}
where $B^{\pm}(\varepsilon, c)$ is as in (\ref{ZiFF6}),
(\ref{ZiFF7}). Let us show now that, for $\pm h \geq 0$,
\begin{equation} \label{zifb4}
\lambda^h \left[B^{\pm}(\varepsilon, c)\right] \geq \lambda
\left[B^{\pm}(\varepsilon, c)\right].
\end{equation}
For $h\geq 0$, let $I (\omega)$ be the indicator function of the set
$C^{+}_\beta (n;c)$, see (\ref{ZiFF5}). For $\delta >0$ and $t\in
\mathbb{R}$, we set
\[
\iota_\delta (t) = \left\{ \begin{array}{ll} 0 &\quad \ \ t\leq c,\\
(t - c)/\delta &\quad \ \  t\in (c, c+\delta], \\ 1 &\quad  \ \
c\geq c+\delta. \end{array} \right.
\]
Thereby,
\[
I_\delta (\omega) \ \stackrel{\rm def}{ =} \ \prod_{k=0}^n
\iota_\delta \left[ \omega(k \beta /n) \right].
\]
By Lebesgue's dominated convergence theorem
\begin{eqnarray} \label{zifb5}
N^h_\beta \lambda^h \left[ C^{+}_\beta (n;c)\right] & = &
\int_{C_\beta} I (\omega)\exp\left(h \int_0^\beta \omega(\tau){\rm
d}\tau \right)\lambda ({\rm d}\omega) \\ & = & \lim_{\delta
\downarrow 0} \int_{C_\beta} I_\delta (\omega)\exp\left(h
\int_0^\beta \omega(\tau){\rm d}\tau \right)\lambda ({\rm d}\omega).
\nonumber
\end{eqnarray}
As the function $I_\delta$ is continuous and increasing, by the FKG
inequality, see Theorem 6.1 in \cite{[RevMF]}, it follows that
\[
\int_{C_\beta} I_\delta (x)\exp\left(h \int_0^\beta \omega(\tau){\rm
d}\tau \right)\lambda ({\rm d}\omega) \geq N^h_\beta \int_{C_\beta}
I_\delta (\omega)\lambda ({\rm d}\omega).
\]
Passing here to the limit we obtain from (\ref{zifb5})
\[
\lambda^h \left[ C^{+}_\beta (n;c)\right] \geq \lambda \left[
C^{+}_\beta (n;c)\right],
\]
which obviously yields (\ref{zifb4}). For $h \leq 0$, one just
changes the signs of $h$ and $\omega$. Thereby, we can rewrite
(\ref{ziffa}) as follows, c.f., (\ref{KK1}),
\[
\big{\langle} \left[ \omega_\ell(0) \right]^2
\big{\rangle}_{\nu_{\Lambda}^{\rm per}} \geq  \varepsilon^2 + [\log
\gamma (m) ]/ \beta J d.
\]
Then one applies the arguments from the very end of the proof of
Lemma \ref{FUlm}.
$\square$

\noindent {\it Proof of Lemma \ref{ZiFFtm}:} Suppose that $h>0$.
Then restricting the integration in (\ref{ZiFF10}) to
$[B^{+}(\varepsilon,c)]^\Lambda$, we get
\begin{eqnarray} \label{ZiFF11}
p^{\rm per}_{\Lambda} (h) & \geq & h \beta \varepsilon + \log
N_\beta + \frac{1}{2}
 \beta \varepsilon^2 \sum_{\ell'\in \Lambda} J^{\Lambda}_{\ell\ell'} + \log \lambda   [B^+(\varepsilon,c)]  \\
 & \geq &  h \beta \varepsilon + \log N_\beta + \log \gamma (m) . \nonumber
\end{eqnarray}
As the right-hand side of the latter estimate is independent of
$\Lambda$, it can be extended to  the limiting pressure $p(h)$. For
any positive $h\in \mathcal{R}^c$, by the convexity of $p(h)$ one
has
\begin{eqnarray*}
M(h) & \geq &  \left[ p(h) - p(0)\right]/ \beta h \\& \geq &
\varepsilon + \frac{1}{\beta h}\left\{ - p(0) + \log N_\beta + \log
\gamma (m)  \right\}.
\end{eqnarray*}
Picking  big enough $h$ we get the positivity stated. The negativity
can be proven in the same way.
$\square$

Now we are at a position to prove the main statement
of this subsection.
\begin{theorem}
\label{phasymtm} Let the model be scalar, translation invariant, and
with the nearest-neighbor interaction. Let also $d\geq 3$. Then for
every $\beta$, there exist $m_*>0$ and $J_*
>0$ such that, for all $m>
m_*$ and $J>J_*$,  there exists $h_*\in \mathbb{R}$, possibly
dependent on $m$, $\beta$, and $J$, such that $p' (h)$ gets
discontinuous at $h_*$, i.e., the model has a first order phase
transition.
\end{theorem}
{\it Proof:}
Let $m_*$ be as in (\ref{ZiFF8}) and $J_*$, $\theta$ be as in Lemma
\ref{Ziplm}. Fix any $\beta>0$ and $m >m_*$. Then, for $J>J_*$, the
estimate (\ref{zifc}) holds, which yields the validity of
(\ref{rp54}) for all boxes $\Lambda$ with such $\beta$, $m$, and
$\nu =1$. Thereby, we increase $J$, if necessary, up to the value at
which (\ref{rp51}) holds. Afterwards, all the parameters, except for
$h$, are set fixed. In this case,  there exists a periodic state
$\mu\in \mathcal{G}^{\rm t}$ such that the first summand in
(\ref{RP6}) is positive; hence, $D_{\ell \ell'}^\mu$ does not decay
to zero as $|\ell - \ell'|\rightarrow +\infty$, see (\ref{RL}) and
(\ref{RP6}). If $p(h)$ is everywhere differentiable, i.e., if
$\mathcal{R}= \emptyset$, then by Lemma \ref{ZiFFtm} there
exists $h_*$ such that $M(h_*)=0$; hence, the state $\mu$ with such
$h_*$ is non-ergodic, which yields $|\mathcal{G}^{\rm t}|>1$ and
hence a first order phase transition. Otherwise, $\mathcal{R}\neq
\emptyset$.
$\square$

\subsection{Comments} \label{ssC3}
\begin{itemize}
\item \emph{Subsection \ref{3.1.ss}:}
According to Definition \ref{phdef}, the phase transition
corresponds to the existence of multiple equilibrium phases at the same values of the model parameters and temperature.
This is a standard definition for theories, which employ  Gibbs
states, see \cite{[Ge]}. In the translation invariant case, a way of
proving phase transitions can be to show the existence of
non-ergodic elements of $\mathcal{G}^{\rm t}$. For classical lattice
systems, it was realized in \cite{[FSS]} by means of infrared
estimates. More or less at the same time, an alternative rigorous
theory of phase transitions in classical lattice spin models based
on contour estimates has been proposed. This is the Pirogov-Sinai theory
elaborated in \cite{[PS]}, see also \cite{[SinaiB]}. Later on, this
theory was essentially extended and generalized into an abstract
sophisticated method, applicable also to classical (but not
quantum) models with unbounded spins, see \cite{[Zah]} and the
references therein.

For quantum lattice models, the theory of phase transitions has
essential peculiarities, which distinguish it from the corresponding
theory of classical systems.  Most of the results in this domain
were obtained by means of quantum versions of the method of infrared
estimates. The first publication in which such estimates were
applied to quantum spin models seems to be the article \cite{[DLS]}.
After certain modifications this method was applied to a number of
models with unbounded Hamiltonians
\cite{[AKKR],[BaK],[BaK0],[DLP],[Kondr],[Pastur]}. In our approach,
the quantum crystal is described as a system of `classical' infinite
dimensional spins. This allows for applying here the original
version of the method of  infrared estimates elaborated in
\cite{[FSS]} adapted to the infinite dimensional case, which has
been realized in the present work.  Among others, the adaptation
consists in employing such tools as the Garsia-Rodemich-Rumsey
lemma, see \cite{[Garsia]}. This our approach is more effective and
transparent than the one used in \cite{[AKKR],[BaK],[BaK0],[Kondr]}.
It also allows for comparing the conditions (\ref{rp40w}),
(\ref{RP4}) with the stability conditions obtained in the next
section.

 In the physical literature, there exist
definitions of phase transitions alternative to Definition
\ref{phdef}, based directly on the thermodynamic properties of the
system. These are the definition employing the differentiability of
the pressure (Definition \ref{landau}, which is applicable to
translation invariant models only), and the definition based on the
long range order. The relationship between the latter two notions is
established by means of the Griffiths theorem, Proposition \ref{Grpn},
the proof of which can be found   in \cite{[DLS]}. For  translation
invariant models  with bounded interaction, non-differentiability of
the pressure corresponds to the non-uniqueness of the Gibbs states,
see \cite{[Israel],[Simon]}. We failed to prove this for our model.

In the language of limit theorems of probability theory, the
appearance of the long range order corresponds to the fact that a
new law of large numbers comes to power, see Theorem \ref{Grpn} and
the discussion preceding Definition \ref{Weberdf}. The critical
point of the model corresponds to the case where the law of large
numbers still holds in its original form (in the translation
invariant case this means absence of the first order phase
transitions), but the central limit theorem holds true with an
abnormal normalization. For a hierarchical version of the model
(\ref{U1}), (\ref{U2}), the critical point was described in
\cite{[Kozak]}.
Algebras of abnormal fluctuation operators were studied in
\cite{[Broi]}. In application to quantum crystals, such operators were discussed
in \cite{[VZ1],[VZ2]}, where the reader can find
a more detailed discussion of this subject as well as the
corresponding bibliography.

\item \emph{Subsection \ref{3.2.ss}:} As was mentioned above, the method of
infrared estimates was originated in \cite{[FSS]}. The version
employed here is close to the one presented in \cite{[KKE]}. We note
that, in accordance with the conditions (\ref{rp40x}),(\ref{rp40w}),
and (\ref{RP4}), the infrared bound was obtained for the Duhamel
function, see (\ref{rp45}), rather than for
\[
\sum_{\ell'\in \Lambda}\langle (\omega_\ell (\tau) , \omega_{\ell'}(
\tau))\rangle_{\nu_\Lambda^{\rm per}} \cdot \cos  (p, \ell - \ell'),
\]
which was used in \cite{[RevMF],[BaK],[BaK0],[Kondr]}.

\item \emph{Subsection \ref{3.3.ss}:} The lower bound (\ref{rp47a})
was obtained in the spirit of \cite{[DLP],[Pastur]}. The estimate
stated in Lemma \ref{FUlm} is completely new; the key element of its
proving is the estimate (\ref{ZiFF4}), obtained by means of
Proposition \ref{grrpn}. The sufficient condition for the phase
transition obtained in Theorem \ref{phrottm} is also new. Its
significant feature is the appearance of a universal parameter
responsible for the phase transition, which includes the particle
mass $m$, the anharmonicity parameter $\vartheta_*$, and the
interaction strength $J$. This is the parameter on the left-hand
side of (\ref{rp52}). The same very parameter will describe the
stability of the model studied in the next section. Theorem
\ref{phrot1tm} is also new.

\item \emph{Subsection \ref{6.3.2.ss}:}
Here we mostly repeat the corresponding results of \cite{[KoT]},
announced in \cite{[KoT1]}.

\item \emph{Subsection \ref{6.3.3.ss}:} The main characteristic
feature of the scalar model studied in
\cite{[AKKR],[BaK],[BaK0],[DLP],[Kondr],[Pastur]}, as well the the
one described by Theorem \ref{phsctm}, was the $Z_2$-symmetry broken
by the phase transition. This symmetry allowed for obtaining
estimates like (\ref{rp55A}), crucial for the method. However, in
classical models, for proving phase transitions by means of the
infrared estimates, symmetry was not especially important, see
Theorem 3.5 in \cite{[FSS]} and the discussion preceding this theorem. There
might be two explanations of such a discrepancy: (a) the symmetry
was the key element but only of the methods employed therein, and,
like in the classical case, its lack does not imply the lack of
phase transitions; (b) the symmetry is crucial in view of e.g.
quantum effects, which stabilize the system, see the next section.
So far, there has been no possibility to check which of these
explanations is true. Theorem \ref{phasymtm} solves this dilemma
 in favor of explanation (a). Its main element is again an estimate,
 obtained by means of the Garsia-Rodemich-Rumsey lemma.
The corresponding result was announced in \cite{[KaK]}.
\end{itemize}

\section{Quantum Stabilization}

\label{4s}

 In physical substances containing light quantum
particles moving in multi-welled potential fields phase transitions
are experimentally suppressed by application of strong hydrostatic
pressure, which makes the wells closer to each other and increases
the tunneling of the particles. The same effect is achieved by
replacing the particles with the ones having smaller mass. The aim
of this section is to obtain a description of such effects in the
framework of the theory developed here and to compare it with the
theory of phase transitions presented in the previous section.

\subsection{The stability of quantum crystals}

\label{stabcr}

Let us look at the scalar harmonic version of the model (\ref{U1})
 -- a quantum harmonic crystal. For this model, the
one-particle Hamiltonian includes first two terms of (\ref{U2})
only. Its spectrum consists of the eigenvalues $E_n^{\rm har} =
(n+1/2)\sqrt{a/m} $, $n \in \mathbb{N}_0$. The parameter  $a>0$ is
the oscillator rigidity. For reasons, which become clear in a while,
we consider the following gap parameter
\begin{equation}
\label{Gap}
 \mathit{\Delta}^{\rm har} = \min_{n \in \mathbb{N}} (E_n^{\rm har} - E_{n-1}^{\rm har}).
\end{equation}
Then
\begin{equation} \label{De2}
  \mathit{\Delta}^{\rm har} = \sqrt{a/m}; \qquad  a = m \mathit{\Delta}_{\rm har}^2.
\end{equation}
The set of tempered Euclidean Gibbs measures of the harmonic crystal
can be constructed similarly as it was done in section \ref{2s}, but
with one exception. Such measures exist only under the stability
condition (\ref{si}), which might now be rewritten
\begin{equation} \label{De2a}
\hat{J}_0 < m \mathit{\Delta}_{\rm har}^2.
\end{equation}
In this case, $\mathcal{G}^{\rm t}$ is a singleton at all $\beta$,
that readily follows from Theorem \ref{httm}. As the right-hand side
of (\ref{De2a}) is independent of $m$, this stability condition is
applicable also to the classical harmonic crystal which is obtained
in the classical limit $m\rightarrow +\infty$, see \cite{[RevMF]}.
According to
 (\ref{a2}) the anharmonic potentials $V_\ell$ have a
super-quadratic growth due to which the tempered Euclidean Gibbs
measures of anharmonic crystals exist for all $\hat{J}_0$. In this
case, the instability of the crystal is connected with phase
transitions. A sufficient condition for some of the models described
in the previous section  to have a phase transition may be derived
from the equation (\ref{rp51}). It is
\begin{equation} \label{De1}
2 \beta J \vartheta_* f ( \beta / 4 m \vartheta_*) > \mathcal{J}
(d),
\end{equation}
which in  the classical limit $m\rightarrow + \infty$ takes the form
\[
2 \beta J \vartheta_*  > \mathcal{J} (d).
\]
The latter condition can be satisfied by picking big enough $\beta$.
Therefore, the classical anharmonic crystals always have phase
transitions -- no matter how small is the interaction intensity. For
finite $m$, the left-hand side of (\ref{De1}) is bounded by $8 m
\vartheta_*^2 J$, and the bound is achieved in the limit $\beta
\rightarrow + \infty$. If for given values of the interaction
parameter $J$, the mass $m$, and the parameter $\vartheta_*$ which
characterizes the anharmonic potential, this bound does not
exceed $\mathcal{J}(d)$, the condition (\ref{De1}) will never  be
satisfied. Although this condition is only
sufficient, one might expect that the phase transition can be
eliminated at all $\beta$ if the compound parameter $8 m
\vartheta_*^2 J$ is small enough. Such an effect, if really exists,
could be called \emph{quantum stabilization} since it is principally
impossible in the classical analog of the model.

\subsection{Quantum rigidity}
\label{4.2.ss} In the harmonic case, big values of the rigidity $a$
ensure the stability. In this subsection, we introduce and stugy
{\it quantum rigidity}, which plays a similar role in the anharmonic
case

Above the sufficient condition (\ref{De1}) for a phase transition to
occur was obtained for a simplified version of the model (\ref{U1}),
(\ref{U2}) -- nearest neighbor interactions, polynomial anharmonic
potentials of special kind (\ref{Iub}), ect. Then the results were
extended to more general models via correlation inequalities.
Likewise here, we start with a simple scalar version of the
one-particle Hamiltonian (\ref{U1}), which we take in the form
\begin{equation} \label{De7}
H_m = \frac{1}{2m} p^2 + \frac{a}{2} q^2 + V(q),
\end{equation}
where  the anharmonic potential is, c.f., (\ref{Iub}),
\begin{equation} \label{De8}
V(q) = b^{(1)} q^2 + b^{(2)} q^4 + \cdots + b^{(r)} q^{2r}, \qquad
b^{(r)} >0, \quad r \in \mathbb{N}\setminus \{1\}.
\end{equation}
The subscript $m$ in (\ref{De7}) indicates the dependence of the
Hamiltonian on the mass. Recall that $H_m$ acts in the physical
Hilbert space $L^2(\mathbb{R})$. Its relevant properties are
summarized in the following
\begin{proposition} \label{spectpn}
The Hamiltonian $H_m$ is essentially self-adjoint on the set
$C_0^\infty(\mathbb{R})$ of infinitely differentiable functions with
compact support. The spectrum of $H_m$ has the following properties:
(a) it consists of eigenvalues $E_n$, $n\in \mathbb{N}_0$ only; (b) to
each $E_n$ there corresponds exactly one eigenfunction $\psi_n\in
L^2(\mathbb{R})$; (c) there exists $\gamma
>1$ such that
\begin{equation} \label{De3}
n^{-\gamma} E_n \rightarrow + \infty, \qquad {\rm as} \ \
n\rightarrow +\infty.
\end{equation}
\end{proposition}
{\it Proof:}
The essential self-adjointness of $H_m$ follows from the Sears
theorem, see Theorem 1.1, page 50 of \cite{[BeS]} or Theorem X.29 of
\cite{[RS2]}. The spectral properties follow from Theorem 3.1, page
57 (claim (a)) and Proposition 3.3, page 65 (claim (b)), both taken
from the book \cite{[BeS]}. To prove claim (c) we employ a classical
formula, see equation (7.7.4), page 151 of the book \cite{[Titch]},
which in our context reads
\begin{equation} \label{De3a}
\frac{2}{\pi}\sqrt{2m} \int_0^{u_n}\sqrt{E_n - V(u)}\ {\rm d} u =
n+\frac{1}{2} + O\left( \frac{1}{n}\right),
\end{equation}
where $n$, and hence $E_n$, are big enough so that the equation
\begin{equation} \label{De3b}
V(u) = E_n
\end{equation}
have the unique positive solution $u_n$. Then
\begin{equation} \label{De3c}
u_n^{r+1} \int_0^1 \sqrt{\phi_n (t) - t^{2r}} \ {\rm d}t =
\frac{\pi}{2 \sqrt{2m b^{(r)}} }\left( n + \frac{1}{2} \right) +
O\left(\frac{1}{n}\right),
\end{equation}
where
\[
\phi_n (t) = \frac{E_n}{b^{(r)} u_n^{2r}} - \frac{u_n^{2 -
2r}}{b^{(r)}} (b^{(1)} + a/2)t^2 - \dots - \frac{u_n^{- 2
}}{b^{(r)}} b^{(r-1)} t^{2(r-1)}.
\]
Note that $\phi_n (1) =1$ for all $n$, which follows from
(\ref{De3b}). Thus,
\begin{equation} \label{De3d}
\frac{E_n}{b^{(r)} u_n^{2r}} \rightarrow 1, \qquad {\rm as} \   \ n
\rightarrow +\infty.
\end{equation}
Thereby, we have
\begin{eqnarray} \label{De3e}
c_n & \stackrel{\rm def}{=} & \int_0^1 \sqrt{\phi_n (t) - t^{2r}} \
{\rm d}t \rightarrow \int_0^1 \sqrt{1 - t^{2r}} \ {\rm d} t \\ & = &
\frac{\Gamma\left(\frac{3}{2}\right)
\Gamma\left(\frac{1}{2r}\right)}{2 r \Gamma\left(\frac{3}{2} +
\frac{1}{2r}\right)}. \nonumber
\end{eqnarray}
Then combining (\ref{De3e}) with (\ref{De3b}) and (\ref{De3d}) we
get
\begin{equation} \label{De8a}
E_n = \left[\frac{b^{(r)}}{(2 m)^r} \right]^{1/(r+1)} \cdot
\left[\frac{ \pi r \Gamma\left(\frac{3}{2} + \frac{1}{2r}
\right)}{\Gamma\left(\frac{3}{2} \right) \Gamma\left( \frac{1}{2r}
\right)}\cdot \left( n + \frac{1}{2} \right)\right]^{\frac{2r}{r+1}}
+ o\left(1\right),
\end{equation}
which readily yields (\ref{De3}) with any $\gamma \in (1, 2r/
(r+1))$.
$\square$
\vskip.1cm
 Thus, in view of the property (\ref{De8a})
we introduce the gap
 parameter
\begin{equation} \label{De4}
\mathit{\Delta}_m  = \min_{n \in \mathbb{N}} (E_n - E_{n-1}),
\end{equation}
and thereby, c.f., (\ref{De2}),
\begin{equation} \label{De5}
\mathcal{R}_m = m \mathit{\Delta}_m^2,
\end{equation}
which can be called \emph{quantum rigidity} of the oscillator. One
might expect that the stability condition for quantum anharmonic
crystals, at least for their scalar versions with the
anharmonic potentials independent of $\ell$, is similar to
(\ref{De2a}). That is, it has the form
\begin{equation} \label{De6}
\hat{J}_0 < \mathcal{R}_m.
\end{equation}

\subsection{Properties of quantum rigidity}

\label{gap}

Below $f\sim g$ means that $\lim (f / g) =1$.
\begin{theorem} \label{gap1tm}
For every $r \in \mathbb{N}$, the gap parameter $\mathit{\Delta}_m$,
and hence the quantum rigidity $\mathcal{R}_m$ corresponding to the
Hamiltonian (\ref{De7}), (\ref{De8}), are continuous functions of
$m$. Furthermore,
\begin{equation} \label{De9}
\mathit{\Delta}_m \sim \mathit{\Delta}_0 m^{-r/(r+1)}, \quad \ \
\mathcal{R}_m \sim \mathit{\Delta}_0^2 m^{-(r-1)/(r+1)}, \quad \ \
m\rightarrow 0,
\end{equation}
with a certain $\mathit{\Delta}_0>0$.
\end{theorem}
{\it Proof:}
Given $\alpha >0$, let $U_\alpha: L^2(\mathbb{R})\rightarrow
L^2(\mathbb{R})$ be the following unitary operator
\begin{equation} \label{De10}
\left(U_\alpha \psi\right) (x) = \sqrt{\alpha} \psi(\alpha x).
\end{equation}
Then by (\ref{cr})
\[
U^{-1}_\alpha p U_\alpha = \alpha p, \qquad U^{-1}_\alpha q U_\alpha
= \alpha^{-1} q.
\]
Fix any $m_0>0$ and  set $\rho = (m /m_0)^{1/(r+1)}$, $\alpha =
\rho^{1/2}$. Then
\begin{equation} \label{De11}
\widetilde{H}_m \ \stackrel{\rm def}{=} \ U^{-1}_\alpha H_m U_\alpha
= \rho^{-r} T(\rho),
\end{equation}
where
\begin{eqnarray} \label{De12}
T(\rho) & = & H_{m_0} + Q(\rho)\\
& = & \frac{1}{2m_0} p^2 + \rho^{r-1} ( b^{(1)} +a/2) q^2 +
\rho^{r-2}
b^{(2)} q^4 + \cdots + b^{(r)} q^{2r}, \nonumber\\
\label{De13} Q(\rho) & = & (\rho-1) \left[ p_{r-1} (\rho)( b^{(1)}
+a/2) q^2 \right. \\ & + & \left. p_{r-2} (\rho) b^{(2)} q^4 +
\cdots + p_{r-s} (\rho) b^{(s)} q^{2s} + \cdots + b^{(r-1)}
q^{2(r-1)}\right], \nonumber
\end{eqnarray}
and
\begin{equation} \label{De14}
p_k (\rho) = 1 + \rho + \rho^2 +\cdots + \rho^{k-1}.
\end{equation}
As the operators $H_m$, $\widetilde{H}_m$, are unitary equivalent,
their gap parameters (\ref{De4}) coincide. The operators
$\widetilde{H}_m$ and $T(\rho)$, $\rho>0$ possess the properties
established by Proposition \ref{spectpn}. In particular, they have
the property (\ref{De3}) with one and the same $\gamma$. Therefore,
there exist $\varepsilon
>0$ and $k\in \mathbb{N}$ such that for $|\rho-1|< \varepsilon$, the
gap parameters (\ref{De4}) for  $\widetilde{H}_m$ and $T(\rho)$ are
defined by the first $k$ eigenvalues of these operators. As an
essentially self-adjoint operator, $T(\rho)$ possesses
a unique self-adjoint extension $\hat{T}(\rho)$, the eigenvalues of
which coincide with those of $T(\rho)$. Furthermore, for complex
$\rho$, $\hat{T}(\rho)$ is a closed operator, its domain
$Dom[\hat{T}(\rho)]$ does not depend on $\rho$. For every $\psi \in
Dom[\hat{T}(\rho)]$, the map $\mathbb{C} \ni \zeta  \mapsto
\hat{T}(\zeta) \psi \in L^2(\mathbb{R})$ is holomorphic. Therefore,
$\{\hat{T} (\rho)\ | \ |\rho - 1|< \varepsilon\}$ is a self-adjoint
holomorphic family. Hence, the eigenvalues $\Theta_n (\rho)$, $n \in
\mathbb{N}_0$ of $\hat{T}(\rho)$ are continuous functions of
$\rho\in (1-\varepsilon, 1+ \varepsilon)$, see Chapter VII, $\S$3 in
the book \cite{[Kato]}. At $\rho =1$ they coincide with those of
$\hat{H}_{m_0}$. Since we have given $k \in \mathbb{N}$ such that,
for all $\rho\in (1-\varepsilon, 1+ \varepsilon)$,
\[
\min_{n \in \mathbb{N}} \left[\Theta_n (\rho) - \Theta_{n-1}
(\rho)\right]=  \min_{n \in \{1, 2, \dots , k\} }\left[\Theta_n
(\rho) - \Theta_{n-1} (\rho)\right],
\]
the function
\begin{equation} \label{De14a}
\widetilde{\mathit{\Delta}} (\rho) \  \stackrel{\rm def}{=} \
\min_{n \in \mathbb{N}} \rho^{-r}\left[\Theta_n (\rho) -
\Theta_{n-1} (\rho)\right]
\end{equation}
is continuous. But by (\ref{De11})
\begin{equation} \label{De14b}
\mathit{\Delta}_m = \widetilde{\mathit{\Delta}}
\left(\left({m}/{m_0}\right)^{1/(r+1)}\right),
\end{equation}
which proves the continuity stated since $m_0>0$ has been chosen
arbitrarily.

To prove the second part of the theorem we rewrite (\ref{De12}) as
follows
\begin{equation} \label{De14c}
T(\rho) = H^{(0)}_{m_0} + R(\rho),
\end{equation}
where
\[
H^{(0)}_{m_0} = \frac{1}{2m_0} p^2 + b^{(r)}q^{2r},
\]
and
\[
R(\rho) = \rho\left( \rho^{r-2} (b^{(1)} + a/2) q^2 + \rho^{r-3}
b^{(2)} q^4 + \cdots + b^{(r-1)} q^{2(r-1)} \right).
\]
Repeating the above perturbation arguments one concludes that the
self-adjoint family $\{\hat{T}(\rho) \ | \ |\rho|< \varepsilon\}$ is
holomorphic at zero; hence, the gap parameter of (\ref{De14c})
tends, as $\rho \rightarrow 0$, to that of $H^{(0)}_{m_0}$, i.e., to
$\mathit{\Delta}_0$. Thereby, the asymptotics (\ref{De9}) for
$\mathit{\Delta}_m$ follows from (\ref{De11}) and the unitary
equivalence of $H_m$ and $\widetilde{H}_m$.
$\square$
\vskip.1cm Our second result in this domain is the quasi-classical
 analysis of the parameters
(\ref{De4}), (\ref{De5}). Here we shall suppose that the anharmonic
potential $V$ has the form (\ref{De8}) with $b^{(s)} \geq 0$ for all
$s=2, \dots, r-1$, c.f., (\ref{Iub}). We remind that in this case
the parameter $\vartheta_*>0$ is the unique solution of the equation
(\ref{Ip3}).
\begin{theorem} \label{gap2tm}
Let $V$ be as in (\ref{Iub}). Then the gap parameter
$\mathit{\Delta}_m$ and the quantum rigidity $\mathcal{R}_m$ of the
Hamiltonian (\ref{De7}) with such $V$ obey the estimates
\begin{equation} \label{De15}
\mathit{\Delta}_m \leq \frac{1}{2 m \vartheta_*}, \qquad
\mathcal{R}_m \leq \frac{1}{4 m \vartheta_*^2}.
\end{equation}
\end{theorem}
{\it Proof:}
Let $\varrho_m$ be the local Gibbs state (\ref{a4}) corresponding to
the Hamiltonian (\ref{De7}). Then by means of the inequality
(\ref{rp47b}) and the Gaussian upper bound  we get, see
(\ref{rp55}),
\[
a + 2 b^{(1)} + \mathit{\Phi} \left(\varrho_m (q^2) \right) \geq 0,
\]
by which
\begin{equation} \label{De16}
\varrho_m (q^2) \geq \vartheta_*.
\end{equation}
Let $\psi_n$, $n \in \mathbb{N}_0$ be the eigenfunctions of the
Hamiltonian $H_m$ corresponding to the eigenvalues $E_n$. By
Proposition \ref{spectpn}, to each $E_n$ there corresponds exactly
one $\psi_n$. Set
\[
Q_{nn'} = (\psi_n , q \psi_{n'})_{L^2(\mathbb{R})}, \quad n, n' \in
\mathbb{N}_0.
\]
Obviously, $Q_{nn} = 0$ for any $n\in \mathbb{N}_0$. Consider
\[
\Gamma (\tau, \tau') = \varrho_m \left[q \exp\left( - (\tau' -
\tau)H_m\right) q \exp\left( - (\tau - \tau')H_m\right)\right],
\quad \tau, \tau' \in [0,\beta],
\]
which is the Matsubara function corresponding to the state
$\varrho_m$ and the operators $F_1 = F_2 = q$.  Set
\begin{equation} \label{De17}
\hat{u} (k) = \int_0^\beta \Gamma (0, \tau) \cos k \tau {\rm d}\tau,
\qquad k \in \mathcal{K} = \{({2\pi}/{\beta}) \kappa \ | \kappa \in
\mathbb{Z}\}.
\end{equation}
Then
\begin{eqnarray} \label{De17a}
\hat{u}(k) & = & \frac{1}{Z_m} \sum_{n, n'=0}^{+\infty} \left\vert
Q_{nn'}\right\vert^2 \frac{E_n - E_{n'}}{k^2 + (E_n - E_{n'})^2} \\
& \times & \left\{ \exp (- \beta E_{n'}) - \exp (- \beta E_{n})
 \right\}, \nonumber
\end{eqnarray}
where $Z_m = {\rm trace} \exp(- \beta H_m)$. The term $(E_n -
E_{n'})^2$ in the denominator can be estimated by means of
(\ref{De4}), which yields
\begin{eqnarray} \label{De18}
\hat{u}(k) & \leq & \frac{1}{k^2 + \mathit{\Delta}_m^2}\cdot
\frac{1}{Z_m} \sum_{n, n'=0}^{+\infty} \left\vert
Q_{nn'}\right\vert^2 (E_n - E_{n'}) \\
& \times & \left\{ \exp (- \beta E_n) - \exp (- \beta E_{n'})
 \right\} \nonumber \\
 & \leq & \frac{1}{k^2 + \mathit{\Delta}_m^2}\cdot
\varrho_m\left(\left[q, \left[H_m, q\right] \right] \right)
\nonumber
\\ & = & \frac{1}{m(k^2 + \mathit{\Delta}_m^2)}. \nonumber
\end{eqnarray}
By this estimate we get
\begin{eqnarray} \label{De19}
\varrho_m (q^2) & = & \Gamma (0,0) = \frac{1}{\beta} \sum_{k \in
\mathcal{K}} u(k) \\ & \leq & \frac{1}{\beta} \sum_{k \in
\mathcal{K}} \frac{1}{m ( k^2 + \mathit{\Delta}_m^2)} = \frac{1}{2m
\mathit{\Delta}_m} \coth \left(\beta \mathit{\Delta}_m/2 \right).
\nonumber
\end{eqnarray}
Combining the latter estimate with (\ref{De16}) we arrive at
\[
\mathit{\Delta}_m \tanh \left(\beta \mathit{\Delta}_m/2 \right) < 1
/ (2 m \vartheta_*),
\]
which yields (\ref{De15}) in the limit $\beta \rightarrow + \infty$.
$\square$
\vskip.1cm Now let us analyze the quantum stability condition
(\ref{De6}) in the light of the latter results. The first conclusion
is that, unlike to the case of harmonic oscillators, this condition
can be satisfied for all $\hat{J}_0$ by letting the mass be small
enough. For the nearest-neighbor interaction, one has $\hat{J}_0 = 2
d J$; hence, if (\ref{De6}) holds, then
\begin{equation} \label{De20}
8 d m \vartheta_*^2 J < 1.
\end{equation}
This can be compared with the estimate
\begin{equation} \label{DeE}
8 d m \vartheta_*^2 J > d \mathcal{J}(d),
\end{equation}
guaranteeing a phase transition, which one derives from (\ref{De1}).

For finite $d$, $d\mathcal{J}(d) > 1$, see Proposition \ref{DLPpn};
hence, there is a gap between the latter estimate and (\ref{De20}),
which however diminishes as $d \rightarrow + \infty$ since
\[
\lim_{d \rightarrow + \infty}d \mathcal{J}(d) = 1.
\]
In the remaining part of this section, we show that for the quantum
crystals, both scalar and vector, a stability condition like
(\ref{De6}) yields a sufficient decay of the pair correlation
function. In the scalar case, this decay guaranties the uniqueness
of tempered Euclidean Gibbs measures. However, in the vector case it
yields a weaker result -- suppression of the long range order and of
the phase transitions of any order in the sense of Definition
\ref{landau}. The discrepancy arises from the fact that the
uniqueness criteria  based on the FKG inequalities are applicable to
scalar models only.

\subsection{Decay of correlations in the scalar case}
\label{7.2.1} In this subsection, we consider the model (\ref{U1}),
(\ref{U2}) which is (a) translation invariant; (b) scalar; (c) the
anharmonic potential is $V(q)=v(q^2)$ with $v$ being convex on
$\mathbb{R}_+$.

Let $\Lambda$ be the box (\ref{box}) and $\Lambda_*$ be its
conjugate (\ref{rp39}). For this $\Lambda$, let
\begin{equation}
\label{CF} K_{\ell\ell'}^\Lambda (\tau, \tau') \ \stackrel{\rm
def}{=} \ \big{\langle} \omega_{\ell} (\tau)
\omega_{\ell'}(\tau')\big{\rangle}_{\nu_\Lambda^{\rm per}}
\end{equation}
be the periodic correlation function. Recall that the periodic
interaction potential $J^\Lambda_{\ell \ell'}$ was defined by
(\ref{A2}). For the one-particle Hamiltonian (\ref{U2}), let
$\hat{u}(k)$ be as in (\ref{De17}).
\begin{theorem} \label{nagumo1}
Let the model be as just describes. If
\begin{equation} \label{De20a}
\hat{u}(0) \hat{J}_0 < 1,
\end{equation}
then
\begin{eqnarray} \label{De21}
K_{\ell\ell'}^\Lambda (\tau, \tau') \leq \frac{1}{\beta |\Lambda|}
\sum_{p\in \Lambda_*} \sum_{k\in \mathcal{K}} \frac{\exp\left[\imath
(p, \ell - \ell') + \imath k(\tau - \tau')\right]}{[\hat{u}(k)]^{-1}
- \hat{J}^\Lambda_0 + \mathit{\Upsilon}^\Lambda (p)},
\end{eqnarray}
where
\begin{equation} \label{De22}
\hat{J}^\Lambda_0 = \sum_{\ell'\in \Lambda}J^\Lambda_{\ell\ell'},
\quad \ \ \mathit{\Upsilon}^\Lambda (p) = \hat{J}^\Lambda_0 -
\sum_{\ell'\in \Lambda} J^\Lambda_{\ell\ell'} \exp[\imath (p , \ell
- \ell')].
\end{equation}
\end{theorem}
{\it Proof:}
Along with the periodic local Gibbs measure (\ref{A5}) we introduce
\begin{eqnarray} \label{De23}
& & \nu_{ \Lambda}^{\rm per}({\rm d} \omega_\Lambda|t)\qquad   \\
& & \ \quad = \frac{1}{N_{ \Lambda}^{\rm per}(t)}
\exp\left\{\frac{t}{2} \sum_{\ell , \ell'\in \Lambda}
J^\Lambda_{\ell\ell'} (\omega_\ell, \omega_{\ell'})_{L^2_\beta} -
\int_0^\beta \sum_{\ell \in \Lambda} V(\omega_\ell (\tau)){\rm
d}\tau \right\}\chi_{\Lambda}({\rm d}\omega_\Lambda), \nonumber
\end{eqnarray}
where $t\in [0,1]$ and $N_{ \Lambda}^{\rm per}(t)$ is the
corresponding normalization factor. Thereby, we set
\begin{equation} \label{De24}
X_{\ell \ell'} (\tau, \tau'|t) = \langle \omega_\ell (\tau)
\omega_{\ell'}(\tau')\rangle_{\nu_{ \Lambda}^{\rm per}(\cdot |t)},
\quad \ell ,\ell' \in \Lambda.
\end{equation}
By direct calculation
\begin{eqnarray} \label{De25}
& & \frac{\partial}{\partial t}X_{\ell \ell'} (\tau, \tau'|t) \\
& & \qquad = \frac{1}{2} \sum_{\ell_1 , \ell_2 \in \Lambda}
J^\Lambda_{\ell_1 \ell_2} \int_0^\beta R_{\ell \ell' \ell_1 \ell_2}
(\tau, \tau' , \tau'' , \tau''|t) {\rm d} \tau'' \nonumber\\ & & \qquad +
\sum_{\ell_1 , \ell_2 \in \Lambda} J^\Lambda_{\ell_1 \ell_2}
\int_0^\beta X_{\ell \ell_1} (\tau, \tau''|t) X_{\ell_2 \ell'}
(\tau'', \tau'|t) {\rm d}\tau'', \nonumber
\end{eqnarray}
where
\begin{eqnarray*}
R_{\ell_1 \ell_2 \ell_3 \ell_4} (\tau_1, \tau_2 , \tau_3 , \tau_4|t)
& = & \langle \omega_{\ell_1} (\tau_1)
\omega_{\ell_2}(\tau_2)\omega_{\ell_3}(\tau_3)\omega_{\ell_4}(\tau_4)\rangle_{\nu_{
\Lambda}^{\rm per}(\cdot |t)} \\ & - & \langle \omega_{\ell_1}
(\tau_1) \omega_{\ell_2}(\tau_2)\rangle_{\nu_{\Lambda}^{\rm
per}(\cdot |t)} \cdot \langle
\omega_{\ell_3}(\tau_3)\omega_{\ell_4}(\tau_4)\rangle_{\nu_{
\Lambda}^{\rm per}(\cdot |t)}  \nonumber \\ & - & \langle
\omega_{\ell_1} (\tau_1) \omega_{\ell_3}(\tau_3)\rangle_{\nu_{
\Lambda}^{\rm per}(\cdot |t)} \cdot \langle
\omega_{\ell_2}(\tau_2)\omega_{\ell_4}(\tau_4)\rangle_{\nu_{
\Lambda}^{\rm per}(\cdot |t)}  \nonumber \\ & - & \langle
\omega_{\ell_1} (\tau_1) \omega_{\ell_4}(\tau_4)\rangle_{\nu_{
\Lambda}^{\rm per}(\cdot |t)} \cdot \langle
\omega_{\ell_2}(\tau_2)\omega_{\ell_3}(\tau_3)\rangle_{\nu_{
\Lambda}^{\rm per}(\cdot |t)}.  \nonumber
\end{eqnarray*}
By the Lebowitz inequality, see \cite{[RevMF]}, we have
\begin{equation} \label{De26}
R_{\ell_1 \ell_2 \ell_3 \ell_4} (\tau_1, \tau_2 , \tau_3 , \tau_4|t)
\leq 0,
\end{equation}
holding
for all values of the arguments. Let us consider (\ref{De25}) as an
integro-differential equation subject to the initial condition
\begin{equation} \label{De27}
X_{\ell \ell'}(\tau , \tau'|0) = \delta_{\ell \ell'} \Gamma (\tau,
\tau') = (\delta_{\ell \ell'}/\beta) \sum_{k \in \mathcal{K}}
\hat{u}(k) \cos k(\tau - \tau').
\end{equation}
Besides, we also have
\begin{equation} \label{De28}
X_{\ell \ell'}(\tau , \tau'|1) = K_{\ell \ell'}^\Lambda(\tau ,
\tau'|p).
\end{equation}
Along with the Cauchy problem (\ref{De25}), (\ref{De27}) let us
consider the following equation
\begin{equation} \label{De29}
\frac{\partial}{\partial t} Y_{\ell \ell'} (\tau , \tau'|t) =
\sum_{\ell_1 , \ell_2\in \Lambda} \left[ J^\Lambda_{\ell_1 \ell_2}
+\frac{\varepsilon}{|\Lambda|}\right] \int_0^\beta Y_{\ell \ell_1}
(\tau , \tau'' |t) Y_{\ell_2 \ell'} (\tau'' , \tau'|t) {\rm d}
\tau'',
\end{equation}
where $\varepsilon >0$ is a parameter, subject to the initial
condition
\begin{eqnarray} \label{De30}
& & Y_{\ell \ell'}(\tau , \tau'|0)  =  X_{\ell \ell'}(\tau ,
\tau'|0)
\\& & \qquad \ \ = (\delta_{\ell
\ell'}/\beta) \sum_{k \in \mathcal{K}} \hat{u}(k) \cos k(\tau -
\tau'). \nonumber
\end{eqnarray}
Let us show that under the condition (\ref{De20a}) there exists
$\varepsilon_0 >0$ such that, for all $\varepsilon \in [0,
\varepsilon_0)$, the problem (\ref{De29}), (\ref{De30}), $t \in
[0,1]$, has the unique solution
\begin{equation} \label{De31}
Y_{\ell \ell'} (\tau , \tau'|t) = \frac{1}{\beta|\Lambda|}
\sum_{p\in \Lambda_*} \sum_{k \in \mathcal{K}} \frac{\exp\left[
\imath (p , \ell - \ell') + \imath k (\tau -
\tau')\right]}{[\hat{u}(k)]^{-1} - t [\hat{J}^\Lambda_0 +
\varepsilon \delta_{p,0}] + t \mathit{\Upsilon}^\Lambda (p)},
\end{equation}
where $\hat{J}_0$, $\mathit{\Upsilon}^\Lambda(p)$ are the same as in
(\ref{De22}) and $\delta_{p,0}$ is the Kronecker symbol with respect
to each of the components of $p$. By means of the Fourier
transformation
\begin{eqnarray} \label{De32}
\qquad \quad Y_{\ell \ell'} (\tau , \tau'|t) & = & \frac{1}{\beta
|\Lambda|}\sum_{p\in \Lambda_*} \sum_{k \in
\mathcal{K}}\widehat{Y}(p,k|t) \exp\left[\imath (p, \ell - \ell') +
\imath k (\tau - \tau') \right], \qquad \\
\widehat{Y}(p,k|t)  & = & \sum_{\ell' \in \Lambda} \int_0^\beta
Y_{\ell \ell'} (\tau , \tau'|t) \exp\left[-\imath (p, \ell - \ell')
- \imath k (\tau - \tau') \right] {\rm d} \tau', \nonumber
\end{eqnarray}
we bring (\ref{De29}), (\ref{De30}) into the following form
\begin{equation} \label{De33}
\frac{\partial}{\partial t} \widehat{Y} (p,k|t) = \left[
\hat{J}^\Lambda (p) + \varepsilon \delta_{p,0} \right]\cdot \left[
\widehat{Y} (p,k|t)\right]^2, \quad \widehat{Y} (p,k|0) =
\hat{u}(k),
\end{equation}
where, see (\ref{De22}),
\begin{equation} \label{De34}
\hat{J}^\Lambda(p) = \sum_{\ell'\in \Lambda} J^\Lambda_{\ell \ell'}
\exp\left[ \imath (p, \ell - \ell')\right] = \hat{J}^\Lambda_0 -
\mathit{\Upsilon}^\Lambda(p).
\end{equation}
Clearly, $\hat{J}^\Lambda_0 \leq \hat{J}_0$, $|\hat{J}^\Lambda(p)|
\leq  \hat{J}^\Lambda_0$, and $\hat{u}(k) \leq \hat{u}(0)$. Then in
view of (\ref{De20a}), one finds $\varepsilon_0
>0$ such that, for all $\varepsilon \in (0, \varepsilon_0)$, the
following holds
\[
\left[\hat{J}^\Lambda (p) + \varepsilon \delta_{p,0} \right]
\hat{u}(k) < 1,
\]
for all $p \in \Lambda_*$ and $k \in \mathcal{K}$. Thus, the problem
(\ref{De33}) can be solved explicitly, which via the transformation
(\ref{De32}) yields (\ref{De31}).

Given $\theta \in (0,1)$, we set
\begin{equation} \label{De35}
Y^{(\theta)}_{\ell \ell'} (\tau , \tau'|t) = Y_{\ell \ell'} (\tau ,
\tau'|t+\theta), \quad t \in [0, 1-\theta].
\end{equation}
Obviously, the latter function obeys the equation (\ref{De29}) on $t
\in [0, 1-\theta]$ with the initial condition
\begin{equation} \label{De36}
Y^{(\theta)}_{\ell \ell'} (\tau , \tau'|0) = Y_{\ell \ell'} (\tau ,
\tau'|\theta) > Y_{\ell \ell'} (\tau , \tau'|0) = X_{\ell \ell'}
(\tau , \tau'|0) .
\end{equation}
The latter inequality is due to the positivity of both sides of
(\ref{De29}). Therefore,
\begin{equation} \label{De37}
Y^{(\theta)}_{\ell \ell'} (\tau , \tau'|t) >0,
\end{equation}
for all $\ell , \ell'\in \Lambda$, $\tau , \tau' \in [0, \beta]$,
and $t \in [0, 1-\theta]$.

Let us show now that under the condition (\ref{De20a}), for all
$\theta \in (0, 1)$ and $\varepsilon \in (0, \varepsilon_0)$,
\begin{equation}\label{De38}
X_{\ell \ell'} (\tau , \tau'|t)< Y^{(\theta)}_{\ell \ell'} (\tau ,
\tau'|t),
\end{equation}
also for all $\ell , \ell'\in \Lambda$, $\tau , \tau' \in [0,
\beta]$, and $t \in [0, 1-\theta]$. To this end we introduce
\begin{equation} \label{De39}
Z_{\ell \ell'}^{\pm} (\tau , \tau'|t) \ \stackrel{\rm def}{=} \
Y^{(\theta)}_{\ell \ell'} (\tau , \tau'|t) \pm X_{\ell \ell'} (\tau
, \tau'|t), \quad t \in [0, 1-\theta].
\end{equation}
Then one has from (\ref{De25}), (\ref{De29})
\begin{eqnarray} \label{De40}
& & \frac{\partial}{\partial t} Z_{\ell \ell'}^{-} (\tau ,
\tau'|t)\\ & & \qquad = \frac{1}{2} \sum_{\ell_1 , \ell_2 \in
\Lambda}J^\Lambda_{\ell_1\ell_2} \int_0^\beta \left\{Z_{\ell
\ell_1}^{+} (\tau , \tau''|t) Z_{\ell' \ell_2}^{-} (\tau' ,
\tau''|t) \right. \nonumber \\ & & \qquad \left. + Z_{\ell
\ell_1}^{-} (\tau , \tau''|t) Z_{\ell' \ell_2}^{+} (\tau' ,
\tau''|t) \right\}{\rm d}\tau'' \nonumber\\
& & \qquad + \frac{\varepsilon}{|\Lambda|} \sum_{\ell_1 ,\ell_2\in
\Lambda} \int_0^\beta Y^{(\theta)}_{\ell \ell_1} (\tau , \tau''|t)
Y^{(\theta)}_{\ell' \ell_2} (\tau' , \tau''|t){\rm d}\tau'' -
S_{\ell \ell'}(\tau, \tau'|t), \nonumber
\end{eqnarray}
where $S_{\ell \ell'}(\tau, \tau'|t)$ stands for the first term on
the right-hand side of (\ref{De25}). By (\ref{De39}) and
(\ref{De36})
\begin{equation} \label{De41}
Z^{-}_{\ell \ell'}(\tau, \tau'|0) = Y_{\ell \ell'}(\tau,
\tau'|\theta) - X_{\ell \ell'}(\tau, \tau'|0) >0,
\end{equation}
which holds for all $\ell , \ell'\in \Lambda$, $\tau , \tau' \in [0,
\beta]$. For every $\ell , \ell'\in \Lambda$, both $Y_{\ell
\ell'}(\tau, \tau'|t)$, $X_{\ell \ell'}(\tau, \tau'|t)$ and, hence,
$Z^{\pm}_{\ell \ell'}(\tau, \tau'|t)$ are continuous functions of
their arguments.  Set
\begin{equation} \label{De42}
\zeta (t) = \inf \left\{ Z_{\ell \ell'}^{-} (\tau , \tau'|t)\ | \
\ell , \ell' \in \Lambda , \ \  \tau , \tau' \in [0, \beta]
\right\}.
\end{equation}
By (\ref{De41}), it follows that $\zeta (0) >0$. Suppose now that
$\zeta (t_0) =0$ at some $t_0 \in [0, 1-\theta]$ and $\zeta (t) >0$
for all $t \in [0, t_0)$. Then by the continuity of $Z^{-}_{\ell
\ell'}$, there exist $\ell , \ell'\in \Lambda$ and $\tau , \tau' \in
[0, \beta]$ such that
\[
Z^{-}_{\ell \ell'}(\tau , \tau'|t_0) = 0 \quad \ \ {\rm and} \quad
Z^{-}_{\ell \ell'}(\tau , \tau'|t) > 0 \quad \ {\rm for} \ \ {\rm
all} \ t < t_0.
\]
For these $\ell , \ell'\in \Lambda$ and $\tau , \tau' \in [0,
\beta]$, the derivative $(\partial / \partial t) Z^{-}_{\ell
\ell'}(\tau , \tau'|t)$ at $t=t_0$ is positive since on the
right-hand side of (\ref{De40}) the third term is positive and the
remaining terms are non-negative. But a differentiable function,
which is positive at $t\in [0, t_0)$ and zero at $t=t_0$, cannot
increase at $t=t_0$. Thus, $\zeta (t) >0$ for all $t\in [0,
1-\theta]$, which yields (\ref{De38}). By the latter estimate, we
have
\begin{eqnarray*}
& & X_{\ell \ell'} (\tau , \tau'|1-\theta) <  Y_{\ell \ell'} (\tau ,
\tau'|1) \\ & & \qquad = \frac{1}{\beta|\Lambda|} \sum_{p\in
\Lambda_*} \sum_{k \in \mathcal{K}} \frac{\exp\left[ \imath (p ,
\ell - \ell') + \imath k (\tau - \tau')\right]}{[\hat{u}(k)]^{-1} -
t [\hat{J}^\Lambda_0 + \varepsilon \delta_{p,0}] + t
\mathit{\Upsilon}^\Lambda (p)}.
\end{eqnarray*}
All the function above depend on $\theta$ and $\varepsilon$
continuously. Hence, passing here to the limit $\theta = \varepsilon
\downarrow 0$ and taking into account (\ref{De28}) we obtain
(\ref{De21}).
$\square$
\vskip.1cm

By means of Proposition \ref{periodtm}, the result just proven can
be extended to all periodic elements of $\mathcal{G}^{\rm t}$. For
$\mu \in \mathcal{G}^{\rm t}$, we set
\begin{equation}
\label{De43} K^\mu_{\ell \ell'} (\tau, \tau') = \big{\langle}
\omega_{\ell} (\tau) \omega_{\ell'}(\tau')\big{\rangle}_{\mu}.
\end{equation}
\begin{theorem} \label{nagumo2}
Let the stability condition (\ref{De6}) be satisfied. Then for every
periodic $\mu \in \mathcal{G}^{\rm t}$, the correlation function
(\ref{De43}) has the bound
\begin{eqnarray} \label{De44}
K^\mu_{\ell \ell'} (\tau, \tau')& \leq & Y_{\ell \ell'}(\tau ,
\tau') \\ & \stackrel{\rm def}{=} & \frac{1}{\beta (2 \pi)^d}\sum_{k
\in \mathcal{K}} \int_{(-\pi, \pi]^d} \frac{\exp\left[\imath (p ,
\ell - \ell') + \imath k(\tau - \tau') \right]}{[\hat{u}(k)]^{-1} -
\hat{J}_0 + \mathit{\Upsilon}(p)} {\rm d} p , \nonumber
\end{eqnarray}
where
\begin{equation} \label{De45}
\mathit{\Upsilon} (p) = \hat{J}_0 - \sum_{\ell'}J_{\ell \ell'}
\exp[\imath (p , \ell - \ell')], \quad p \in (-\pi, \pi]^d.
\end{equation}
 The same bound has also the correlation
function $K^{\mu_0}_{\ell \ell'}(\tau, \tau')$, where $\mu_0\in
\mathcal{G}^{\rm t}$ is the same as in Proposition \ref{MA1tm}.
\end{theorem}
\begin{remark} \label{nagumor}
By (\ref{De18}), $[\hat{u}(k)]^{-1} \geq m(\mathit{\Delta}_m^2 +
k^2)$. The upper bound in (\ref{De44}) with $[\hat{u}(k)]^{-1}]$
replaced by $m ([\mathit{\Delta}^{\rm har}]^2 + k^2)$ turns into the
infinite volume correlation function for the quantum harmonic
crystal discussed at the beginning of subsection \ref{stabcr}. Thus,
under the condition (\ref{De20a})  the decay of the correlation
functions in the periodic states  is not less than it is in the
stable quantum harmonic crystal. As we shall see in the next
subsection, such a decay stabilizes also anharmonic ones.
\end{remark}

For $\mathit{\Upsilon} (p) \sim \mathit{\Upsilon}_0 |p|^2$,
$\mathit{\Upsilon}_0 >0$, as $p \rightarrow 0$, the asymptotics of
the bound in (\ref{De44}) as $\sqrt{|\ell - \ell'|^2 + |\tau -
\tau'|^2} \rightarrow + \infty$ will be the same as for the $d+1$-dimensional free field,
which is well known, see claim (c) of Proposition 7.2.1, page 162 of
\cite{[GJ]}. Thus, we have the following
\begin{proposition} \label{nagumo3}
If the function (\ref{De45}) is such that $\mathit{\Upsilon} (p)
\sim \mathit{\Upsilon}_0 |p|^2$, $\mathit{\Upsilon}_0 >0$, as $p
\rightarrow 0$, the upper bound in (\ref{De44}) has an exponential
spacial decay.
\end{proposition}

\subsection{Decay of correlations in the vector case}

\label{vectc}

In the vector case, the eigenvalues of the Hamiltonian (\ref{De7})
are no longer simple; hence, the parameter (\ref{De4})  definitely
equals zero. Therefore, one has to pick another parameter, which can
describe the quantum rigidity in this case. If the model is rotation
invariant, its dimensionality $\nu$ is just a parameter. Thus, one
can compare the stability of such a model with the stability of the
model with $\nu=1$. This approach was developed in \cite{KozZ}, see
also \cite{[RevMF],[Kargol]}. Here we present the most general
result in this domain, which  is then used to study the quantum
stabilization in the vector case.

We begin by  introducing the corresponding class of functions. A
function $f:\mathbb{R}\rightarrow \mathbb{R}$ is called polynomially
bounded if $f(x) / (1 + |x|^k)$ is bounded for some
$k\in\mathbb{N}$. Let $\mathcal{F}$ be the set of continuous
polynomially bounded $f:\mathbb{R}\rightarrow \mathbb{R}$ which are
either odd and increasing or even and positive.
\begin{proposition} \label{sdpn}
Suppose that the model is rotation invariant and for all $\ell \in
\Lambda$, $\Lambda \Subset \mathbb{L}$, $V_\ell (x) = v_\ell
(|x|^2)$ with $v_\ell$ being convex on $\mathbb{R}_+$. Then for any
$\tau_1 , \dots, \tau_n\in [0, \beta]$, $\ell_1 , \dots , \ell_n \in
\Lambda$, $j = 1 , \dots , \nu$, $f_1 , \dots f_n \in \mathcal{F}$,
\begin{equation} \label{SD}
\langle f_1 (\omega^{(j)}_{\ell_1} (\tau_1)) \cdots f_n
(\omega^{(j)}_{\ell_n} (\tau_n)) \rangle_{\nu_\Lambda}   \leq
\langle f_1 (\omega_{\ell_1} (\tau_1)) \cdots f_n (\omega_{\ell_n}
(\tau_n)) \rangle_{\tilde{\nu}_\Lambda}   ,
\end{equation}
where $\tilde{\nu}_\Lambda$ is the Euclidean Gibbs measure
(\ref{a27}) of the scalar model with the same $J_{\ell\ell'}$ as the
model considered and with the anharmonic potentials $V_\ell (q) =
v_\ell (q^2)$.
\end{proposition}
By this statement one immediately gets the following fact.
\begin{theorem} \label{sdtm}
Let the model be translation invariant and such as in Proposition
\ref{sdpn}. Let also $\mathit{\Delta}_m$ be the gap parameter
(\ref{De4}) of the scalar model with the same interaction
intensities $J_{\ell\ell'}$ and with the anharmonic potentials $V
(q) = v(q^2)$. Then if the stability condition (\ref{De6}) is
satisfied, the longitudinal correlation function
\begin{equation} \label{SD0}
K^\mu_{\ell \ell'} (\tau, \tau') = \langle
\omega^{(j)}_{\ell'}(\tau)\omega^{(j)}_{\ell}(\tau')\rangle_\mu ,
\quad \ \ j = 1, 2, \dots , \nu,
\end{equation}
corresponding to any of the periodic states $\mu\in \mathcal{G}^{\rm
t}$, as well as to any of the accumulation points of the family
$\{\pi_\Lambda (\cdot |0)\}_{\Lambda \Subset \mathbb{L}}$, obeys the
estimate (\ref{De44}) in which $\hat{u}(k)$ is calculated according
to (\ref{De17a}) for the one-dimensional anharmonic oscillator of
mass $m$ and the anharmonic potential $v(q^2)$.
\end{theorem}

\subsection{Suppression of phase transitions}
\label{sps}

From the `physical'  point of view, the decay of correlations
(\ref{De44}) already corresponds to the lack of any phase
transition. However, in the mathematical theory, one should show
this as a mathematical fact basing on the definition of a phase
transition. The most general one is Definition \ref{phdef} according
to which the suppression of phase transitions corresponds to the
uniqueness of tempered Euclidean Gibbs states. Properties like the
differentiability of the pressure, c.f., Definition \ref{landau}, or
the lack of the order parameter, see Definition \ref{rppdf}, may
also indicate the suppression of phase transitions, but in a weaker
sense. The aim of this section is to demonstrate that the decay of
correlations  caused by the quantum stabilization yields the
two-times differentiability of the pressure, which in the scalar
case yields the uniqueness. This result is then extended to the
models which are not necessarily translation invariant.

In the scalar case, the most general result is the following
statement, see Theorem 3.13 in \cite{[KoT]}.
\begin{theorem} \label{7.1tm}
Let  the anharmonic potentials $V_\ell$ be even and such that there
exists a convex function $v:\mathbb{R}_+ \rightarrow \mathbb{R}$,
such that, for any $V_\ell$,
\begin{equation} \label{De51}
V_\ell (x_\ell) - v(x_\ell^2) \leq V_\ell (\tilde{x}_\ell) -
v(\tilde{x}_\ell^2) \quad {\rm whenever} \ \ x_\ell^2 <
\tilde{x}_\ell^2.
\end{equation}
For such $v$, let $\mathit{\Delta}_m$ be the gap parameter of the
one-particle Hamiltonian (\ref{U1}) with the anharmonic potential
$v(q^2)$. Then the set of tempered Euclidean Gibbs measures of this
model is a singleton if the stability condition (\ref{De6})
involving $\mathit{\Delta}_m$ and the interaction parameter
$\hat{J}_0$ of this model is satisfied.
\end{theorem}
The proof of this theorem is conducted by comparing the model with
the translation invariant reference model with  the anharmonicity
potential $V(q) = v(q^2)$. By Proposition \ref{MAtm}, for  the model
considered and the reference model, there exist maximal elements,
$\mu_+$ and $\mu_{+}^{\rm ref}$, respectively. By means of the
symmetry $V_\ell (q) = V_{\ell}(-q)$ and the FKG inequality, one
proves that, for both models, the uniqueness  occurs if
\begin{equation} \label{SD1}
\langle \omega_\ell (0) \rangle_{\mu^{\rm ref}_{+}} = 0, \qquad \ \
\langle \omega_\ell (0) \rangle_{\mu_{+}} = 0, \quad \ \ {\rm for} \
\ {\rm all} \  \ \ell.
\end{equation}
By  the GKS inequalities, the condition (\ref{De51}) implies
\begin{equation} \label{SD2}
0 \leq \langle \omega_\ell (0) \rangle_{\mu_{+}} \leq \langle
\omega_\ell (0) \rangle_{\mu^{\rm ref}_{+}},
\end{equation}
which means that the reference model is less stable with respect to
the phase transitions than the initial model. The reference model is
translation invariant. By means of a technique employing this fact,
one proves that the decay of correlations in the reference model
which occurs under the stability condition (\ref{De6}) yields,
see Theorem \ref{nagumo1},
\[
\langle \omega_\ell (0) \rangle_{\mu^{\rm ref}_{+}} = 0,
\]
and therefrom (\ref{SD1}) by (\ref{SD2}). The details can be found
in \cite{[KoT]}.

As was mentioned above, in the vector case we did not manage to
prove that the decay of correlations implies the uniqueness. The
main reason for this is that the proof of Theorem \ref{7.1tm} was
based on the FKG inequality, which can be proven for scalar models
only. In the vector case, we get a weaker result, by which the decay
of correlations yields the normality of thermal fluctuations. To
this end we introduce \emph{the fluctuation operators}
\begin{equation} \label{De64}
Q^{(j)}_\Lambda = \frac{1}{\sqrt{|\Lambda|}} \sum_{\ell \in \Lambda}
q^{(j)}_\ell, \qquad  \Lambda \Subset \mathbb{L}, \ \ \ j =
1, \dots, \nu.
\end{equation}
Such operators correspond to \emph{normal fluctuations}.
\begin{definition} \label{normalf}
The fluctuations of the displacements of oscillators are called
normal if the Matsubara functions (\ref{a9}) for the operators $F_1=
Q^{(j_1)}, \dots , F_n = Q^{(j_n)}$, remain bounded as $\Lambda
\nearrow \mathbb{L}$.
\end{definition}
If $\Lambda$ is a box, the parameter (\ref{gri6}) can be written
\begin{equation} \label{De65}
P^{(\alpha)}_\Lambda = \frac{1}{\beta^2
|\Lambda|^\alpha}\sum_{j=1}^\nu \int_0^\beta\int_0^\beta
\Gamma^{\beta , \Lambda}_{Q^{(j)}_\Lambda, Q^{(j)}_\Lambda} (\tau ,
\tau') {\rm d} \tau {\rm d}\tau'.
\end{equation}
Thus, if the fluctuations are normal, phase transitions of the
second order (and all the more of the first order) do not occur.

Like in the proof of Theorem \ref{sdtm}, the model is compared with the scalar ferromagnetic model with the same
mass and the anharmonic potential $v(q^2)$. Then the gap parameter
$\mathit{\Delta}_m$ is the one calculated
for the latter model.
\begin{theorem} \label{nagumo5}
Let the model be the same as in Theorem \ref{sdtm} and let the
stability condition involving the interaction parameter $\hat{J}_0$
of the model and the gap parameter $\mathit{\Delta}_m$ corresponding
to its scalar analog be satisfied. Then the fluctuations of the
displacements of the oscillators remain normal at all temperatures.
\end{theorem}

\subsection{Comments}
\begin{itemize}

\item \emph{Subsection \ref{stabcr}:} In an ionic crystal, the ions usually
form massive complexes the dynamics of which determine the physical
properties of the crystal, including its instability with respect to
structural phase transitions, see \cite{[BC]}. Such massive
complexes can be considered as classical particles; hence, the phase
transitions are described in the framework of classical statistical
mechanics. At the same time, in a number of ionic crystals
containing localized light ions certain aspects of the phase
transitions are apparently unusual from the point of view of
classical physics. Their presence can only be explained in a
quantum-mechanical context, which points out on the essential role
of the light ions. This influence of the quantum effects on the
phase transition was detected experimentally already in the early
1970's. Here we mention the data presented in \cite{[Blinc],[12]} on
the KDP-type ferroelectrics and in \cite{[KMueller]} on the
YBaCuO-type superconductors. These data were then used for
justifying the corresponding theoretical models and tools of their
study. On a theoretical level, the influence of quantum effects on
the structural phase transitions in ionic crystals was first
discussed in the paper \cite{[9]}, where the particle mass was
chosen as the only parameter responsible for these effects. The
conclusion, obtained there was that the long range order, see
Definition \ref{rppdf}, gets impossible at all temperatures if the
mass is sufficiently small. Later on, a number of rigorous studies
of quantum effects inspired by this result as well as by the
corresponding experimental data have appeared, see
\cite{[Minlos],[VZ]} and the references therein. Like in \cite{[9]},
in these works the reduced mass (\ref{In}) was the only parameter
responsible for the effects. The result obtained was that the long
range order is suppressed at all temperatures in the light mass
limit $m\rightarrow 0$. Based on the study of the quantum crystals
performed in \cite{[AKK1],[AKK2],[AKKR],[CRAS],[AKKRNN]}, a
mechanism of quantum effects leading to the stabilization against
phase transitions was proposed, see \cite{[AKKRPRL]}.

\item \emph{Subsection \ref{4.2.ss}:} According to \cite{[AKKRPRL]}
the key parameter responsible for the quantum stabilization  is
$\mathcal{R}_m = m\mathit{\Delta}_m^2$, see (\ref{De5}). In the
harmonic case, $m \mathit{\Delta}_m^2$ is merely the oscillator
rigidity and the stability of the crystal corresponds to large
values of this quantity. That is why the parameter $m
\mathit{\Delta}_m^2$ was called quantum rigidity and the effect was
called quantum stabilization. If the tunneling between the wells
gets more intensive (closer minima), or if the mass diminishes, $m
\mathit{\Delta}_m^2$ gets bigger and the particle `forgets' about
the details of the potential energy in the vicinity of the origin
(including instability) and oscillates as if its equilibrium at zero
is stable, like in the harmonic case.

\item \emph{Subsection \ref{gap}:} Theorems \ref{gap1tm} and \ref{gap2tm} are new.
 Preliminary
results of this kind were obtained  in \cite{[AKK2],[Koz4]}.

\item \emph{Subsection \ref{7.2.1}:} Theorems \ref{nagumo1}, \ref{nagumo2}, \ref{nagumo3}
were proven in \cite{[KK]}.

\item \emph{Subsection \ref{vectc}:} Various scalar domination estimates were obtained in
\cite{[KDres],Koz,KozZ}.

\item \emph{Subsection \ref{sps}:} Theorem \ref{7.1tm} was proven in \cite{[KoT]}.
The proof of Theorem \ref{nagumo5} was done in \cite{KozZ}.
The suppression of abnormal fluctuations in the hierarchical version
of the model (\ref{U1}), (\ref{U2}) was proven in \cite{[AKK1]}.

\end{itemize}

\section*{Acknowledgments}
The authors are grateful to M. R\"ockner and T. Pasurek for valuable
discussions. The financial support by the DFG through the project
436 POL 113/115/0-1 and through SFB 701 ``Spektrale Strukturen und
topologische Methoden in der Mathematik" is cordially acknowledged.
A. Kargol is grateful for the support by the KBN under the  Grant N N201 0761 33.


\end{document}